\documentclass[a4paper,11pt]{article}
\pdfoutput=1 

\usepackage{jheppub} 
\usepackage[T1]{fontenc} 

\usepackage{graphicx,multirow,subfigure}
\usepackage{float}
\usepackage{bm}
\usepackage{amsmath}
\usepackage{amssymb}
\usepackage{amscd}
\usepackage{latexsym}
\usepackage{slashed}
\usepackage{color}
\usepackage{graphicx}
\usepackage{ulem}
\usepackage{color}
\usepackage{verbatim}

\def\bea{\begin{eqnarray}}
\def\eea{\end{eqnarray}}
\def\nn{\nonumber}

\title{\centering \boldmath $Z'$, Higgses and heavy neutrinos in  $U(1)'$ models:\\
from the LHC to the GUT scale}

\author[a]{Elena Accomando,}
\author[b,c]{Claudio Corian\`o}
\author[a]{Luigi Delle Rose,}
\author[a]{Juri Fiaschi,}
\author[c]{Carlo Marzo}
\author[a]{and Stefano Moretti}

\affiliation[a]{School of Physics and Astronomy, University of Southampton, \\Highfield, Southampton SO17 1BJ, UK}
\affiliation[b]{STAG Research Centre and Mathematical Sciences, University of Southampton, \\Highfield, Southampton SO17 1BJ, UK}
\affiliation[c]{Dipartimento di Matematica e Fisica ``Ennio De Giorgi'', Universit\`a del Salento and INFN-Lecce, \\Via Arnesano, 73100 Lecce, IT}

\emailAdd{E.Accomando@soton.ac.uk}
\emailAdd{Claudio.Coriano@le.infn.it}
\emailAdd{L.Delle-Rose@soton.ac.uk}
\emailAdd{Juri.Fiaschi@soton.ac.uk}
\emailAdd{Carlo.Marzo@le.infn.it}
\emailAdd{S.Moretti@soton.ac.uk}

\abstract{\noindent
We study a class of 
non-exotic minimal $U(1) '$ extensions of the Standard Model, which includes all scenarios that are anomaly-free with the 
ordinary fermion content augmented by one Right-Handed neutrino per generation, wherein the new Abelian gauge
group is spontaneously broken by the non-zero Vacuum Expectation Value  of an additional Higgs singlet field,
in turn providing mass to a $Z'$ state. By adopting the $B-L$ example,  whose results can be recast
into those pertaining to the whole aforementioned class, and allowing for both scalar and gauge
mixing, we first extract the surviving parameter space in presence
of up-to-date theoretical and experimental constraints. Over the corresponding parameter configurations, we
 then delineate the high energy behaviour of such constructs in terms of their stability and perturbativity. Finally,
we highlight key production and decay channels of the new states entering the spectra of this class of models, i.e.,
heavy neutrinos, a second Higgs state and the $Z'$, which are amenable to experimental investigation at the Large Hadron
Collider. We therefore set the stage to establish a direct link between measurements obtainable at the 
Electro-Weak scale and the dynamics of the underlying model up to those where a Grand Unification Theory embedding a
$U(1)'$ can be realised.
}

\begin{document} 
\maketitle
\flushbottom

\section{Introduction}
\label{sec:introduction}

The evolution of the three Standard Model (SM) gauge couplings through the Renormalisation Group (RG) equations shows a remarkable convergence, although only approximate, around $10^{15}$ GeV. This feature is even more evident in a supersymmetric context and represents one of the most solid hints in favour of a Grand Unification Theory (GUT). One of the main predictions of such theories is the appearance of an extra $U(1)'$ gauge symmetry which can be broken at energies accessible at the Large Hadron Collider (LHC). If this new Abelian group is broken around the TeV scale, the associated neutral gauge boson $Z'$ could be observed as a resonance in the di-lepton channel or elsewhere. 
The search for extra neutral currents at the LHC is for sure one of the most 
important topics of current interest and  is expected to be so also in the future. In fact, stronger and stronger exclusion limits on the mass of an hypothetical 
extra neutral gauge boson, which have been and are being generated by the large amount of data collected at Run 1 and Run 2 by the ATLAS and CMS collaborations, do not actually rule out the possibility of discovering such neutral currents altogether. This is due to the structure of the parameter space of the class of $U(1)'$ models embedding these states which, even in their most economical form, can easily comply with experimental constraints. \\
There are several realisations of GUTs that may predict $Z'$ bosons (see, for instance, \cite{Langacker:1980js,Hewett:1988xc}), such as  string theory motivated $E_6$, $SO(10)$ and Left-Right (LR) symmetric models. In fact, $Z'$ bosons may also appear in effective theories from string compactifications \cite{Faraggi:1990ita,Faraggi:2015iaa,Faraggi:2016xnm} or from Kaluza-Klein constructions \cite{Randall:1999ee}. Each of these is characterised by distinct mass spectra, scalar potential and  flavour structures. 
Even in simpler bottom-up approaches, where the issue of the unification of the underlying gauge symmetry into groups of higher rank is not addressed, several possibilities remain wide open and allow the search to continue. In more general scenarios based on GUTs, however, experimental searches must confront with a larger number of allowed gauge bosons, with the appearance of several scales of sequential breakings of the gauge symmetry and, last but not least, of new matter belonging to higher representations. 

In this work we address a class of minimal models which are the most economical Abelian and renormalisable extension of the SM with only few additional free parameters \cite{Appelquist:2002mw,Chankowski:2006jk,Langacker:2008yv,Erler:2009jh,Salvioni:2009mt}. 
At the Electro-Weak (EW) scale, these Abelian extensions of the SM, in which the gauge symmetry is described by the $G_{\rm SM} \times U(1)'\equiv (SU(3)_C\times SU(2)_L\times U(1)_Y)\times U(1)'$ group, are also characterised by a new complex scalar field, heavier than the SM Higgs, whose Vacuum Expectation Value (VEV) can lie in the TeV range and provides the mass for the $Z'$. Moreover, an enlarged flavour sector is also present.
Indeed, in this class of models, the cancellation of the $U(1)'$ gauge and gravitational anomalies naturally predicts right-handed neutrinos \cite{Appelquist:2002mw}. The same heavy scalar develops, thorough its spontaneous symmetry breaking, a Majorana mass at the TeV scale for the SM singlet fermions, dynamically realising a low-scale seesaw mechanism. We consider three right-handed neutrinos, one for each family, as predicted by an anomaly free and flavour universal Abelian extension. See \cite{Martinez:2013qya,Martinez:2014ova,Martinez:2014rea,Martinez:2015wrp} for some non-universal examples.

The special case of a ``pure'' $U(1)_{B-L}$ extension, where the conserved charge of the extra Abelian symmetry is the $B-L$ number, with $B$ and $L$ the baryon and lepton numbers, respectively, has been extensively scrutinised in the literature, see \cite{Khalil:2007dr,Basso:2008iv,Basso:2010hk,Basso:2010pe,Basso:2010yz,Basso:2010jm,Accomando:2010fz,Basso:2011na,Basso:2012sz,Basso:2012ux,Accomando:2013sfa,Accomando:2015cfa,Accomando:2015ava, Okada:2016gsh}. 
Here, we surpass these previous studies in several directions. 

Firstly, the $B-L$ model setup therein is 
 characterised by the condition of vanishing mixing $\tilde g$ between the two $U(1)_Y$ and $U(1)_{B-L}$ gauge groups at the EW scale, an assumption that is relaxed in our analysis. Indeed, in the framework discussed in this work, an additional $U(1)'$ factor can always be described by a linear combination of the hypercharge and of the $B-L$ quantum number, with coefficients parameterised by the new Abelian gauge coupling $g'_1$ and the mixing $\tilde g$. Therefore, choosing the $U(1)_{B-L}$ as a reference gauge group extension, we can explore an entire class of minimal Abelian models through the ratio of the gauge couplings $\tilde g / g'_1$.
Furthermore, the addition of the mixing parameter allows to explore the phenomenology of different scenarios which are characterised by distinctive features with respect to the ``pure'' $B-L$ case. For instance, new decay channels for the $Z'$ will be opened which, albeit being interesting on their own, will increase the size of the $Z'$ total width and, therefore, the effect of the interference term in the Drell-Yan (DY) process \cite{Accomando:2013sfa}, ultimately resulting in significantly altered experimental
constraints. 

Secondly, we will establish a direct connection between the yield of a variety of possible LHC measuments and the high scale structure
of our construction, thereby directing experimental investigation towards key analyses enabling one to make an assessment
of the model stability based on their low energy spectra. Both the enlarged boson and fermion content of our $B-L$
setup afford one with a variety of specific signatures. For example, 
the extended flavour sector allows for the possibility of a $Z'$ decaying into long-lived heavy neutrinos with very clear multi-leptonic signatures \cite{Basso:2008iv, Kang:2015uoc}, which highlight the striking features of these minimal Abelian extensions providing an optimal channel, in addition to the di-lepton one, for their identification and characterisation. 
Furthermore, 
the presence of an heavy scalar, which can mix with the SM Higgs boson, offers the possibility \cite{Basso:2010yz} to
search for its decay into a pair of heavy neutrinos,  which is a completely novel signature
with respect to many beyond the SM scenarios. Then, the non-zero scalar mixing angle also provides the decay of a heavier Higgs boson state into two light ones, which represents a unique way to probe the scalar sector and the mechanism of spontaneous EW Symmetry Breaking (EWSB).

In essence, with extended scalar, gauge and flavour sectors, it is natural to ask whether the vacuum of the this class of models is in a stable configuration both at the classical and at the quantum level and what is the impact of the new physics on the SM EW ground state \cite{Basso:2010yz,Basso:2010jm,Basso:2013vla,Datta:2013mta,Chakrabortty:2013zja,Coriano:2014mpa,DiChiara:2014wha,Oda:2015gna,Das:2015nwk,Coriano:2015sea,Das:2016zue}. 
Indeed, the extrapolation of the SM to high energy scales, through the RG equations, exhibits a scalar potential with a 
non-trivial structure:  its minimum does not correspond to the EW vacuum which is found to be in a metastable configuration, very close to a phase transition \cite{Bezrukov:2012sa,Buttazzo:2013uya}. This scenario, dubbed near-criticality, has been clarified by the Higgs discovery and by the measurement of its mass at the LHC Run 1. For the measured value of the Higgs mass, the scalar quartic coupling is driven, during the RG evolution, to a negative value by the Yukawa coupling of the top quark. This implies that the EW vacuum is only a local minimum and can possibly decay into the true ground state through  quantum tunnelling. Fortunately, this dangerous situation is safely avoided because the lifetime of our vacuum is found to be much larger than the age of the Universe. This result, however, critically depends on the pole mass of the top quark and on its extraction from experimental measurements \cite{Masina:2012tz,Alekhin:2012py,Bezrukov:2014ina}. 
This picture strictly holds under the {\it big desert} condition, namely, the absence of significant new physics effects between the EW and Planck scales and it is therefore natural to ask which are the variations induced by new degrees of freedom, such as those embedded in our $B-L$ model. Clearly, an extra Abelian symmetry with an augmented scalar and flavour sectors allows one to conceive the possibility of new scenarios in which the issue of stability is resolved and the criticality due to the top mass fades away.

In this analysis we will enforce the requirements of stability of the EW vacuum and perturbativity of the couplings through the RG evolution in order to identify the allowed regions in the parameter space of these minimal Abelian extensions.  For this purpose we will adopt two-loop RG equations equipped with one-loop matching conditions at the EW scale \cite{Coriano:2015sea}. These results will be combined with bounds from di-lepton analyses at the LHC at 8 TeV and with exclusion limits from Higgs searches, therefore delineating the viable parameter space from both a phenomenological perspective and its theoretical consistency. Ultimately, we will assess which of the aforementioned LHC signatures ought to be accessed and when, so as to establish a direct connection between accessible EW scale spectra and a potential underlying GUT structure.

The work is organised as follows. In section \ref{sec:model} we briefly introduce the model and comment on its main aspects. Then, in sections \ref{sec:bounds} and \ref{sec:rg} we look at current experimental bounds and how they impinge on the RG analysis, respectively. 
In section \ref{sec:pheno1} we discuss the most relevant phenomenological features of our $B-L$ scenario at the LHC. Finally, we draw our conclusions in section \ref{sec:conclusions}.

\section{The model}
\label{sec:model}

The model under study is described by the SM gauge group augmented by a single Abelian gauge factor $U(1)'$ which we choose to be $U(1)_{B-L}$.
We investigate a minimal renormalisable Abelian extension of the SM with the only minimal matter content necessary to ensure the consistency of the theory.
In practise, in order to satisfy the cancellation of the gauge and the gravitational anomalies, we enlarge the fermion spectrum by a Right-Handed (RH) neutrino, one for each generation (we assume universality between the flavour families), which has $B-L = -1$ charge and is singlet under the SM gauge group. 
Concerning the scalar sector, in addition to the SM-like Higgs doublet $H$, we introduce a complex scalar field $\chi$ to achieve the spontaneous breaking of the extra Abelian symmetry. The new scalar field is a SM singlet with charge $B-L = 2$ and its vacuum expectation value $x$, which we choose in the TeV range, provides the mass to the $Z'$ gauge boson and to the RH neutrinos. The latter acquire a Majorana mass through the Yukawa interactions and thus naturally implement the type-I seesaw mechanism. More complex matter configurations, usually characterised by extra fermionic degrees of freedom or by non-universal charge assignments, can also be realised but require, consequently, a different charge assignment for the extra complex scalar.

The details of the model can be found in \cite{Basso:2008iv,Coriano:2015sea}, here we restrict the discussion on its salient features. 
The most general kinetic Lagrangian for the two Abelian gauge fields, allowed by  gauge invariance, admits a kinetic mixing between the corresponding field strengths
\bea
\mathcal L = - \frac{1}{4} F^{\mu\nu}F_{\mu\nu}   - \frac{1}{4} F'^{\mu\nu}F'_{\mu\nu}  - \frac{\kappa}{2} F^{\mu\nu}F'_{\mu\nu} \,.
\eea
By a suitable redefinition of the gauge fields it is possible to remove the kinetic mixing $\kappa$, thus recasting the kinetic Lagrangian into a canonical form. In this basis the Abelian part of the covariant derivative acting on the charged field is 
non-diagonal and reads as
\bea
\mathcal D_\mu = \partial_\mu + i g_1 \, Y \, B_\mu + i ( \tilde g \, Y + g'_1 \, Y_{B-L} ) B'_\mu + \ldots,
\eea
where $B_\mu$ and $B'_\mu$ are the gauge fields of the $U(1)_Y$ and $U(1)_{B-L}$ gauge groups, respectively, while $g_1$, $Y$ and $g'_1$, $Y_{B-L}$ are the corresponding couplings and charges. The mixing between the two Abelian groups is described by the new coupling $\tilde g$. For the sake of simplicity we choose to work with a canonically normalised kinetic Lagrangian and a non-diagonal covariant derivative but other parameterisations are nevertheless equivalent. \\
Usually, an effective coupling and charge are introduced as $g_E \, Y_E \equiv \tilde g \, Y + g'_1 \, Y_{B-L}$ and some specific benchmark models can be recovered by particular choices of the two gauge couplings at a given scale (usually the EW one). For instance, the pure $B-L$ model is obtained enforcing $\tilde g = 0$ ($Y_E = Y_{B-L}$) which implies the absence of  mixing at the EW scale. Moreover, the Sequential SM (SSM) is reproduced by requiring $g'_1 = 0$ ($Y_E = Y$), the $U(1)_R$ extension is realised by the condition $\tilde g = - 2 g'_1$ while the $U(1)_\chi$ arising from $SO(10)$ unification is described by $\tilde g = - 4/5 g'_1$. \\
Therefore, a continuous variation of the mixing coupling $\tilde g$ allows to span over the entire class of anomaly-free Abelian extensions of the SM with three RH neutrinos. We stress that there is no loss of generality in choosing the $U(1)_{B-L}$ gauge group to parameterise this class of minimal $Z'$ models because the charges of an arbitrary $U(1)'$ symmetry can always be written as a linear combination of $Y$ and $Y_{B-L}$, as a result of the anomaly cancellation conditions.

The scalar sector is described by the potential
\bea
V(H,\chi) = m_1^2 H^\dag H + m_2^2 \chi^\dag \chi + \lambda_1 (H^\dag H)^2 + \lambda_2 (\chi^\dag \chi)^2 + \lambda_3 (H^\dag H)(\chi^\dag \chi),
\eea
which is the most general renormalisable scalar potential of a $SU(2)$ doublet $H$ and a complex scalar $\chi$. The stability of the vacuum, which ensures that the potential is bounded from below, is achieved by the following conditions
\bea
\label{eq:StabilityConds}
\lambda_1 > 0\,, \quad \lambda_2 >0 \,, \quad 4 \lambda_1 \lambda_2 - \lambda_3^2 > 0 \,,
\eea
obtained by requiring the corresponding Hessian matrix to be definite positive at large field values. \\
After spontaneous EWSB, the mass eigenstates $H_{1,2}$ of the two scalars are defined by the orthogonal transformation
\bea
\left( \begin{array}{c} H_1 \\ H_2 \end{array} \right) = \left( \begin{array}{cc} \cos \alpha & - \sin \alpha \\  \sin \alpha & \cos \alpha \end{array} \right)  \left( \begin{array}{c} H  \\ \chi \end{array} \right),
\eea 
where the scalar mixing angle $\alpha$ is given by
\bea
\label{eq:ScalarAngle}
\tan 2 \alpha = \frac{\lambda_3 \, v \, x}{\lambda_1 \, v^2 - \lambda_2 \, x^2}
\eea
and the masses of the physical scalars are
\bea
\label{eq:ScalarMasses}
m_{H_{1,2}}^2 = \lambda_1 v^2 + \lambda_2 x^2 \mp \sqrt{\left( \lambda_1 v^2 - \lambda_2 x^2\right)^2 + \left( \lambda_3 v x \right)^2} \,,
\eea
with $m_{H_{2}} > m_{H_{1}}$ and $H_1$ identified with the 125.09 GeV Higgs boson. \\
Eqs.~(\ref{eq:ScalarAngle})--(\ref{eq:ScalarMasses}) can  easily be inverted as
\bea
\label{lambdas}
\lambda_1 &=& \frac{m_{H_1}^2}{4v^2}(1+\cos 2\alpha) + \frac{m_{H_2}^2}{4 v^2}(1-\cos 2 \alpha) \,, \nn \\
\lambda_2 &=& \frac{m_{H_1}^2}{4x^2}(1-\cos 2\alpha) + \frac{m_{H_2}^2}{4 x^2}(1+\cos 2 \alpha) \,, \nn \\
\lambda_3 &=& \sin 2 \alpha \left( \frac{m_{H_2}^2 - m_{H_1}^2}{2 v x}\right) \,,
\eea
relations which can be used to define the initial conditions on the quartic couplings through the physical masses $m_{H_{1,2}}$, the VEVs $v, x$ and the mixing angle $\alpha$. Notice that the light (heavy) Higgs boson couples to SM particles proportionally to $\cos \alpha$ ($\sin \alpha$), while the interaction with the $Z'$ and  heavy neutrinos is provided by the complementary angle $\sin \alpha$ ($\cos \alpha$).

When the two scalars acquire non-vanishing VEVs,  the neutral component of the gauge sector becomes massive and with mass eigenstates determined by two mixing angles: the usual Weinberg angle $\theta_w$ and a new mixing angle $\theta'$, for which
\bea
\tan 2 \theta' = \frac{2 \tilde g \sqrt{g_1^ 2 + g_2^2}}{\tilde g^2 + (4 g'_1 x / v)^2 - g_1^ 2 - g_2^2} \,,
\eea
with values in the interval $- \pi/4 \leq \theta' \leq \pi/4$.
In contrast, the charged gauge bosons are unaffected by the presence of the extra Abelian factor and their masses remain as in the SM. \\
The mixing angle is completely defined in terms of the mass of the $Z'$, through the VEV $x$ of the singlet scalar, and of its gauge couplings. In general it is always non-vanishing unless $\tilde g = 0$ which corresponds to the pure $B-L$ model. The EW Precision Tests (EWPTs) considerably constrain the mixing angle to small values, namely, $|\theta'| \lesssim 10^{-3}$ \cite{Cacciapaglia:2006pk,Salvioni:2009mt}, in which case 
\bea
\label{thetapexpandend}
\theta ' \simeq \tilde g \frac{M_Z \, v/2}{M_{Z'}^2 - M_Z^2} \,.
\eea
This relation can be satisfied provided either $\tilde g \ll 1$ or $M_Z / M_{Z'} \ll 1$, the latter condition allowing a generous range of values for $\tilde g$.

Finally, the Yukawa Lagrangian is
\bea
\mathcal L_Y =  \mathcal L_Y^{SM}  - Y_\nu^{ij} \, \overline{L^i} \, \tilde H \, {\nu_R^j}   - Y_N^{ij} \, \overline{(\nu_R^i)^c} \, {\nu_R^j} \, \chi + \, h.c. 
\eea
where $\mathcal L_Y^{SM}$ is the SM contribution. Notice that a Majorana mass term for the RH neutrinos, $M = \sqrt{2}\, x Y_N$, is dynamically generated by the VEV $x$ of the complex scalar $\chi$ and, therefore, the type-I seesaw mechanism is automatically implemented through spontaneous symmetry breaking. The light physical neutrinos emerge as a combination of the left-handed SM neutrinos and a highly damped sterile RH component, while the heavier ones are mostly RH. The damping term in such combinations is proportional to the ratio of the Dirac and Majorana masses.

\section{Constraints from EWPTs and  LHC searches}
\label{sec:bounds}
\begin{figure}
\centering
\includegraphics[scale=0.18]{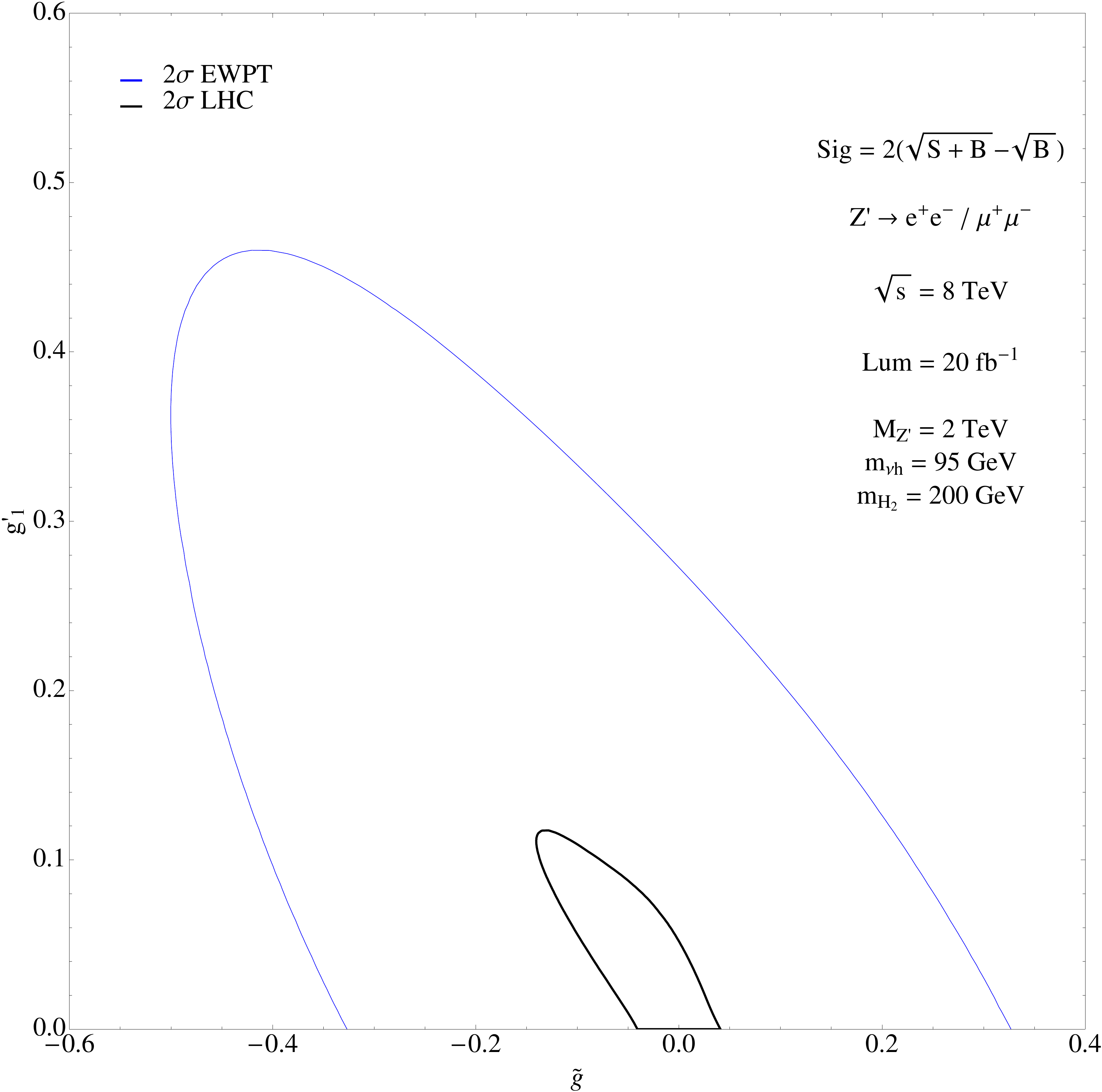} 
\includegraphics[scale=0.18]{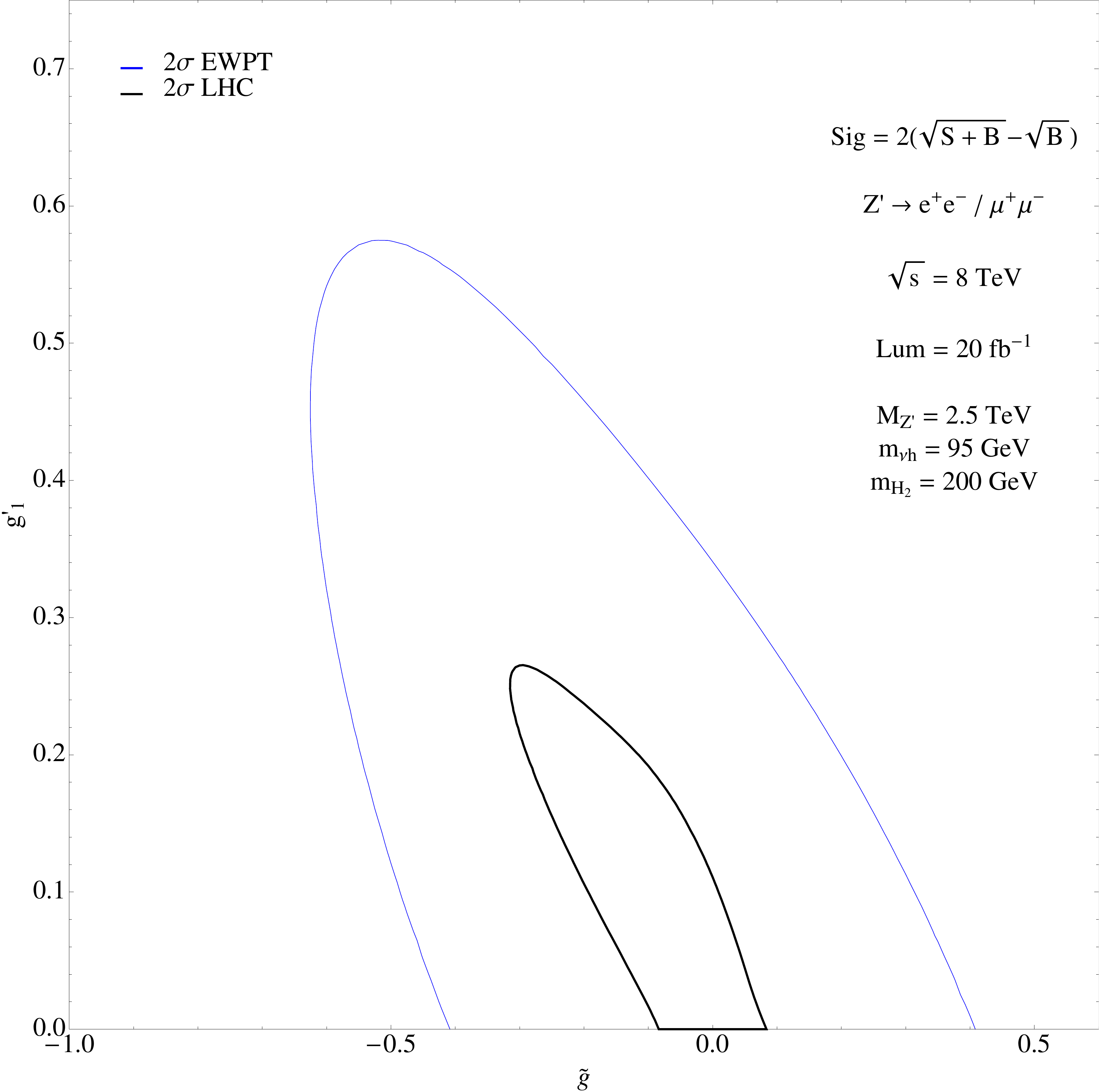} 
\includegraphics[scale=0.18]{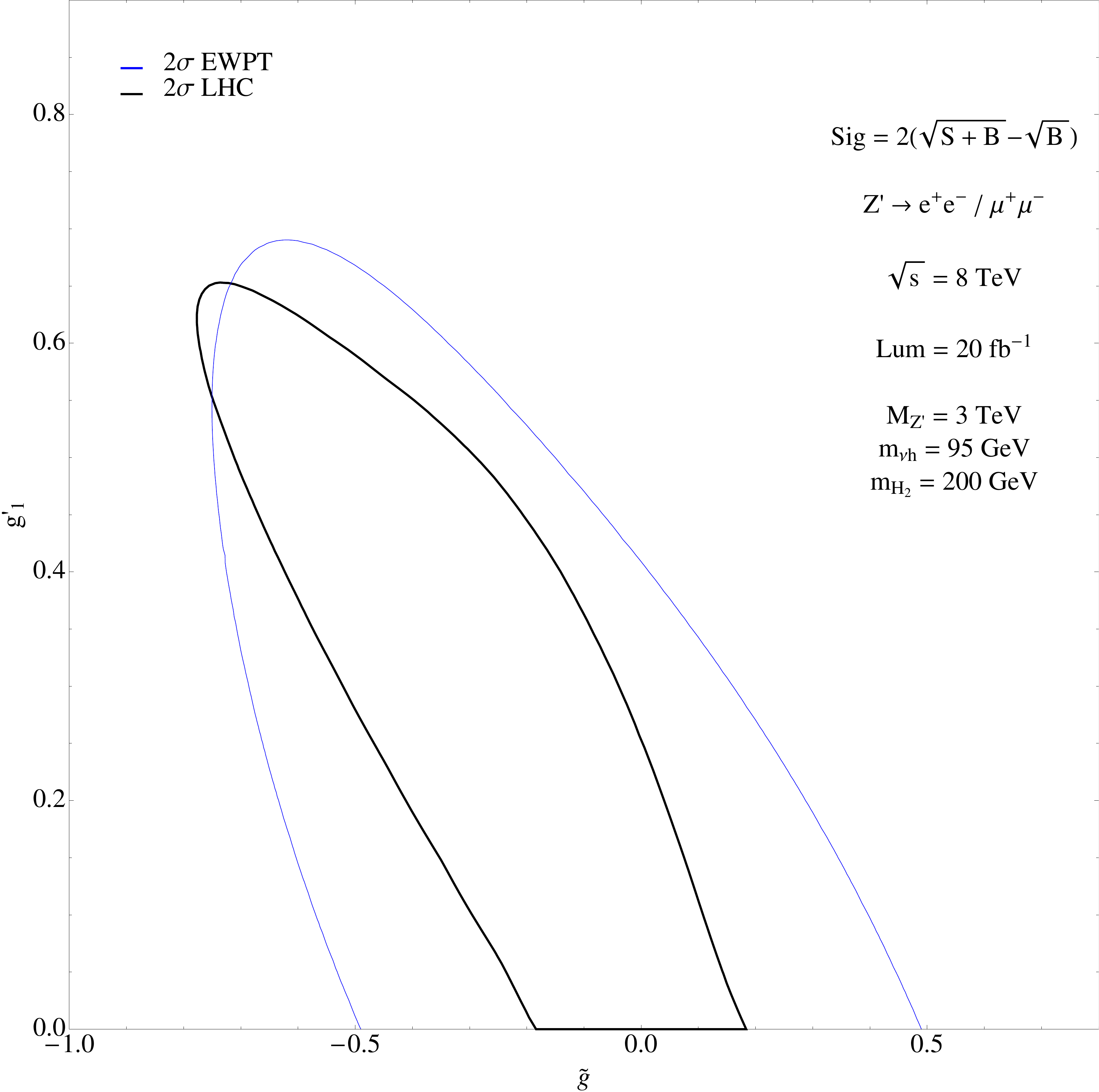} 
\caption{EWPTs vs bounds from a significance analysis at the LHC for different $Z'$ masses. \label{EWPTvsDY} }
\end{figure}

The $(\tilde g, g'_1)$ parameter space is subjected to well established bounds coming from EWPTs extracted from LEP2 data. These bounds can be recast into constraints for a well-defined set of higher-dimensional operators \cite{Cacciapaglia:2006pk} which describe the effects of new physics. For the Abelian extension under study, these operators have been computed in \cite{Salvioni:2009mt} in terms of the $Z'$ mass and gauge couplings $\tilde g$, $g'_1$ neglecting, however, the impact of the heavy neutrinos and of the extended scalar sector.
To these constraints we add the one drawn from the more recent data of the first Run of LHC at 8 TeV and $\mathcal L = 20$ fb$^{-1}$, based on a signal-to-background analysis for the di-lepton (electrons and muons) channel. Next-to-Next-to-Leading-Order (NNLO) Quantum Chromo-Dynamics (QCD) effects are taken into account through a $k$-factor correction. 
We show in fig.~\ref{EWPTvsDY} the exclusion limit at 95\% Confidence Level (CL) from both EWPTs and DY studies for three values of the $M_{Z'}$, namely $M_{Z'} = 2, 2.5$ and 3 TeV. 
For the masses of the $Z'$ under our investigation, the LHC studies represent a strong improvement with respect to the EW related ones, with the only comparable case being the one with  $M_{Z'} = 3$ TeV. Consequently, we will employ all such tight bounds in the following sections. \\
The sequential ($g'_1=0$) and  pure $B-L$ ($\tilde{g}=0$) models are strongly constrained while the leptophobic direction in which the $Z'$ coupling to leptons is minimal $(g'_1/\tilde g \simeq -3/4)$ obviously represents the least bounded charge assignment. Moreover, we have explicitly verified that the bounds from the DY analysis are not considerably modified by the values of the heavy neutrino mass and the parameters of the extra scalar sector, such dependence entering only in the total width of the $Z'$ boson. 

\begin{figure}
\centering
\subfigure[]{\includegraphics[scale=0.5]{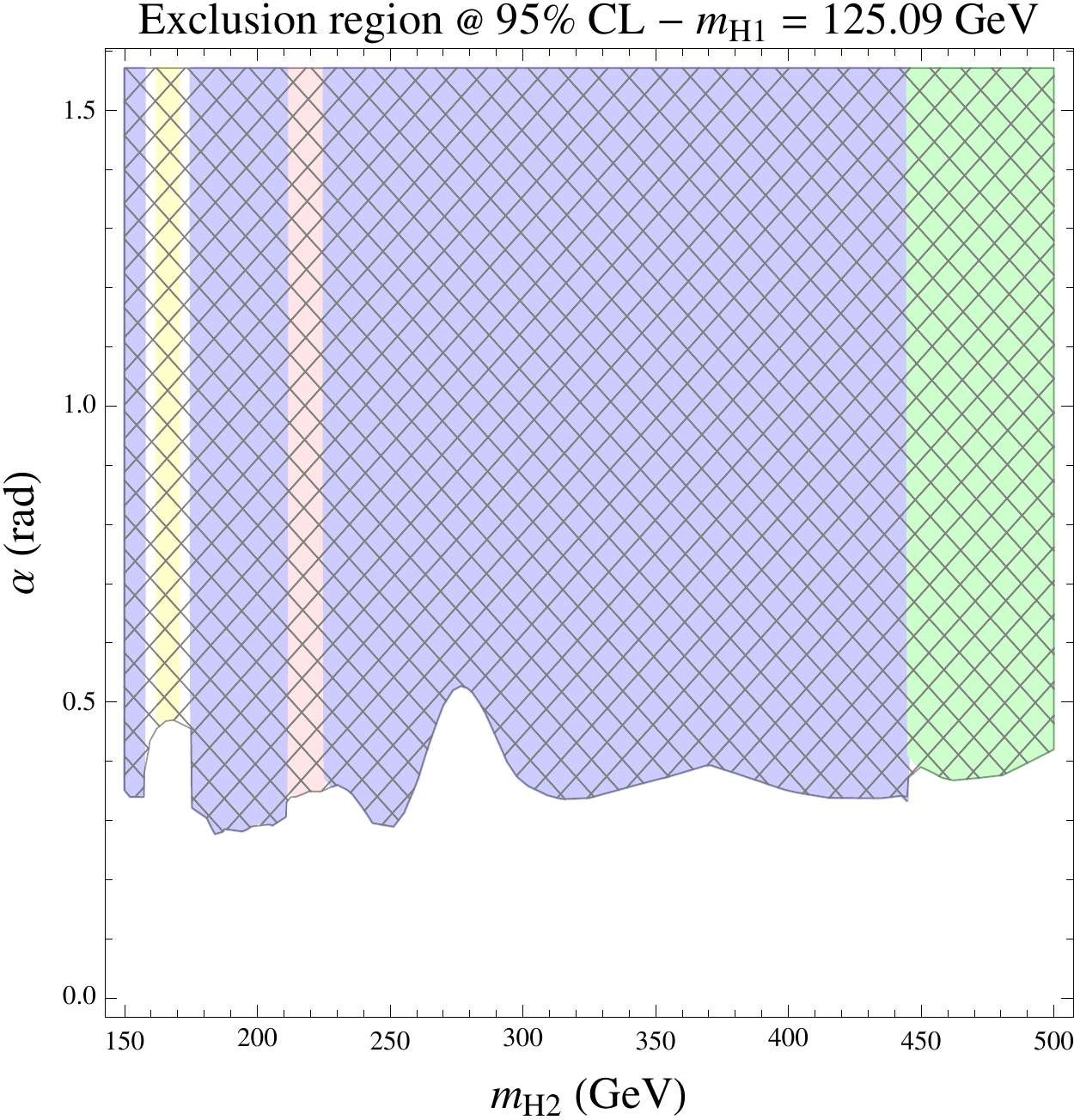}}
\subfigure[]{\includegraphics[scale=0.7]{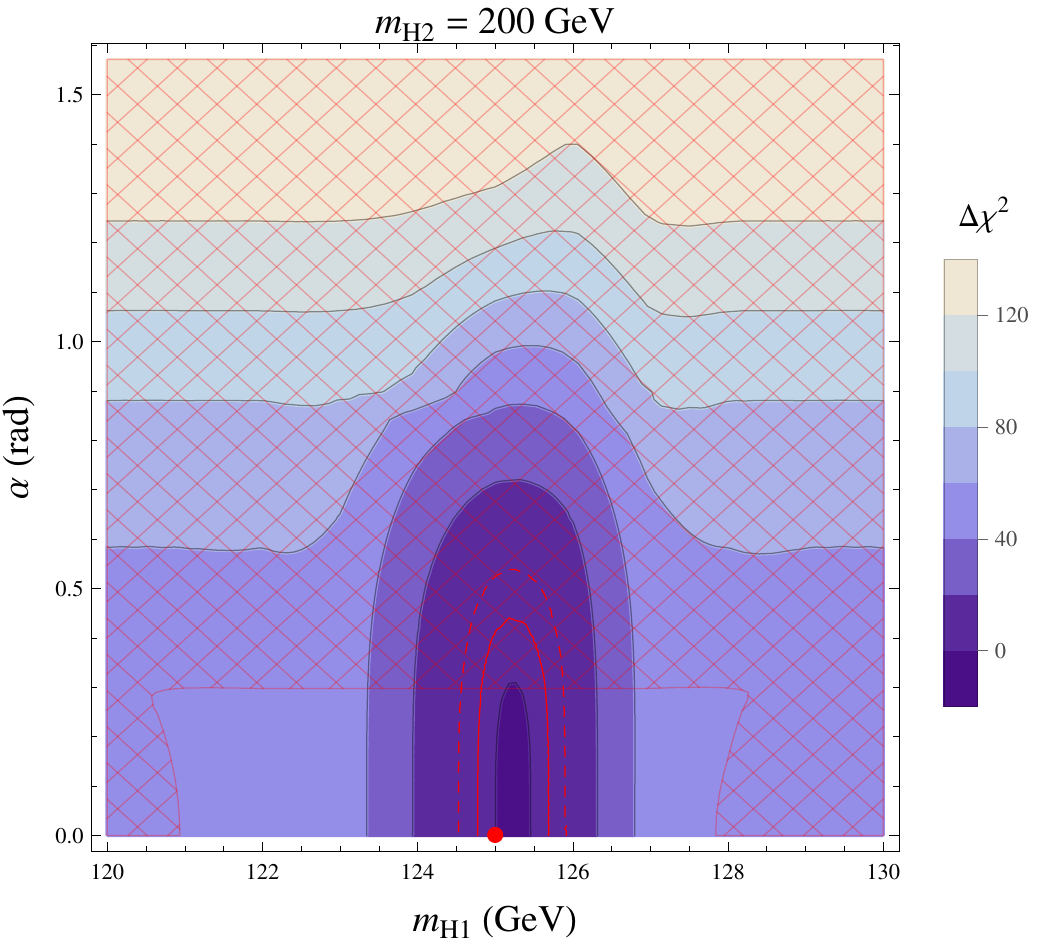}}
\caption{(a) Excluded region by LEP+Tevatron+LHC in the $(m_{H2}, \alpha)$ plane for fixed $m_{H_1} = 125.09$ GeV, $m_{\nu_h} = 95$ GeV and $M_{Z'} = 2$ TeV obtained using \texttt{HiggsBounds}. The most sensitive exclusion channels are the four leptonic decay of two $Z$ bosons \cite{CMS:xwa} (red region), the full leptonic decay of two $W^\pm$ bosons \cite{CMS:bxa} (yellow region), the heavy Higgs decays into two $Z$s or $W^\pm$ s\cite{Khachatryan:2015cwa} (blue region) and  a combined search in five decay modes: $\gamma\gamma$, $ZZ$, $W^+W^-$, $\tau\tau$ and $bb$ \cite{CMS:aya} (green region). (b) Fit results using \texttt{HiggsSignals} with $m_{H_2} = 200$ GeV. The colours indicate levels of $\Delta \chi^2$ from the best fit point, $\chi^2/ndf = 97.5/89$ (red point corresponding to the SM Higgs: $m_{H1} = 125.09$ GeV, $\alpha = 0$).  Solid (dashed) red line corresponds to 1$\sigma$ (2$\sigma$) contours. The hatched region is excluded at 95$\%$ CL. \label{HiggsBounds1}}
\end{figure}

The extra scalar sector is strongly constrained by Higgs searches at LEP, Tevatron and LHC experiments. The present exclusion limits are enforced using \texttt{HiggsBounds} \cite{arXiv:0811.4169,arXiv:1102.1898,arXiv:1301.2345,arXiv:1311.0055,arXiv:1507.06706} and the agreement of the model with the signal strength measurements of the discovered 125.09 GeV Higgs scalar is taken into account via \texttt{HiggsSignals} \cite{Bechtle:2013xfa}. The results in the $(m_{H_2}, \alpha)$ plane are reported in fig.~\ref{HiggsBounds1}. The most sensitive exclusion channels are depicted with different colours depending on the $H_2$ mass (fig.~\ref{HiggsBounds1}(a)). The most effective exclusion search, covering almost all the $m_{H_2}$ mass interval from 150 GeV to 450 GeV, is of a Higgs boson decaying into a pair of $W$ and $Z$ bosons \cite{Khachatryan:2015cwa} (blue region). In particular, the fully leptonic and semileptonic decay channels have been considered for  $H\rightarrow W^+W^-$ while for  $H \rightarrow ZZ$ the final states containing four charged leptons, two charged leptons and two quarks or two neutrinos have been studied. Finally, in fig.~\ref{HiggsBounds1}(b), we show a $\chi^2$ compatibility fit with the Higgs signal measurements in the $(m_{H_1}, \alpha)$ plane. We have chosen a fixed reference value for the $H_2$ mass, namely $m_{H_2} = 200$ GeV, and for the heavy neutrino mass, $m_{\nu_h} = 95$ GeV, so that only SM-like decay channels are open for the lightest scalar $H_1$. The requirement of a compatibility at $2\sigma$ results anyway into a weaker bound with respect to the exclusion limits that we have taken into account in all the following analyses.

\section{The RG analysis}
\label{sec:rg}
The parameter space available at the EW scale can be further constrained through the requirements of  perturbativity of the couplings and  stability of the vacuum through the RG evolution. In this analysis we use RG equations at two-loop order and matching relations at one-loop level  in perturbation theory. The latter supply the initial conditions of the RG running at a particular scale (we choose the top pole mass $M_t$ as a starting point). These relations provide the initial value of the running couplings computed in the $\overline{\rm{MS}}$ renormalisation scheme as a function of the physical on-shell parameters, namely, the pole masses and the physical couplings such as the Fermi constant. The details of the implementation of the matching conditions for various Abelian extensions of the SM can be found in \cite{Coriano:2015sea}. 
Before moving into the details of the analysis, we stress that the role of the two-loop corrections on the $\beta$ functions and of the one-loop matchings on the initial conditions is crucial in order to reach definitive conclusions on the dynamics of these $U(1)'$ extensions of the SM. Indeed, it has been shown in \cite{Coriano:2015sea} that they can lead to sizeable effects on the RG evolution and usually improve the stability of the scalar potential above and beyond the SM conditions.

\begin{figure}
\centering
\includegraphics[scale=0.4]{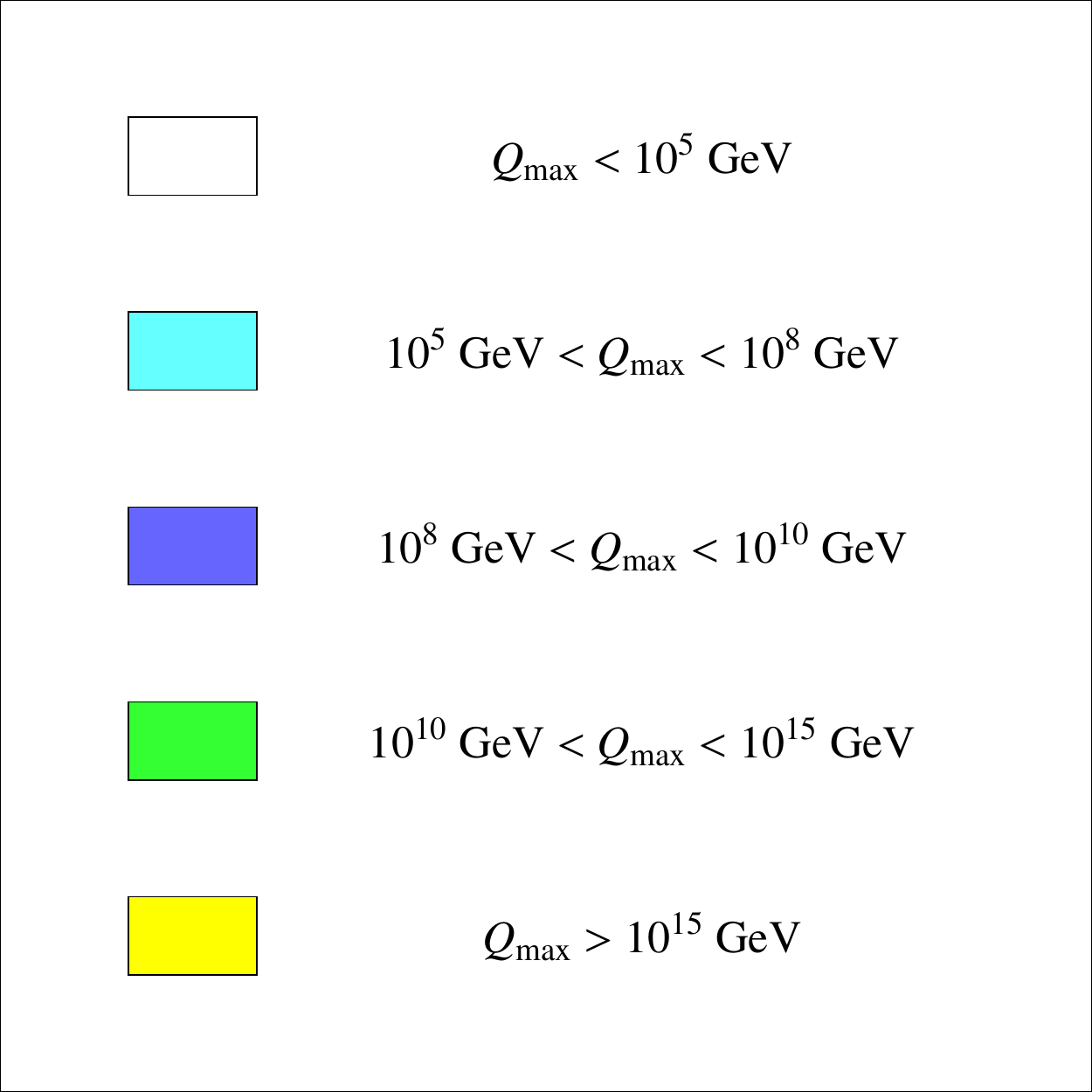}
\caption{Legend of the different regions defined by the maximum scale $Q_{\rm max}$ up to which the model is stable and perturbative. \label{fig.stablegend}}
\end{figure}
We identify the regions in which the class of the Abelian extensions under study possesses a stable vacuum (described by the conditions given in Eq.~(\ref{eq:StabilityConds})) and is characterised by a weakly coupled regime (couplings are required to be less than $\sqrt{4\pi}$\footnote{The parameters upon which the perturbative expansion is performed are usually in the form of $\sqrt{\alpha} = g/\sqrt{4 \pi}$ rather than $g$. Different and less conservative choices are also viable as the one that can be inferred applying Naive Dimensional Analisys (NDA). This latter case would shape a perturbative regime with couplings less then $4 \pi / \sqrt{N}$ with $N$ related to the degrees of freedom inside the loops. Both options have been explored in our analysis with very rare occasions of noticeable discrepancies among the two choices.}) along the RG evolution, up to some given scales which will be specified below. These regions are defined in the space of the new parameters, evaluated at the EW scale, introduced by these minimal Abelian extensions. We will focus, in particular, on the impact of the Abelian gauge couplings $\tilde g$, $g'_1$,  of the scalar mixing angle $\alpha$ and of the masses of the heavy Higgs $H_2$ and $Z'$ boson. \\
For ease of reference, the legend of the stability and perturbativity regions, according to the maximum scale $Q_{\rm max}$ up to which the vacuum is stable and the model remains perturbative, is depicted in fig.~\ref{fig.stablegend}. In the cyan region the new parameters of these Abelian extensions are such that the stability and/or the perturbativity is lost at a scale $Q_{\rm max}$ lower than the instabilitiy scale of the SM. A $Z'$ model with gauge couplings lying in this region of the parameter space worsen the high energy behaviour of the SM and clearly calls, with more urgency, for an embedding into a complex scenario, such as GUT unification, already appearing below the $10^8$ GeV. In the blue region the $U(1)'$ extensions behave, from the stability point of view, as the SM, whose instability scale $\Lambda_{\rm SM} \sim 10^{8-10}$ GeV. In contrast, the green and yellow regions delineate portions of the parameter space in which  this class of models is more stable than the SM, up to the GUT scale and above, thus identifying them as compelling extensions of the EW theory. 
\begin{figure}
\centering
\subfigure[]{\includegraphics[scale=0.39]{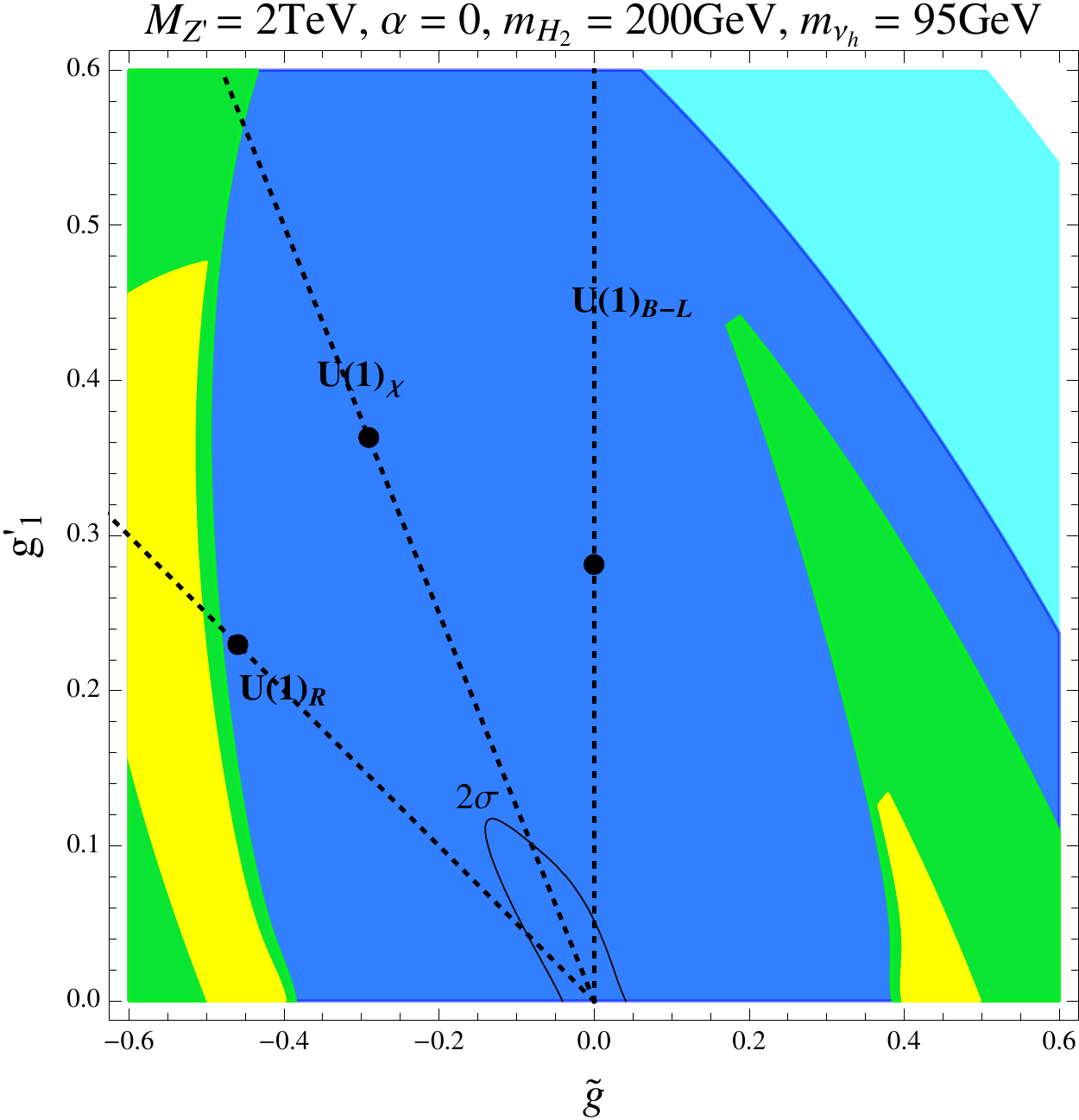}}
\subfigure[]{\includegraphics[scale=0.39]{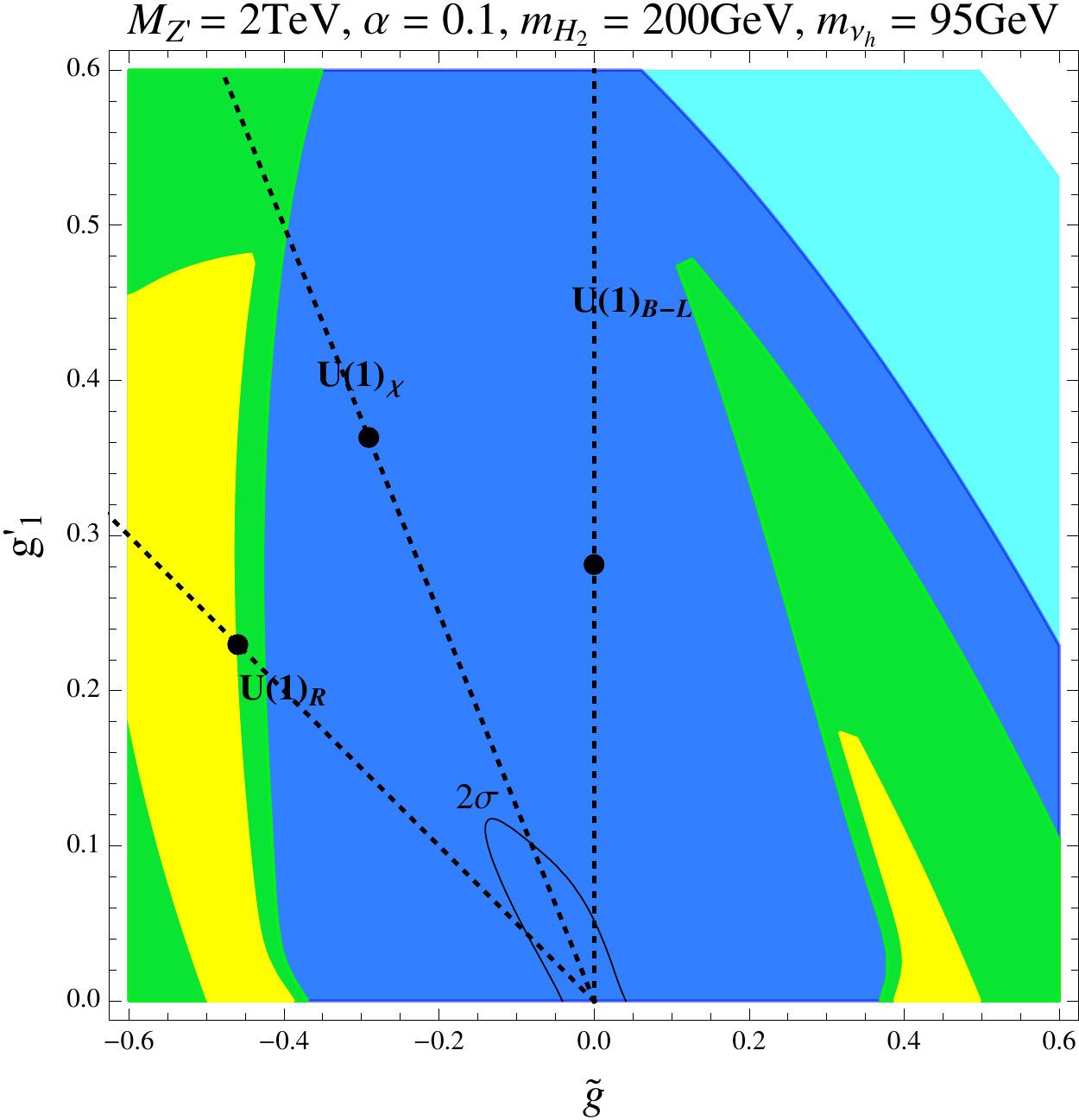}}
\subfigure[]{\includegraphics[scale=0.39]{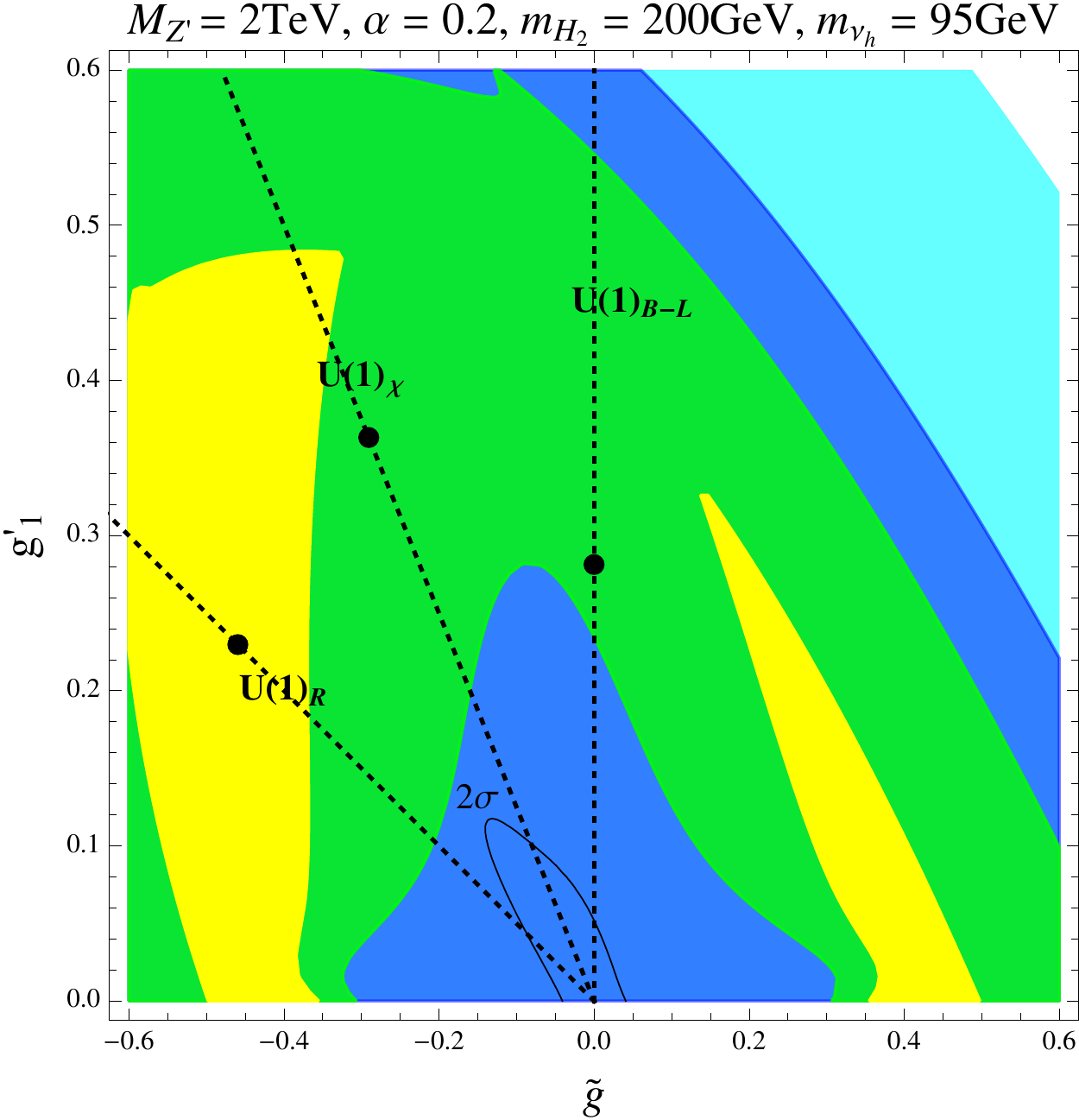}} \\
\subfigure[]{\includegraphics[scale=0.39]{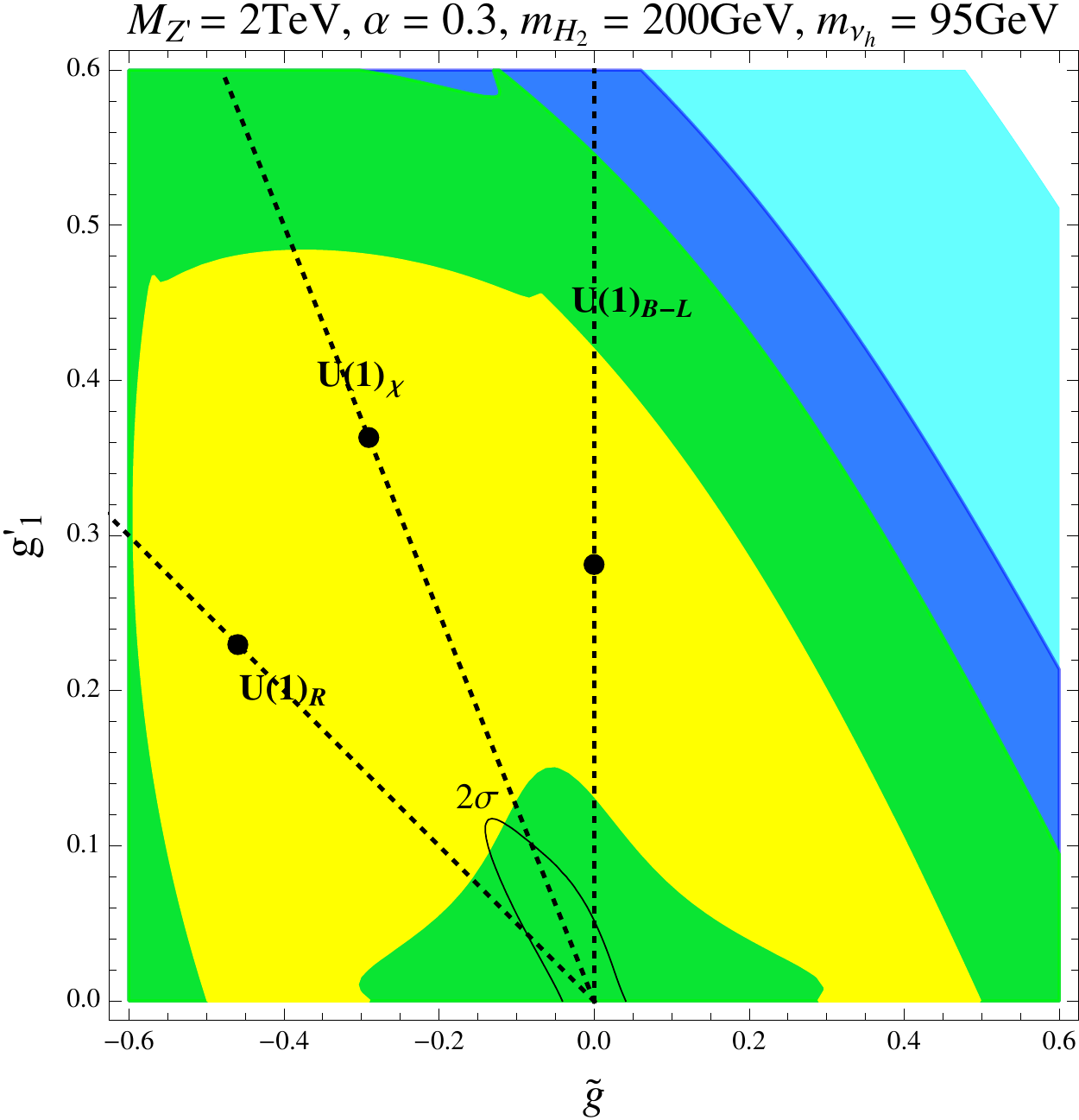}}
\subfigure[]{\includegraphics[scale=0.39]{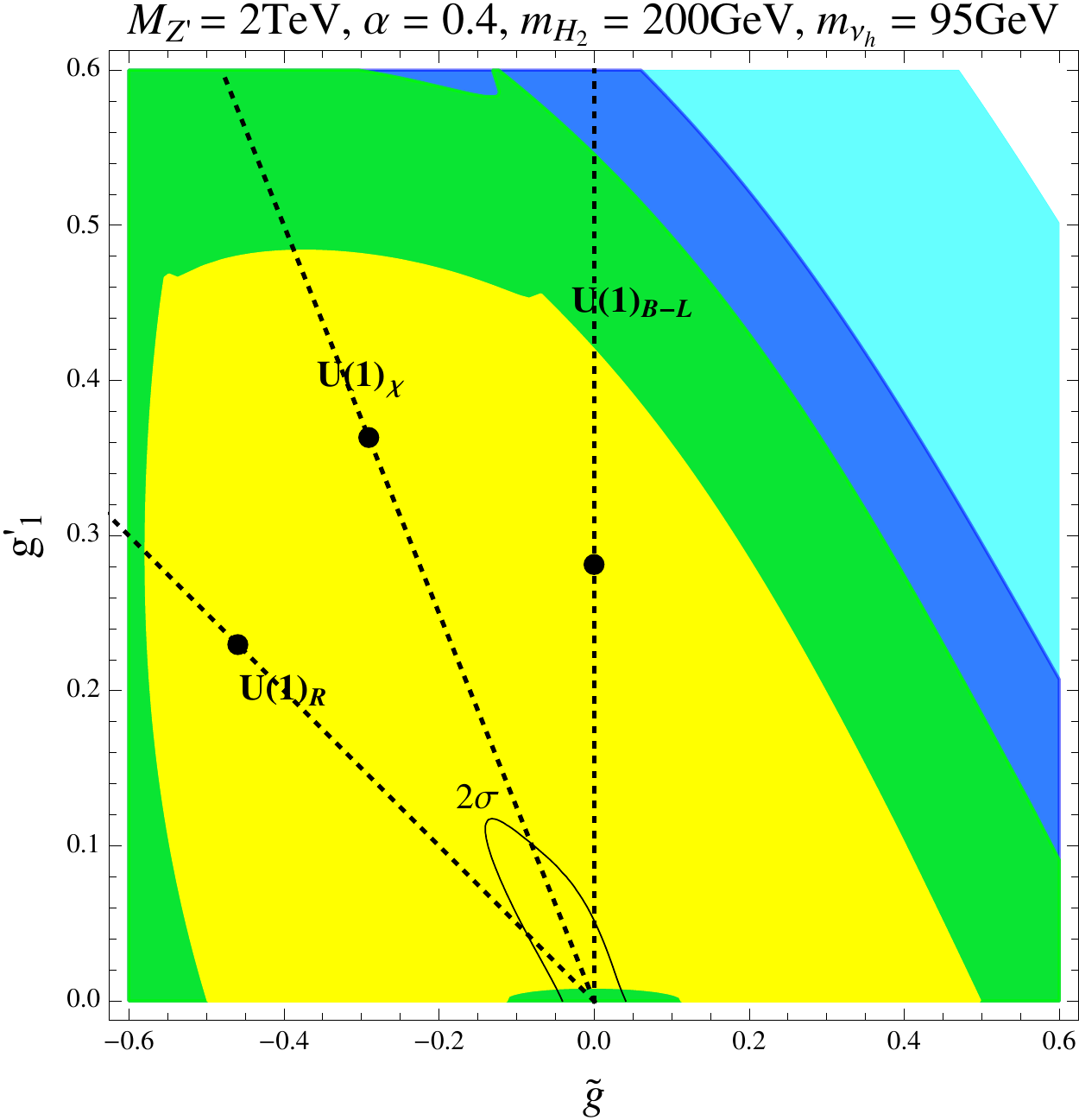}}
\subfigure[]{\includegraphics[scale=0.39]{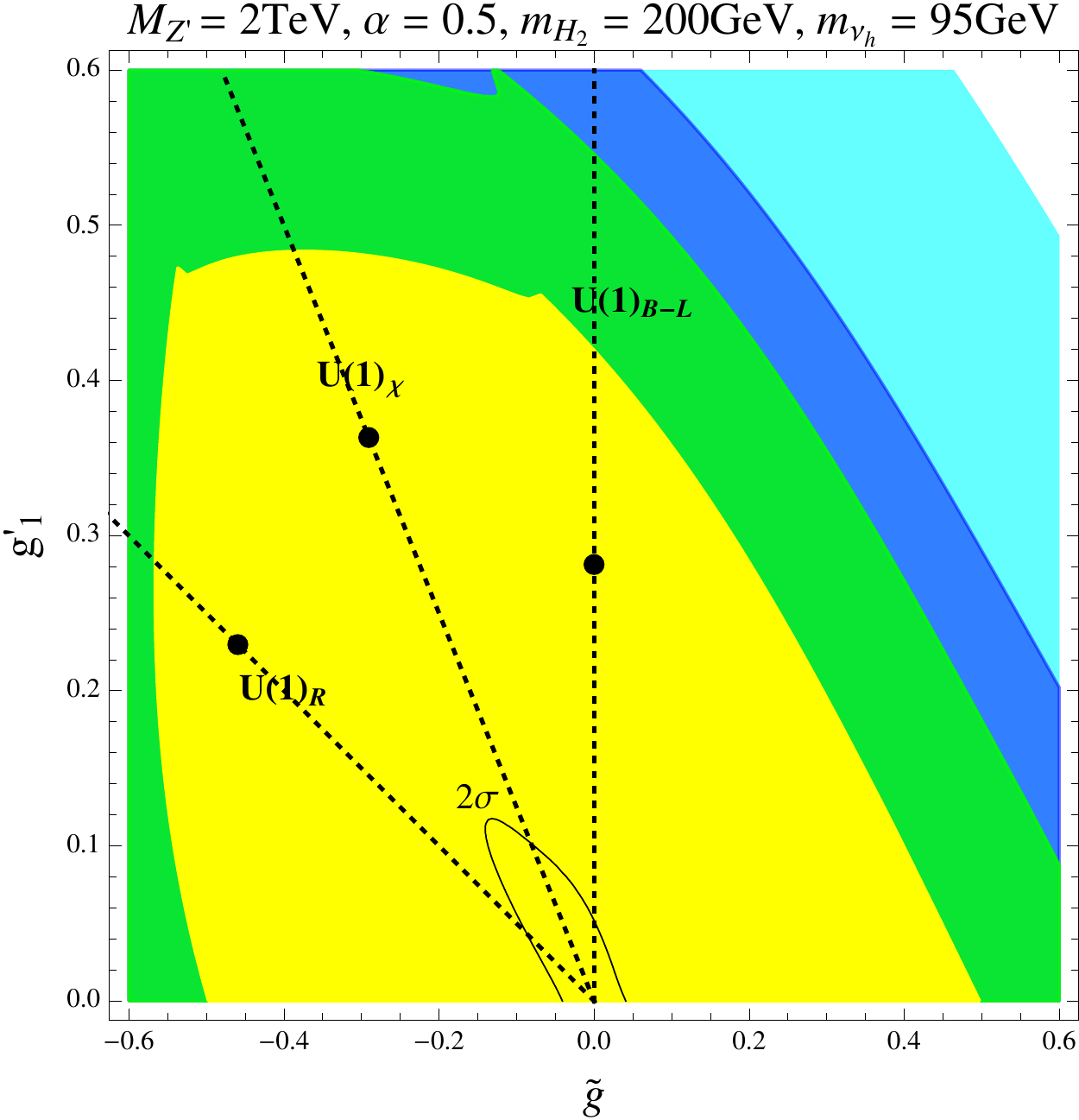}} 
\caption{Stability and perturbativity regions in the $(\tilde g, g_1')$ plane for different values of the scalar mixing angle $\alpha$. The coloured regions are defined according to $Q_{\rm max}$, the maximum value of the stability and perturbativity scale reached by the model. The corresponding legend is depicted in fig.~\ref{fig.stablegend}. \label{fig.stab-gt-g1p}}
\end{figure}
In fig.~\ref{fig.stab-gt-g1p} we show the regions of stability and perturbativity, up to some given scale $Q_{\rm max}$, as a function of the two Abelian couplings $\tilde g$,  $g'_1$ and for different values of the scalar mixing angle $\alpha$.
These results have been obtained for $M_{Z'} = 2$ TeV and $m_{\nu_h} = 95$ GeV which corresponds to a Yukawa coupling $Y_N$ of order $10^{-2}$. This value is too small to affect the RG evolution of the quartic scalar couplings, therefore the destabilising effect of new fermionic degrees of freedom is completely suppressed. Indeed, a $Y_N \gtrsim 0.3$ is, at least, required to appreciate the impact of the heavy RH neutrinos in the running of the scalar sector \cite{Coriano:2015sea}. This roughly corresponds to $m_{\nu_h} \simeq 0.2 \, (M_{Z'}/g'_1)$ for $M_{Z'} \gg M_Z$.\\
The constraints coming from  di-lepton searches at the LHC with $\sqrt{s} = 8$ TeV and $M_{Z'} = 2$ TeV strongly restrict the allowed parameter space in the $(\tilde g, g'_1)$ plane completely leaving out the cyan regions and therefore only selecting the configurations in which the model is at least as stable as the SM.
The dashed lines correspond to three particular and very common $U(1)'$ extensions which can be described, in our conventions, by straight lines in the $(\tilde g, g'_1)$ plane. These are, in anti-clockwise direction, the pure $U(1)_{B-L}$, the $U(1)_\chi$ and the $U(1)_R$ extensions, while the sequential SM lies on the $g'_1$ axis. The black dots represent the benchmark models usually addressed in the literature in which the Abelian gauge couplings are fixed to specific values. Notice also that these reference points, although allowed by EWPTs, are excluded by LHC data if $M_{Z'} = 2$ TeV. \\
As one can easily notice from fig.~\ref{fig.stab-gt-g1p}, the effect of the mixing angle $\alpha$ of the two scalars is crucial for identifying the regions in which the vacuum is stable. Indeed, scalar degrees of freedom, contrary to the fermionic ones, usually drive the instability scale towards higher values improving the stability of the potential. In the $\alpha = 0$ case (which corresponds to $\lambda_3 = 0$), the extra scalar sector is decoupled from the SM Higgs doublet and the RG evolution of the $U(1)'$ extension shares the same behaviour of the SM if the new Abelian gauge couplings are sufficiently small. If $\alpha$ moves away from zero, the two scalar sectors begin to communicate and the stability effect of the complex scalar $\chi$ becomes quickly significant, preventing the decay of the vacuum up to the GUT scale and above.  
\begin{figure}
\centering
\subfigure[]{\includegraphics[scale=0.39]{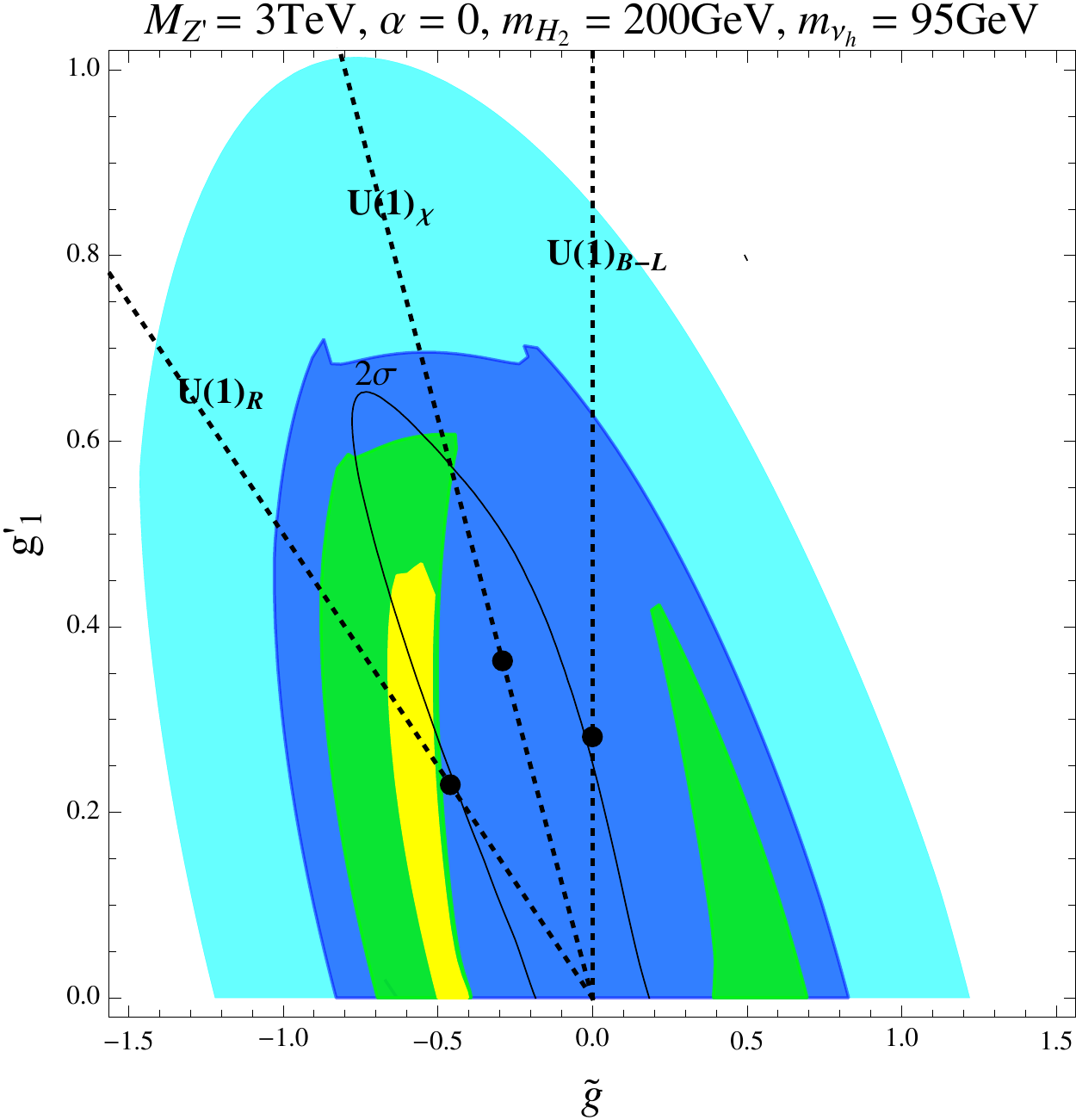}}
\subfigure[]{\includegraphics[scale=0.39]{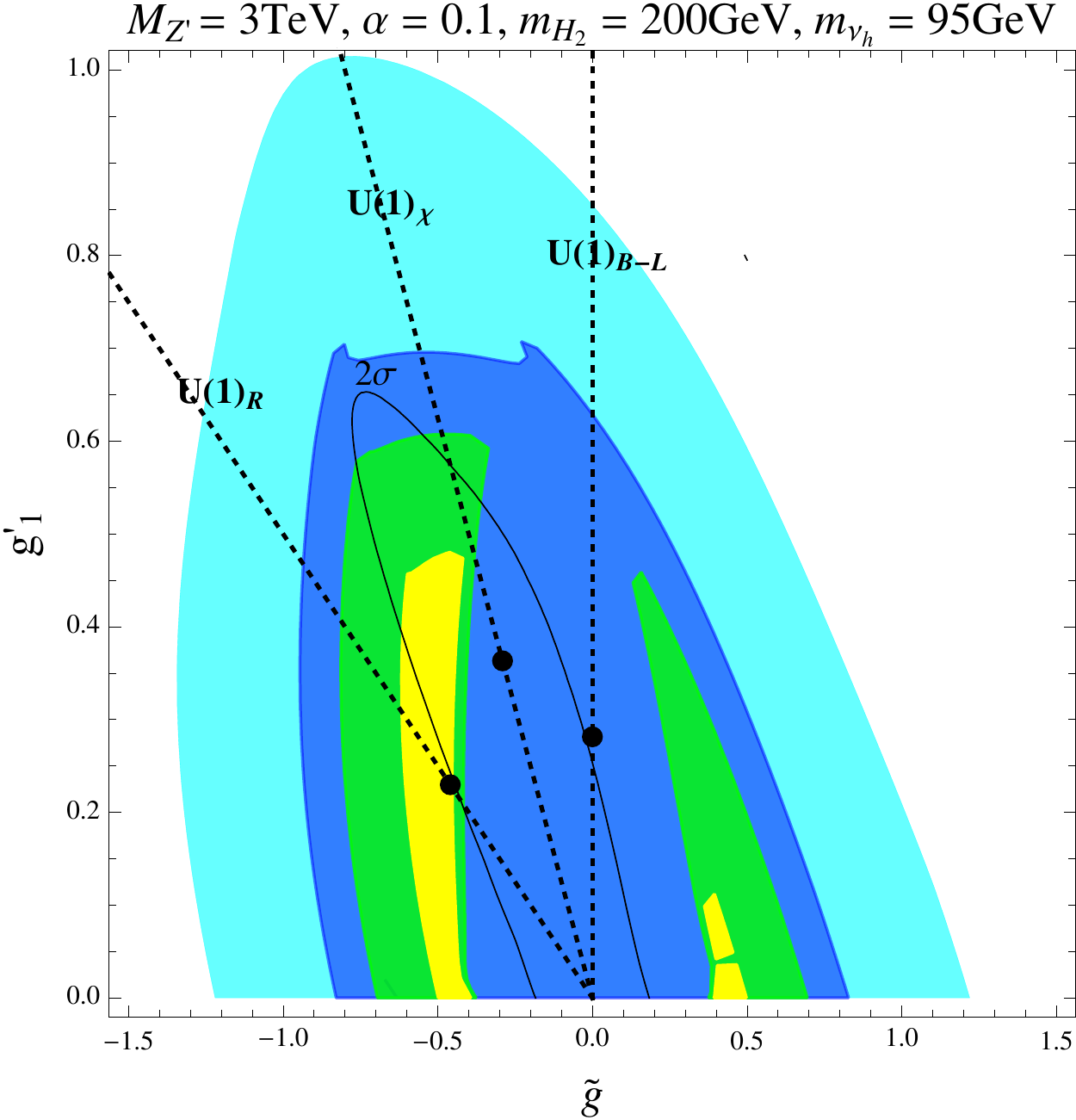}}
\subfigure[]{\includegraphics[scale=0.39]{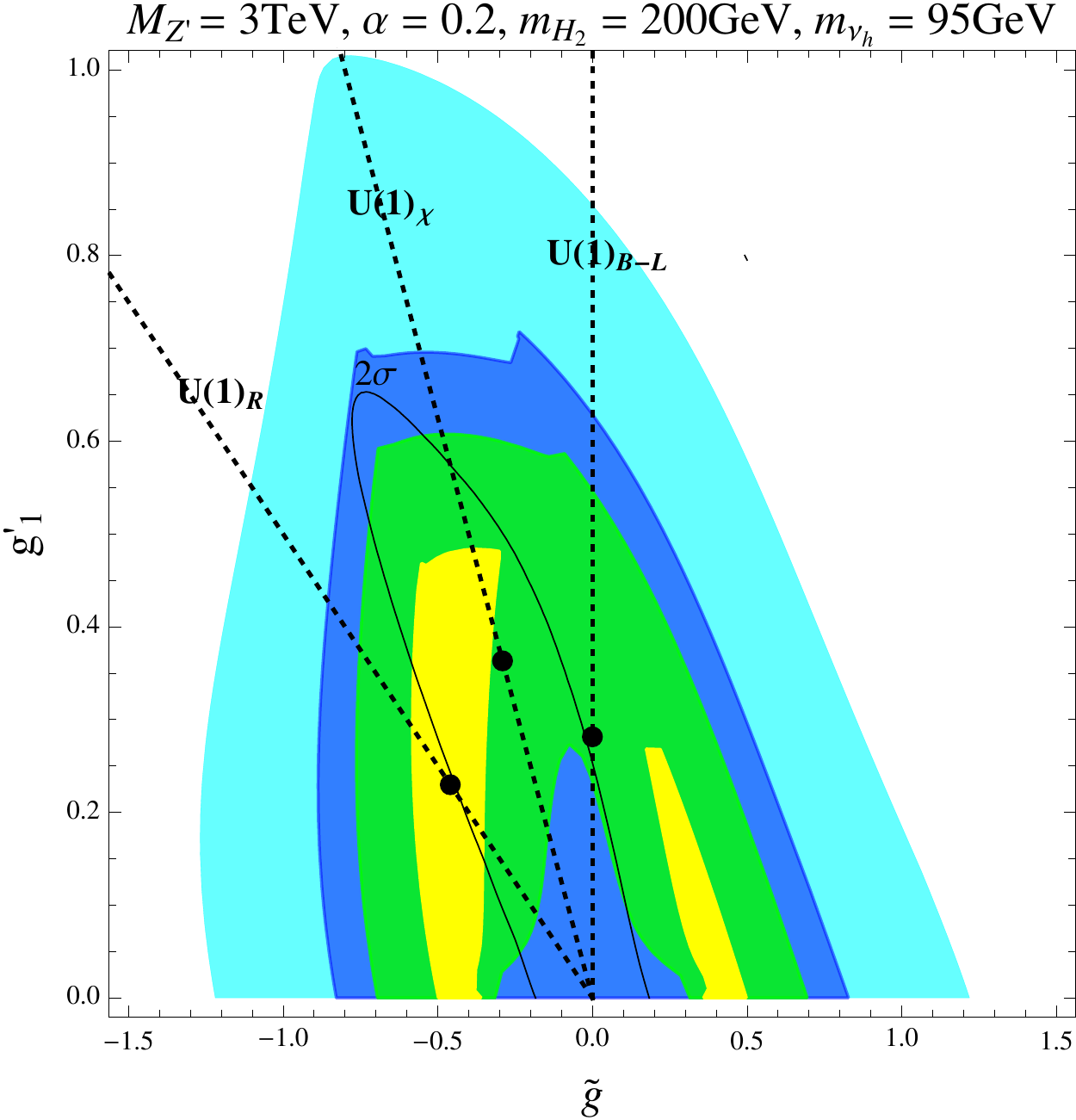}} \\
\subfigure[]{\includegraphics[scale=0.39]{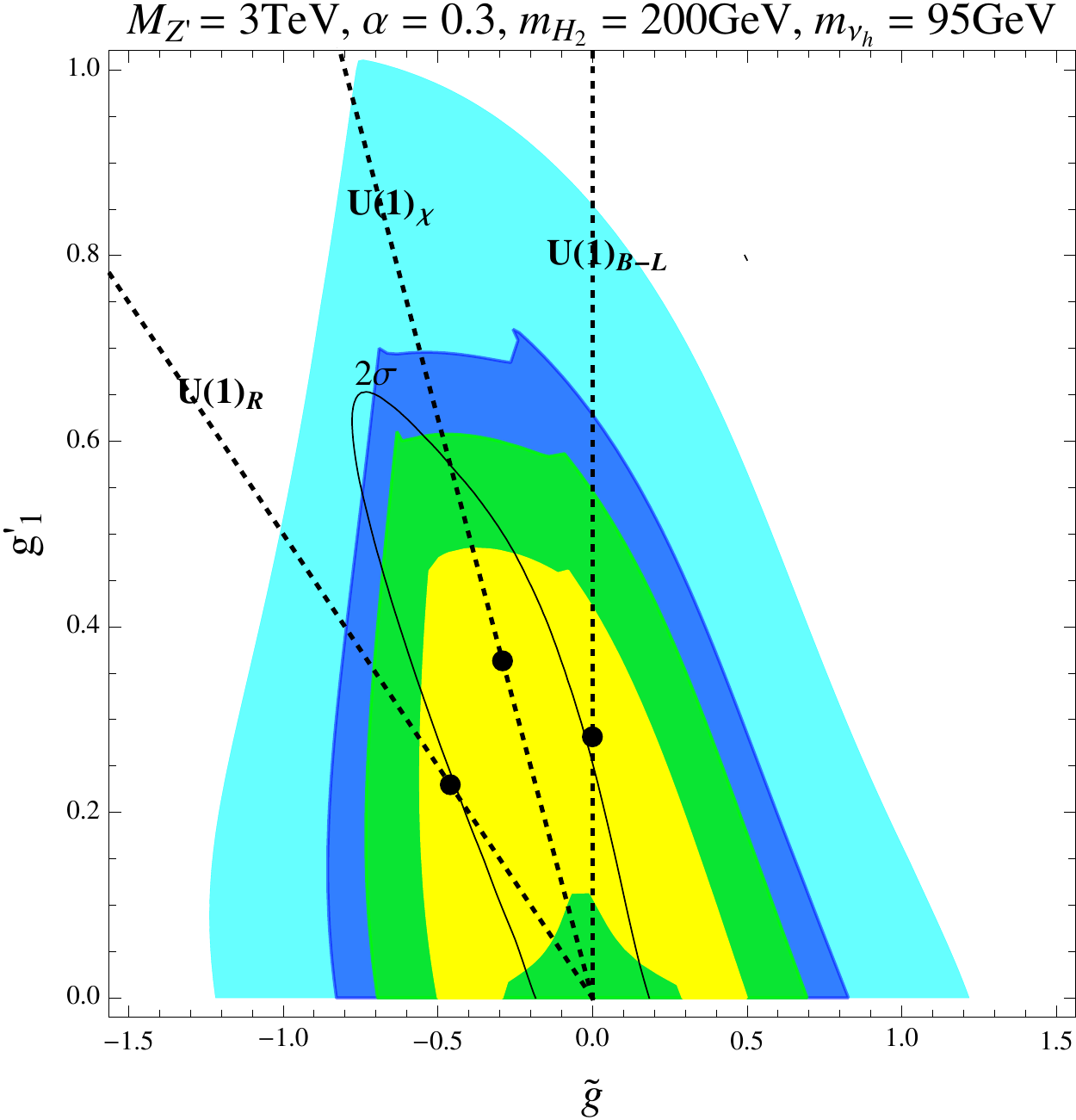}}
\subfigure[]{\includegraphics[scale=0.39]{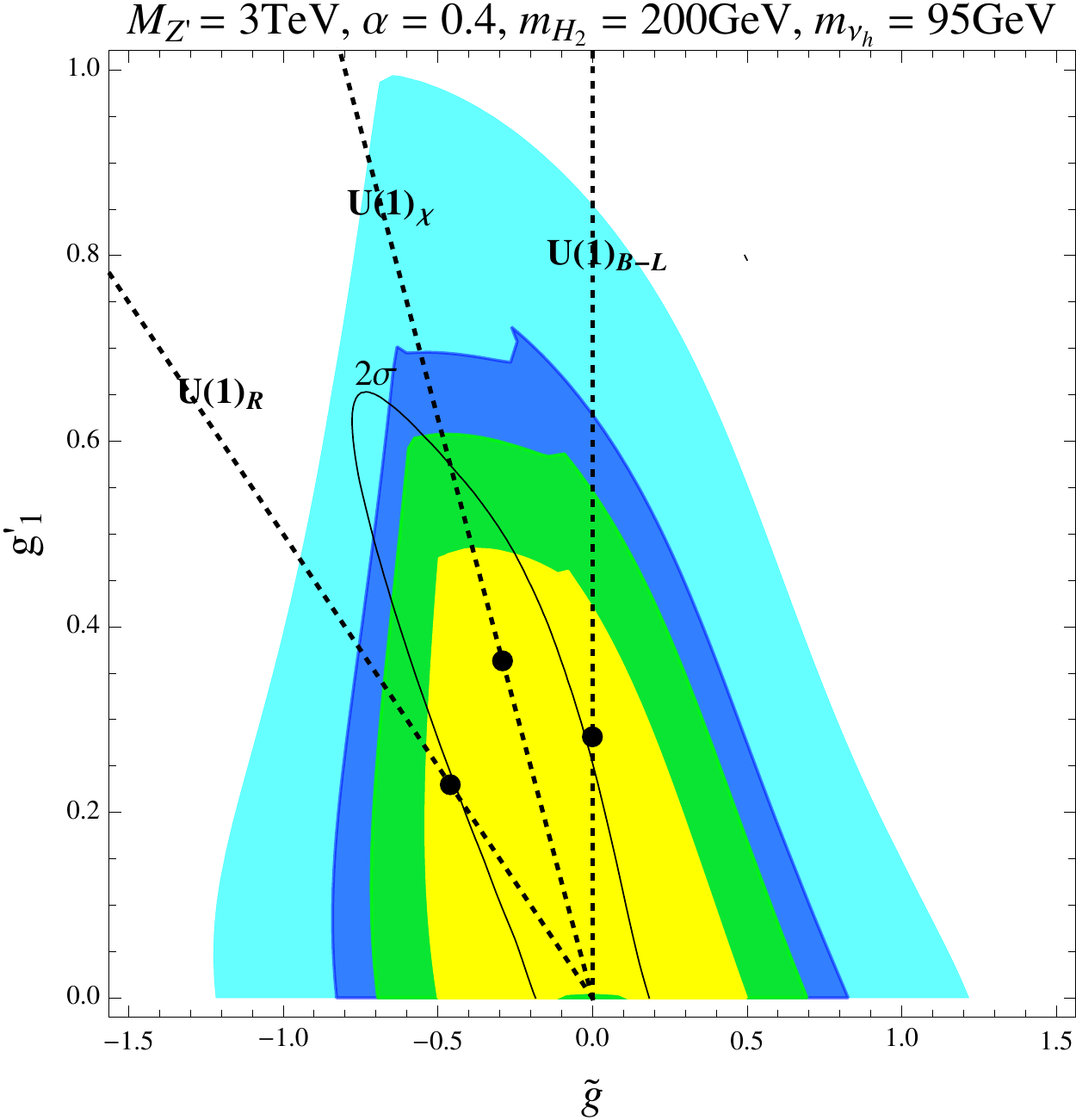}}
\subfigure[]{\includegraphics[scale=0.39]{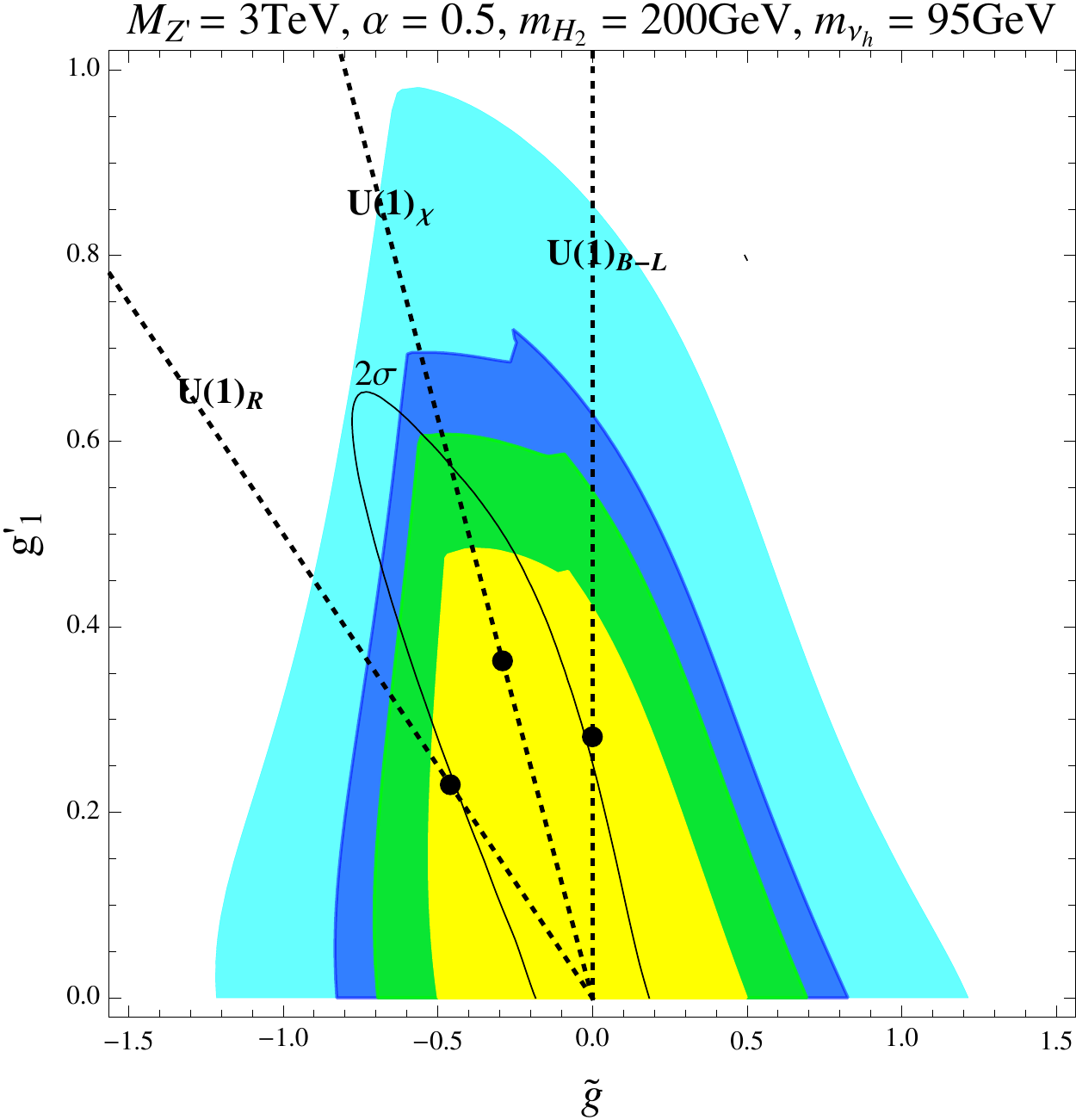}} 
\caption{Same as fig.~\ref{fig.stab-gt-g1p} with $M_{Z'} = 3$ TeV. \label{fig.stab-gt-g1p-3TeV}}
\end{figure}

In fig.~\ref{fig.stab-gt-g1p-3TeV} we show the same study for $M_{Z'} = 3$ TeV. The regions defined by the RG analysis are unchanged with respect to the $M_{Z'} = 2$ TeV case but the LHC bounds become looser. This allows to explore bigger values of the Abelian gauge couplings which can even fall in a region in which the perturbativity is spoilt (cyan), although only for a small slice of the parameter space. For these values, a bigger $\alpha$ is ineffective to increase $Q_{\rm max}$ because the poor behaviour of the model is due to the loss of perturbativity in the Abelian sector and not to the instability of the vacuum. For heavier $Z'$s, the constraints from di-lepton searches at the LHC are overtaken by EWPTs which still enclose this Abelian extension in a configuration almost as stable as the SM, provided $m_{\nu_h} \lesssim M_{Z'}$.

\begin{figure}
\centering
\subfigure[]{\includegraphics[scale=0.59]{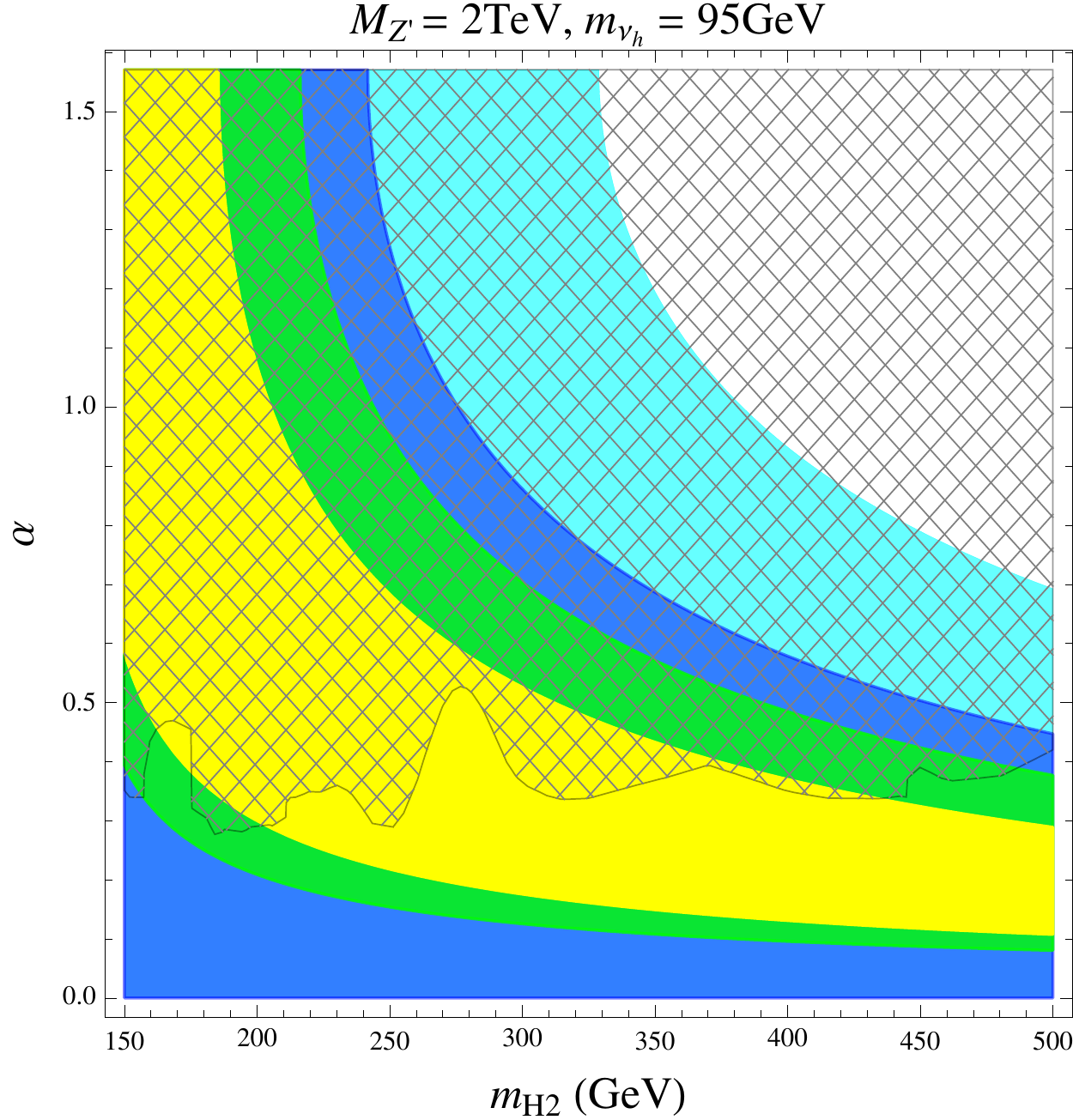}}
\subfigure[]{\includegraphics[scale=0.59]{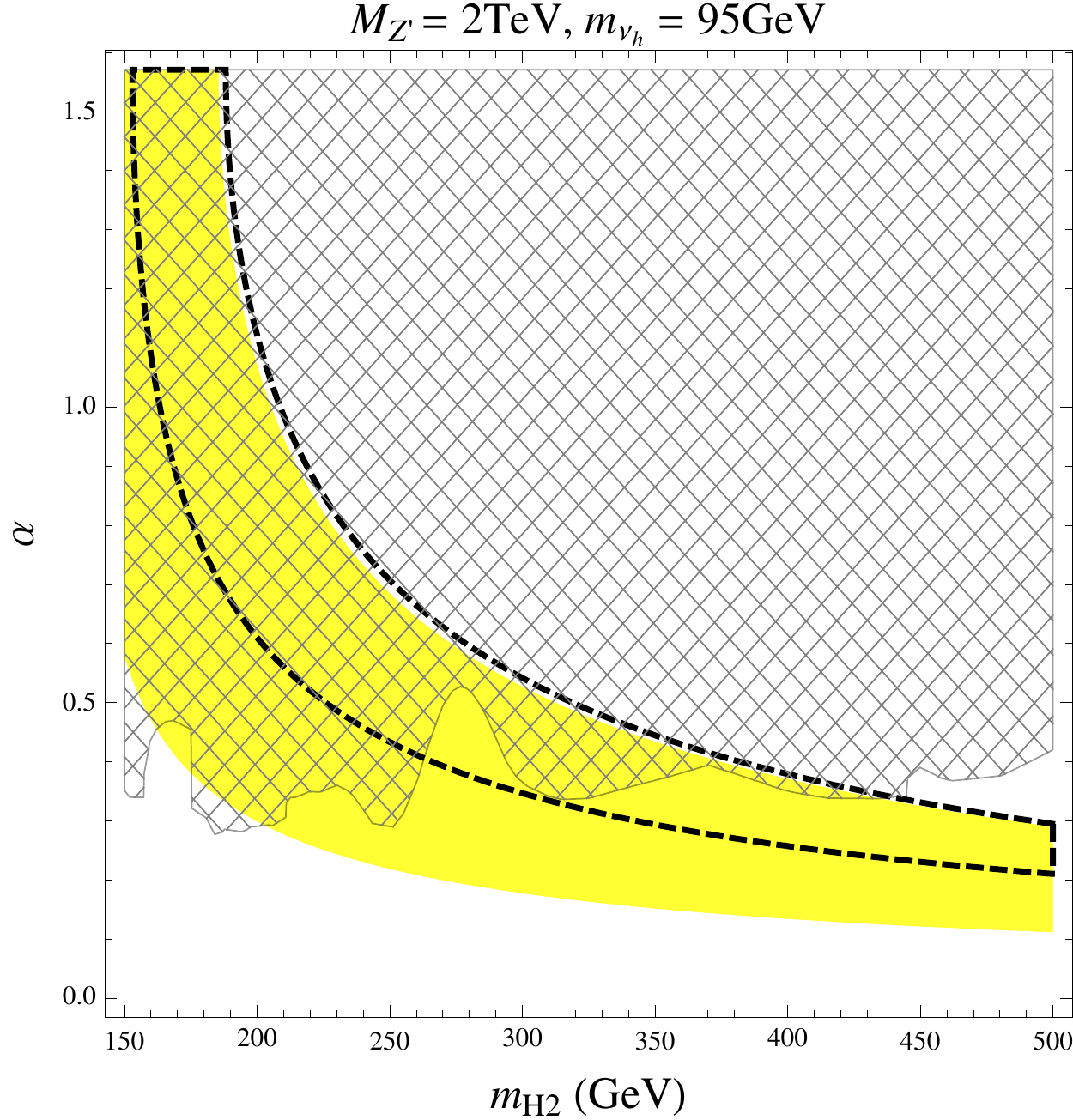}}
\caption{(a) Stability and perturbativity regions in the $(m_{H_2}, \alpha)$ plane according to the colour legend described in fig.~\ref{fig.stablegend}. (b) Comparison between NLO (yellow region) and LO (region delimited by dashed line) results for the requirements of stability and perturbativity up to the GUT scale. The hatched area is excluded by the \texttt{HiggsBounds} analysis.  \label{fig.stab-mh2-alpha}}
\end{figure}

A similar study is carried out in the $(m_{H_2}, \alpha)$ plane in order to emphasise the impact of the extended scalar sector. The results are presented in fig.~\ref{fig.stab-mh2-alpha}(a) where the hatched area has been excluded by LHC data using the \texttt{HiggsBounds} tool. The $U(1)'$ Abelian gauge couplings used for this particular analysis are $\tilde g = -0.13$ and $ g'_1 = 0.11$, which have been chosen on the $2\sigma$ contour line. We have explicitly verified that different values on the same curve do not lead to any qualitative change. Interestingly, the cyan region, in which the RG behaviour of these models worsens with respect to the SM case, is completely disallowed for $m_{H_2} \lesssim 500$ GeV. 
Notice also that, for $m_{H_2} \lesssim 250$ GeV, both stability and perturbativity are satisfied, up to the GUT scale and above, mainly for a highly-mixed scalar sector while, for heavier $H_2$, the mixing angle is bounded from above and the same regions extend horizontally. These regions will eventually shrink at bigger values of the heavy Higgs mass due to a loss of perturbativity of the $\lambda_2$ quartic coupling. \footnote{The parameter space studied in fig. 6(a) is the unique case showing sensitivity to our different definitions of the perturbativity condition. Using the NDA prescription with $N = 1$ we found a broadening of the stability and perturbativity regions at high values of $\alpha$, excluded by Higgs searches, with the cyan region completely covering the white space. The portion of the parameter space allowed by the HiggsBounds analysis remains unaffected by the perturbativity condition being dominated by the stability requirement.} \\
A similar analysis, focusing on the role of an extra real scalar field, has been presented in \cite{Falkowski:2015iwa} in which analogue results have been obtained concerning the one-loop stability of the scalar potential up to the Planck scale in the $(m_{H_2}, \alpha)$ plane. 
The scenario presented in \cite{Falkowski:2015iwa} is not embedded into an extended gauge sector and therefore the VEV  of the extra scalar is unconstrained, while in our case its value is intimately related to the $Z'$ mass\footnote{Moreover, we require a complex scalar since it has to provide the longitudinal degree of freedom to the new massive gauge boson.}. Nevertheless, the source of the main differences is in the perturbative order of the RG evolution. Indeed, 
as emphasised in \cite{Coriano:2015sea}, the role of a NLO RG study, with two-loop $\beta$ functions and one-loop matching equations, is to improve the stability of the potential and therefore it is found to be necessary to draw conclusive statements on the high-energy behaviour of the vacuum. To highlight the impact of a NLO analysis we show in fig.~\ref{fig.stab-mh2-alpha}(b) the region of stability and perturbativity up to the GUT scale at NLO (yellow region) in comparison to a LO only (region enclosed in the dashed curve) study in which only one-loop $\beta$ functions and tree-level matching conditions are employed. It is evident that, in a NLO analysis, the parameter space providing a well-behaved theory up to high energies broadens towards smaller values of the scalar mixing angle $\alpha$, which are, quite interestingly, in the region allowed by Higgs searches at the LHC.

\begin{figure}
\centering
\subfigure[]{\includegraphics[scale=0.59]{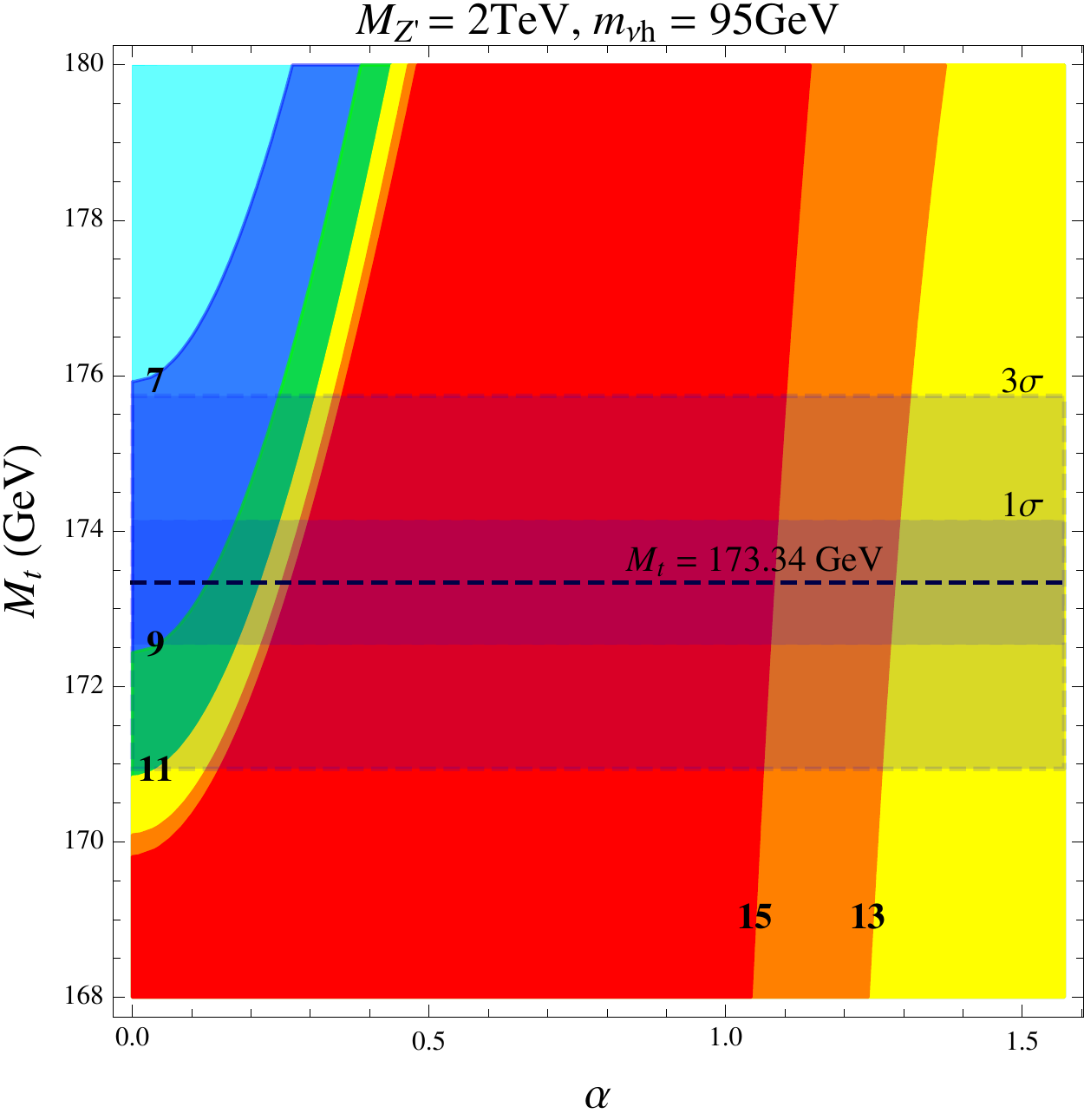}}
\subfigure[]{\includegraphics[scale=0.59]{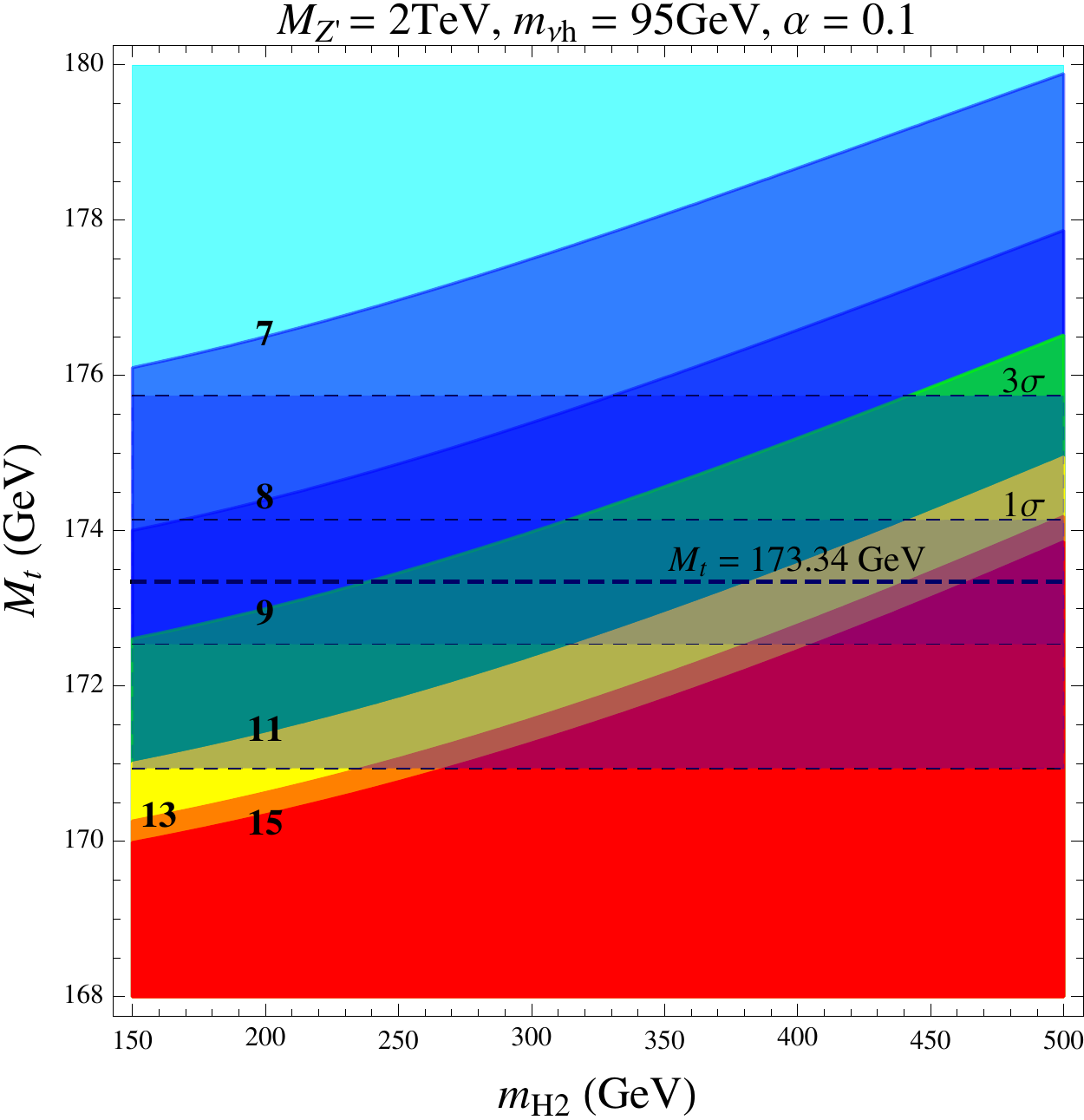}}
\caption{Stability and perturbativity regions in the (a) $(\alpha, M_t)$ and (b) $(m_{H_2}, M_t)$ spaces. The regions are enclosed by $Q_{\rm max} = 10^x$ GeV with $x = 7,8,9,11,13,15$. \label{fig.mtop}}
\end{figure}

In fig.~\ref{fig.mtop} we show regions of stability and perturbativity as a function of the top pole mass $M_t$ and of $\alpha$ (fig.~\ref{fig.mtop}(a) with $m_{H_2} = 200$ GeV) or $m_{H2}$ (fig.~\ref{fig.mtop}(b) with $\alpha = 0.1$). The bold numbers $x$ on the boundaries of the different coloured regions represent the maximum scale of stability and perturbativity $Q_{\rm max} = 10^x$ GeV. The dashed line corresponds to the central value of the top mass $M_t = 173.34 \pm 0.76$ GeV \cite{ATLAS:2014wva} which is an average from the combined analysis of ATLAS, CMS, CDF and D0, extracted through Monte Carlo (MC) modelling of production and decay of the top quark in hadronic collisions. Due to its origin, the measurement leads to the so-called MC mass which does  represent neither the pole mass nor the $\overline{\rm{MS}}$ mass. Usually, one assumes that the MC mass is sufficiently close to the pole mass with differences estimated of the order of 1 GeV \cite{Buckley:2011ms,Juste:2013dsa,Frixione:2014ala} and then extracts its $\overline{\rm{MS}}$ value using matching conditions at the EW scale. The corresponding Yukawa coupling  $Y_t$ is then determined and fed to the RG equations. Unfortunately, this procedure is plagued by many sources of uncertainty and therefore it would be much better, due to its critical role \cite{Bezrukov:2012sa,Buttazzo:2013uya,Masina:2012tz,Alekhin:2012py,Bezrukov:2014ina}, if the MC event generators were defined directly in terms of the $\overline{\rm{MS}}$ Yukawa parameter. The analysis presented in fig.~\ref{fig.mtop}, far from being exhaustive, has the only purpose of showing how the impact of the top mass, investigated in a window of 1 and 3 $\sigma$ according to \cite{ATLAS:2014wva}, is affected by the parameters of the enlarged scalar potential. As one naively expects, the mixing angle $\alpha$ weakens the destabilising effect of the top (fig.~\ref{fig.mtop}(a)) and eventually completely overcomes it for $\alpha \gtrsim 0.4$. The restoration of the vacuum stability, for a fixed value of the top mass, also appears as one increases the mass $m_{H_2}$ of the heavy Higgs (fig.~\ref{fig.mtop}(b)). Contrary to $\alpha$, the effect of $m_{H_2}$ is softened and, in the range 150 \rm{GeV} $\le m_{H_2} \le$ 500 \rm{GeV} with $\alpha = 0.1$, only shifts the instability induced by the top quark to higher values of its mass.

\section{LHC Phenomenology}
\label{sec:pheno1}
In this section we explore the possible experimental signatures that characterise the class of $Z'$ models encompassed by our general analysis.
\subsection{$Z'$ production and decay}
\begin{figure}
\centering
\subfigure[]{\includegraphics[scale=0.39]{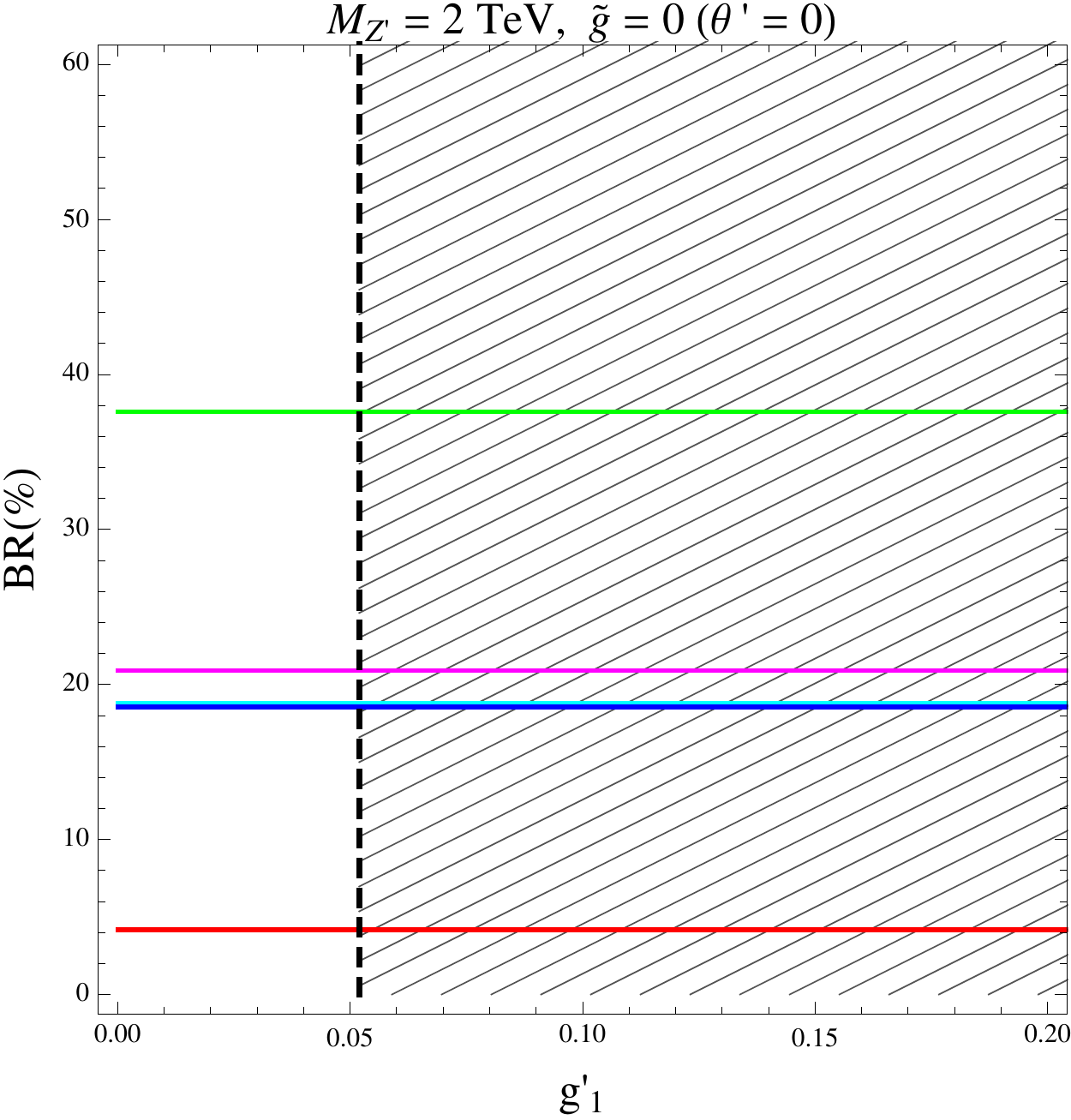}} 
\subfigure[]{\includegraphics[scale=0.39]{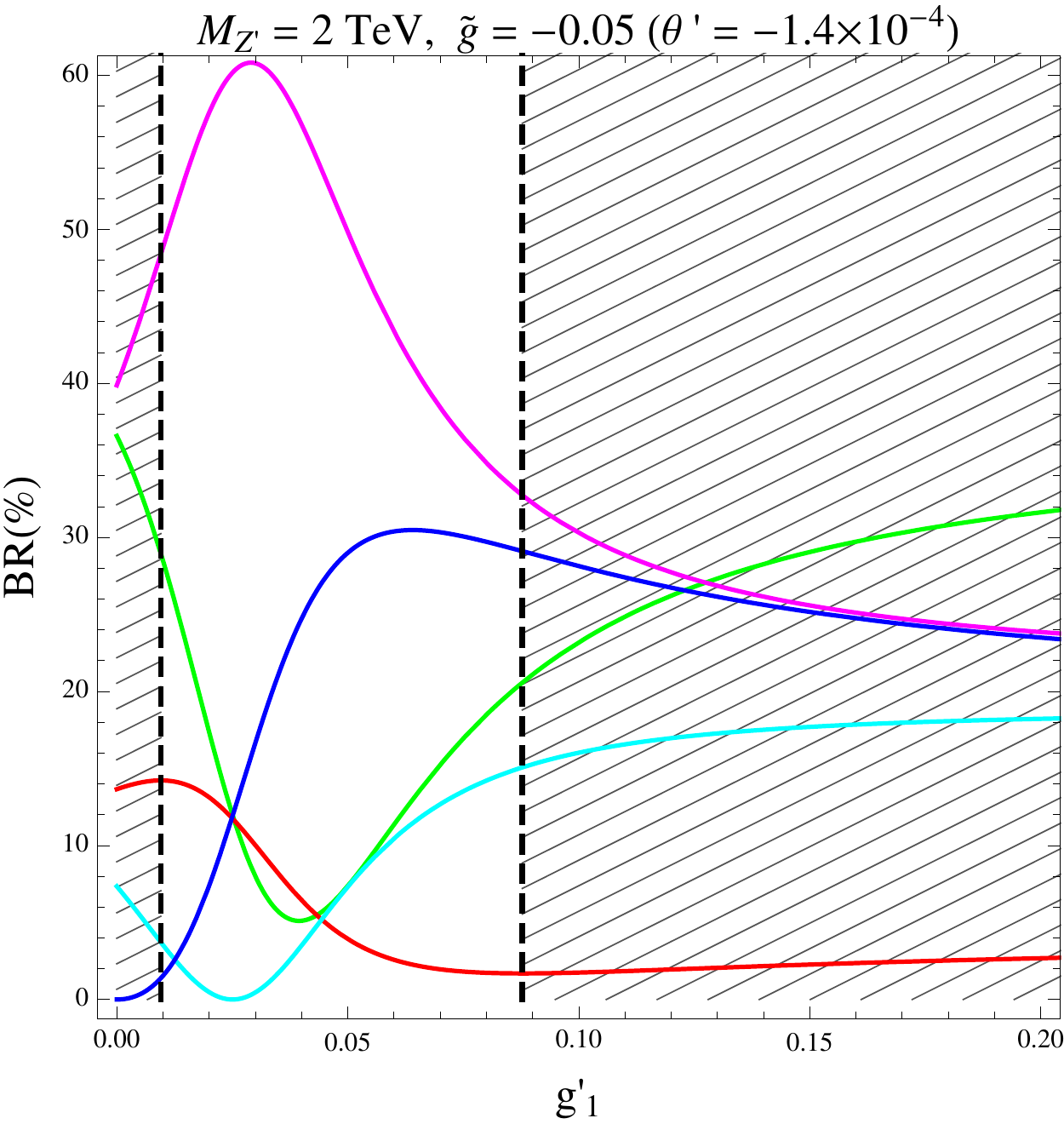}} 
\subfigure[]{\includegraphics[scale=0.39]{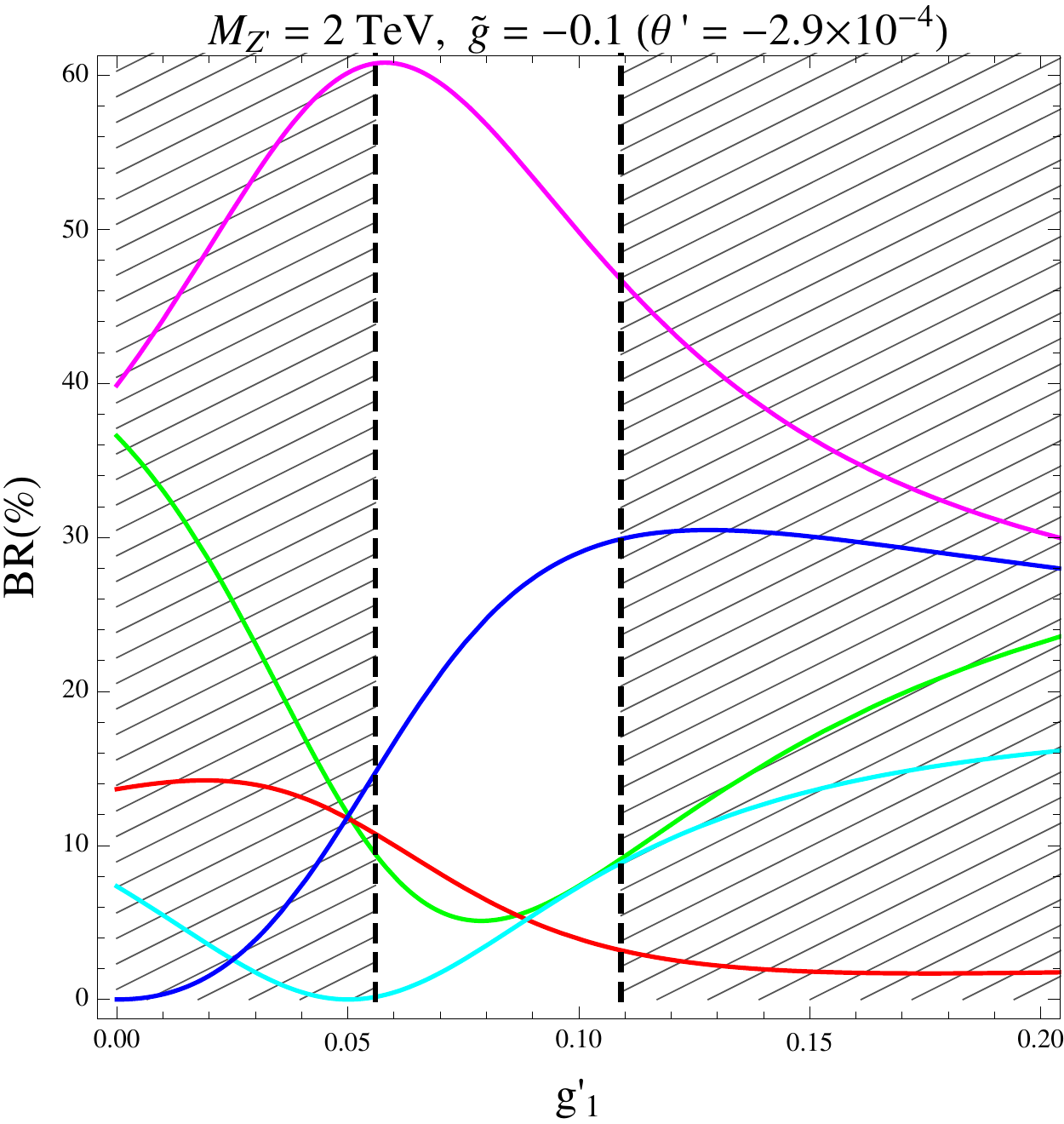}} \\
\subfigure[]{\includegraphics[scale=0.39]{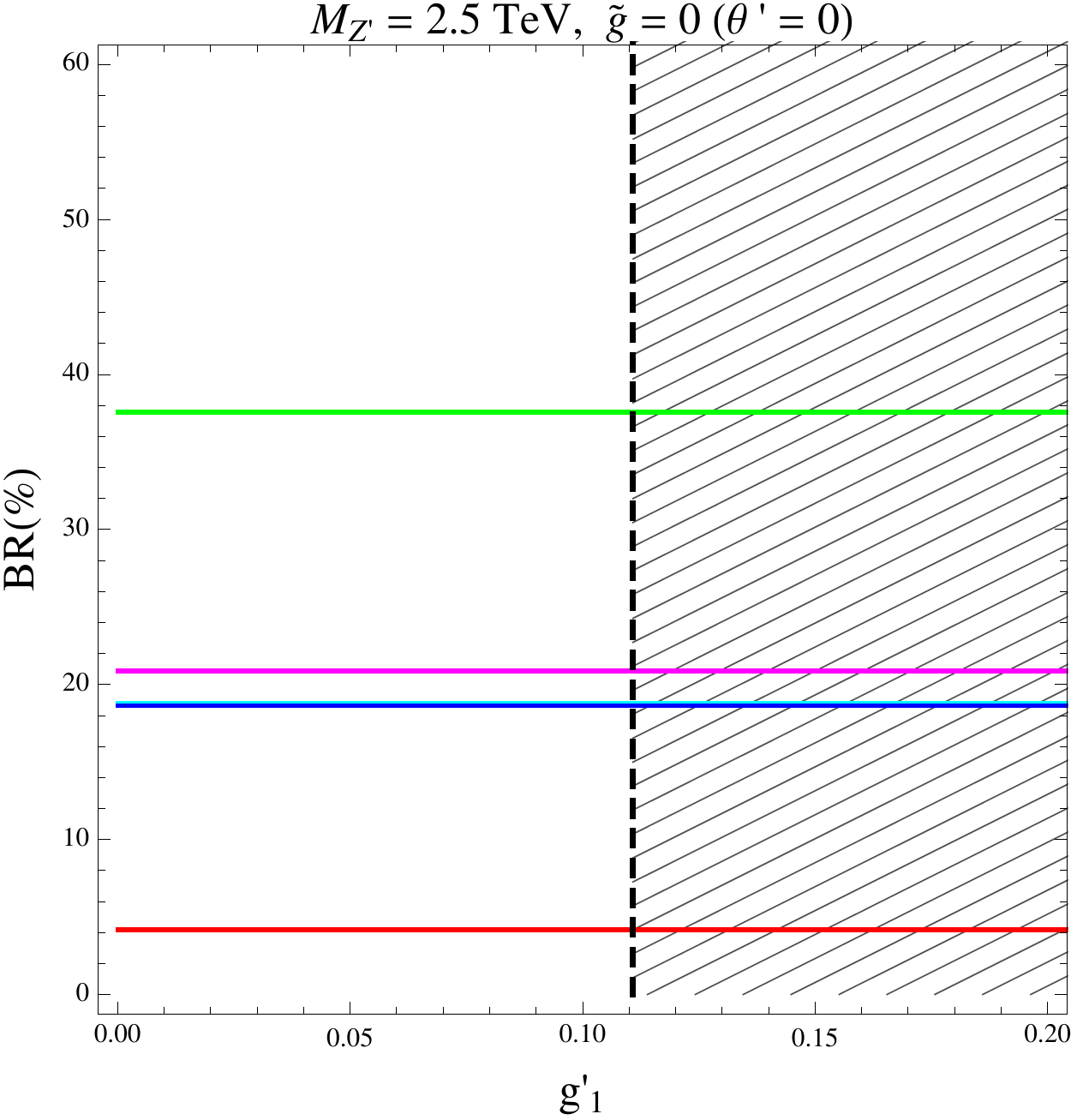}} 
\subfigure[]{\includegraphics[scale=0.39]{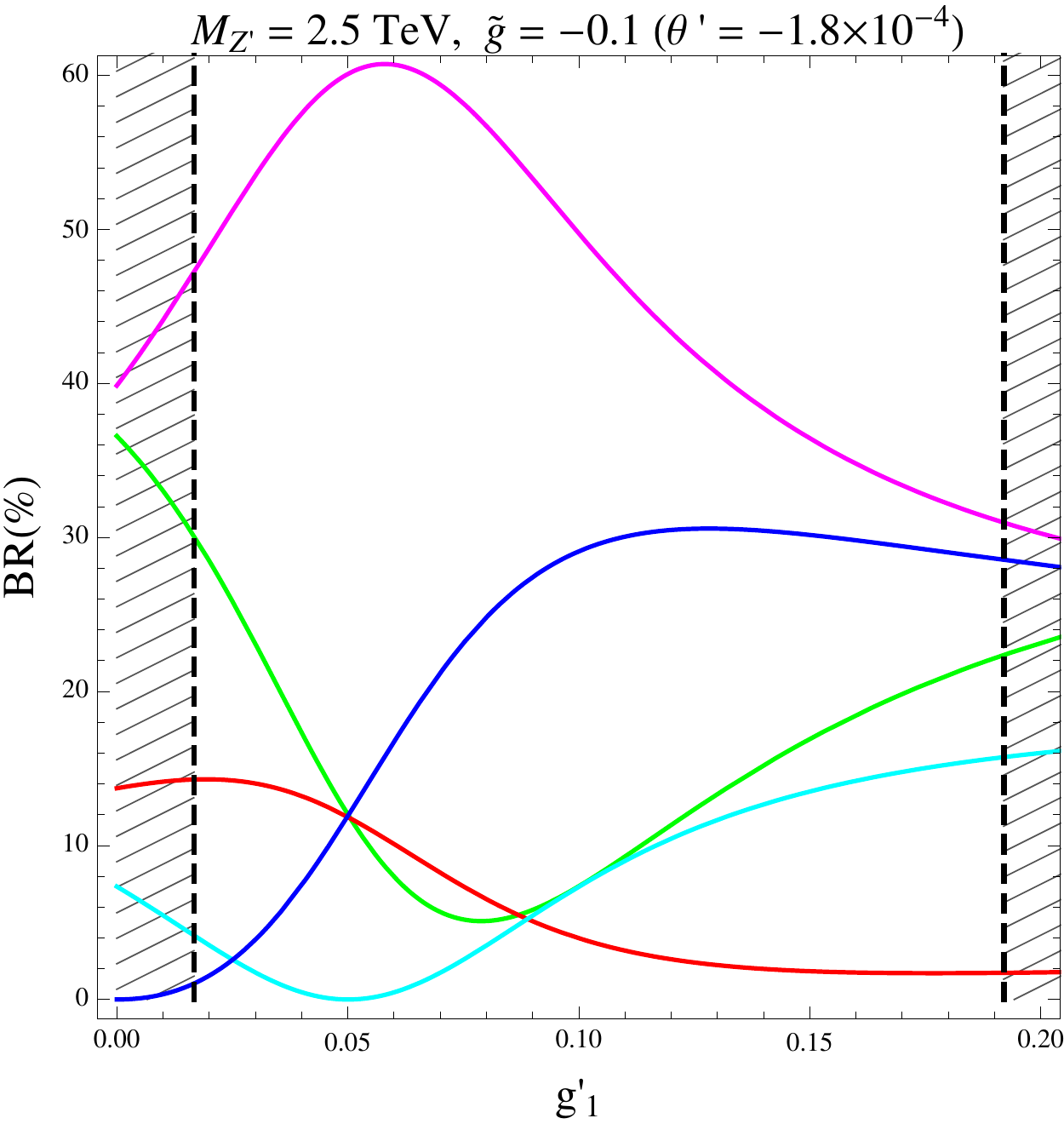}} 
\subfigure[]{\includegraphics[scale=0.39]{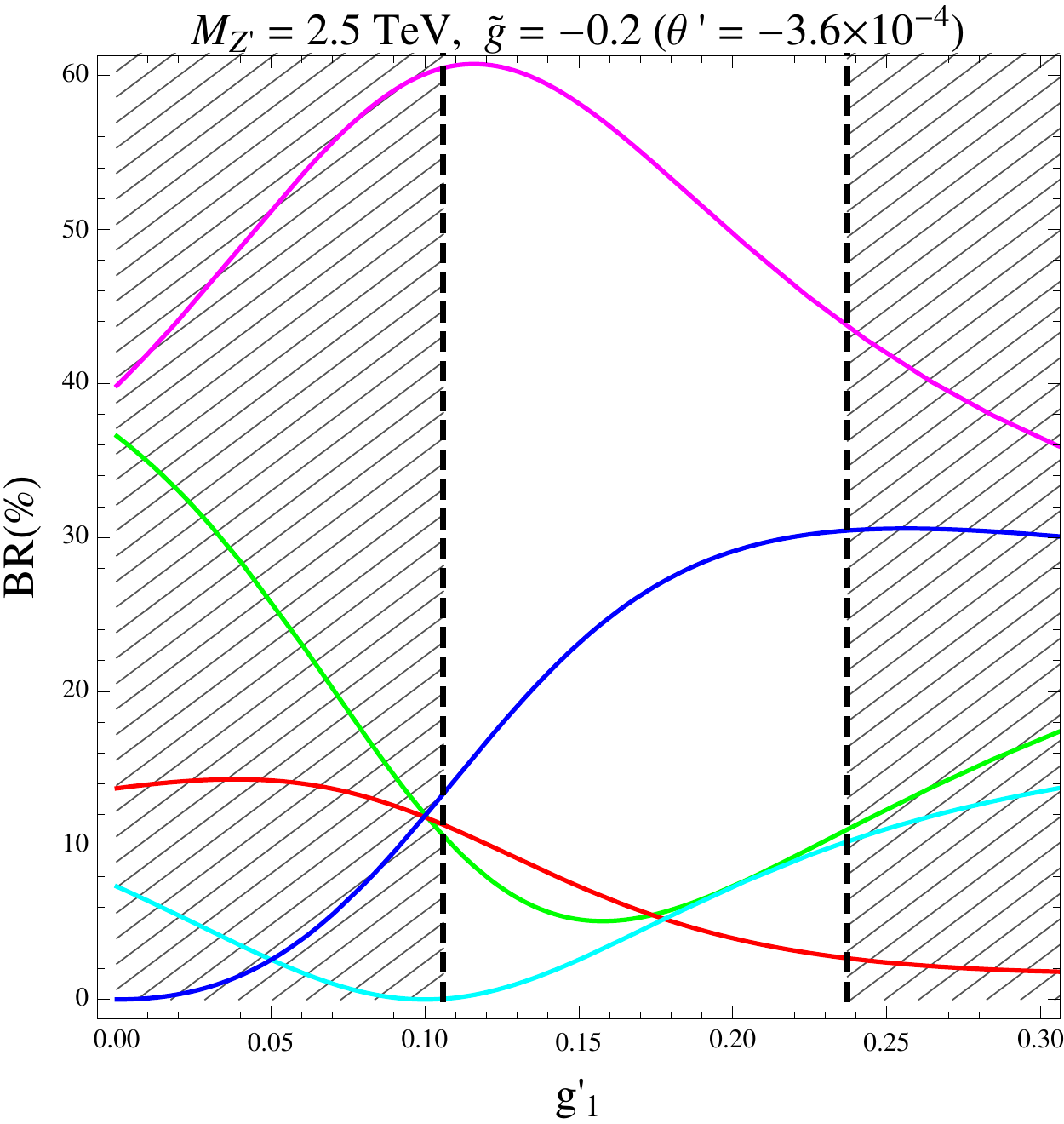}} \\
\subfigure[]{\includegraphics[scale=0.39]{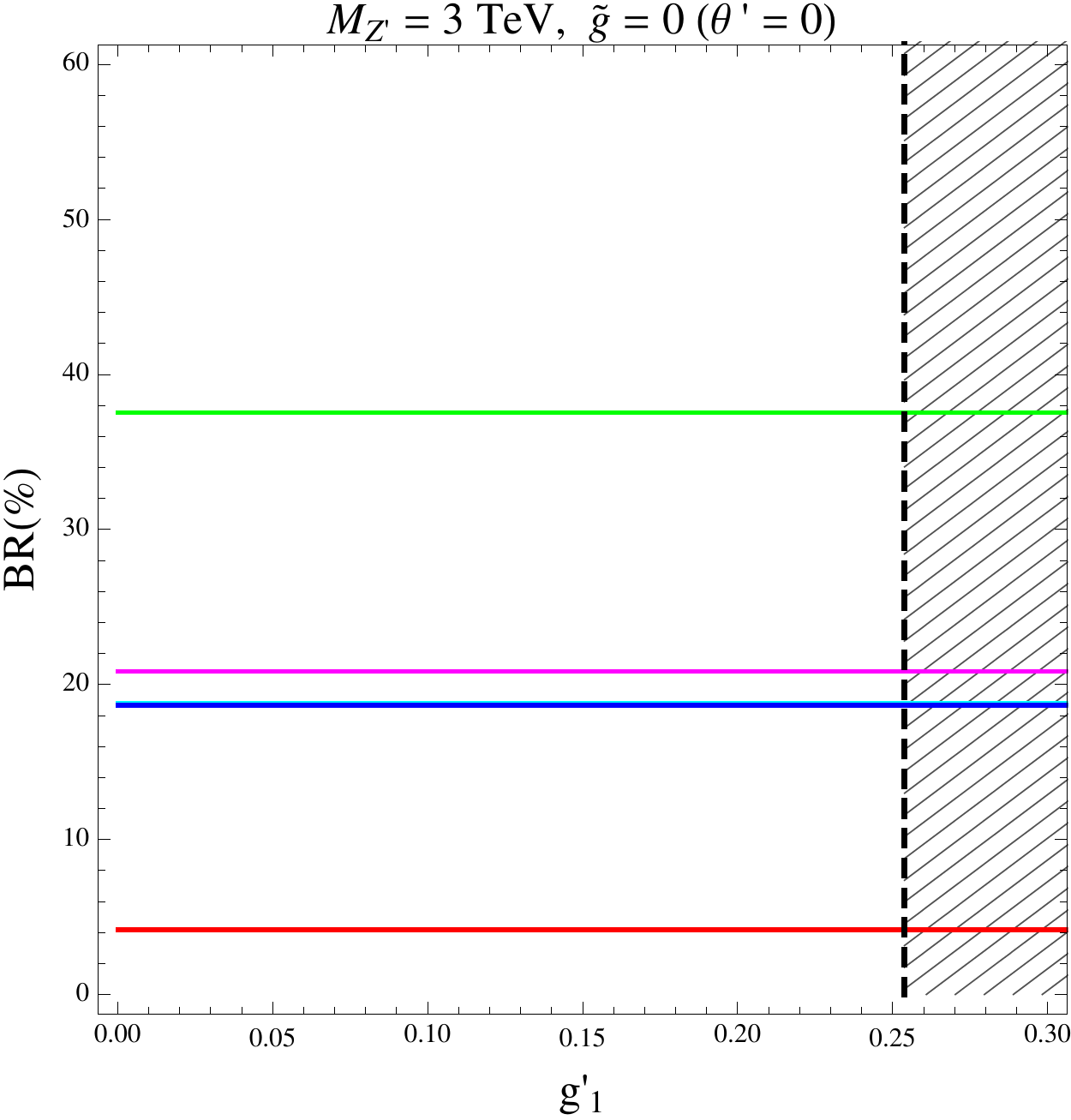}} 
\subfigure[]{\includegraphics[scale=0.39]{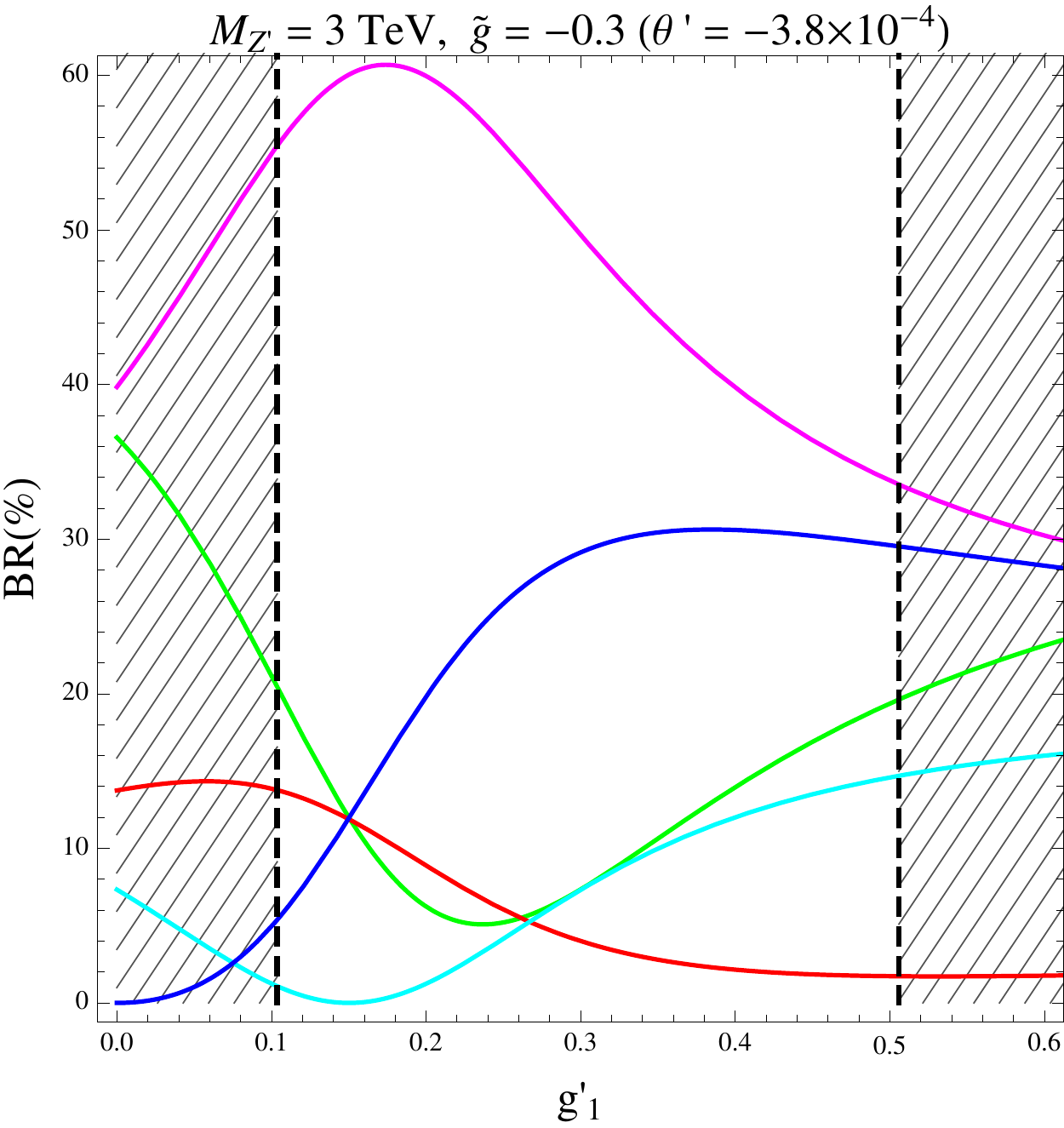}} 
\subfigure[]{\includegraphics[scale=0.39]{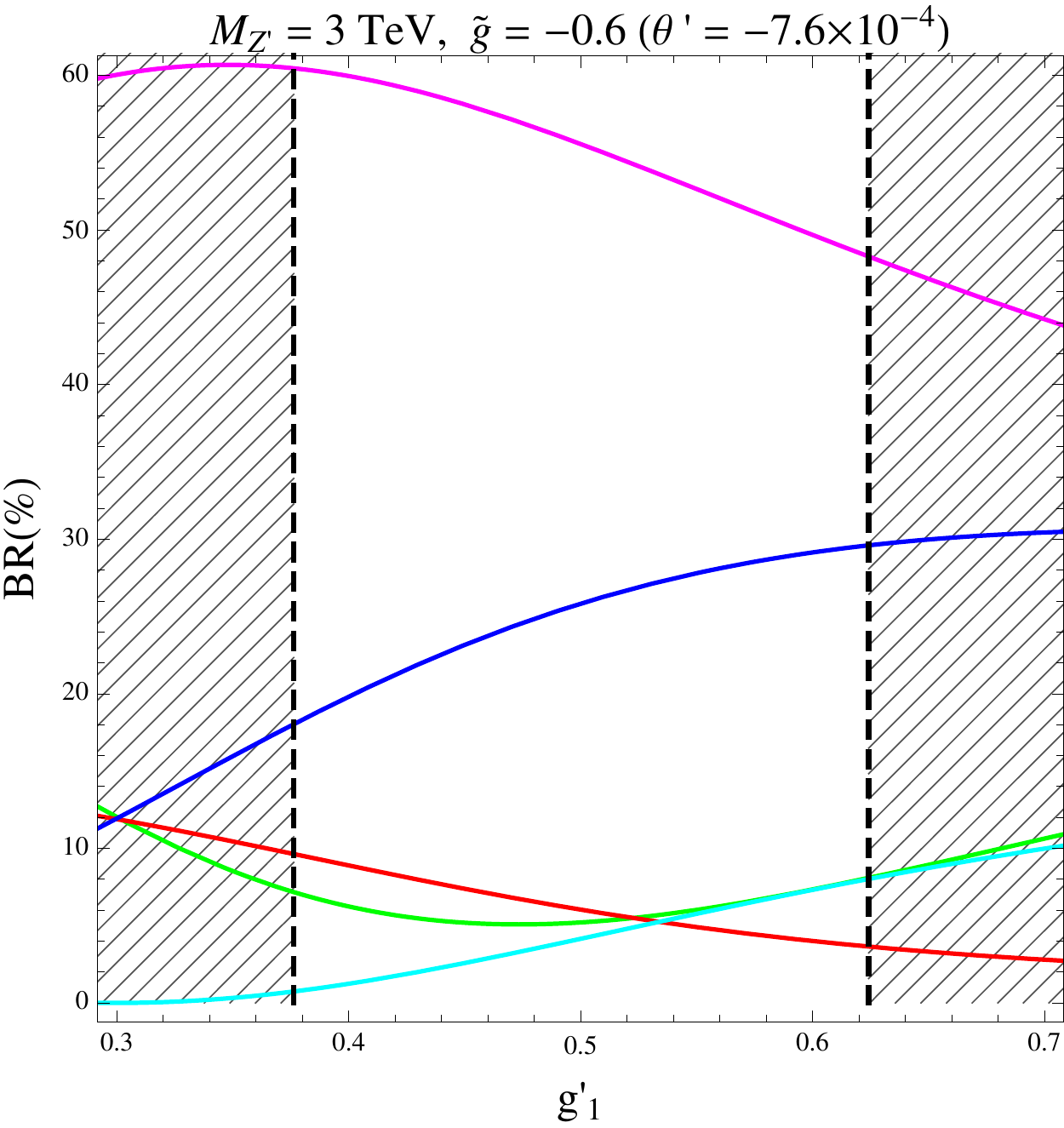}} 
\caption{$Z'$ BRs into fermionic final states as a function of $g_1'$ for several values of $\tilde g$ and for $M_{Z'} = 2, 2.5, 3$ TeV. 
  The $\tilde g = 0$ case corresponds to the pure $B-L$. Dashed regions are excluded by LHC Run 1 data at 95\% CL. 
  The green, cyan, purple, red and blue lines correspond to the $Z'$ decay into two charged leptons, light neutrinos, light quarks, 
  top quarks and heavy neutrinos, respectively.\label{Fig.BRs1}}
\end{figure}

In fig.~\ref{Fig.BRs1} the different branching ratios of a $Z'$ decaying in fermions are displayed for the values of $M_{Z'} = 2, 2.5$ and $3$ TeV and for different 
$\tilde{g}$ (dashed regions are excluded according to fig.~\ref{EWPTvsDY}). The partial decay width of the $Z'$ decaying into leptons and quarks is
\bea
\Gamma(Z' \rightarrow f \bar f) = \frac{M_{Z'}}{12 \pi} C_f \sqrt{1 - \frac{4 m_f^2}{M_{Z'}^2}} \left[ \frac{C_{f, L}^2 + C_{f, R}^2}{2} \left( 1 - \frac{m_f^2}{M_{Z'}^2} \right) + 3 C_{f, L} \, C_{f, R} \frac{m_f^2}{M_{Z'}^2} \right],
\eea
where $C_f$ is the colour factor while $C_{f, L/R}$ are the left/right-handed couplings of the fermion $f$ to the $Z'$ boson. These are given by
\bea
C_{f,L} = - e \frac{s'}{s_W c_W} \left( T^3_f - s_W^2 Q_f\right) + \bar g_{f, L} \, c' \,, \qquad C_{f, R} = e \frac{s_W \, s'}{c_W} Q_f + \bar g_{f, R} \, c',
\eea 
where we have used the short-hand notation $s_W \equiv \sin \theta_W$, $c_W \equiv \cos \theta_W$, $s' \equiv \sin \theta'$ and $c' \equiv \cos \theta'$, with $T^3_f$ being the third component of the weak isospin, $Q_f$ the electric charge in unit of $e$ and $\bar g_{f, L/R} = \tilde g \, Y_{f, L/R} + g'_1 \, z_{f, L/R}$. The hypercharge $Y_{f, L/R}$ is normalised as $Y_{f} = Q_f - T^3_f$ while $z_f$ is the $U(1)'$ charge which we have identified with the $B-L$ quantum number. The decay width of the $Z'$ into heavy neutrinos is
\bea
\Gamma(Z' \rightarrow \nu_h \nu_h) = \frac{M_{Z'}}{24 \pi}  (z_{\nu_R} \, g_1' \, c' )^2 \left( 1- \frac{4 m_{\nu_h}^2}{M_{Z'}^2} \right) \sqrt{1- \frac{4 m_{\nu_h}^2}{M_{Z'}^2}} \,.
\eea
It is clear how the favourite channel for the pure $B-L$ is in two charged leptons \cite{Basso:2008iv}. 
This decay mode provides nearly  40\% of the width, the remainder being almost equally shared by the
decays into light quarks, heavy and light neutrinos (note that we considered the branching for charged leptons, light and heavy neutrino states summed over generations). 
When we turn our attention to the gauge mixing, the decay mode hierarchy is drastically changed.  
In the limit of a sequential $Z'$, which is recovered for $g'_1 = 0$, a preference for light quark decays reaches the highest value of
60\% for the Branching Ratio (BR). The leptonic decay mode is sub-dominant in this range but starts becoming sizeable with increasing $g'_1$. This is to be expected given the restoration
of the pure $B-L$ case in the limit of large $g'_1$ (equivalent to $\tilde g \rightarrow 0$).
The BR in heavy neutrinos shows a variability with $g'_1$. Moving from zero in the sequential limit it reaches the highest contribution at $\sim 30\%$ of the BR in the pure $B-L$ case. Indeed, the partial width $Z'\rightarrow \nu_h \nu_h$ is independent of $\tilde g$ and it is solely controlled by the Abelian coupling $g'_1$.

\begin{figure}
\centering
\subfigure[]{\includegraphics[scale=0.39]{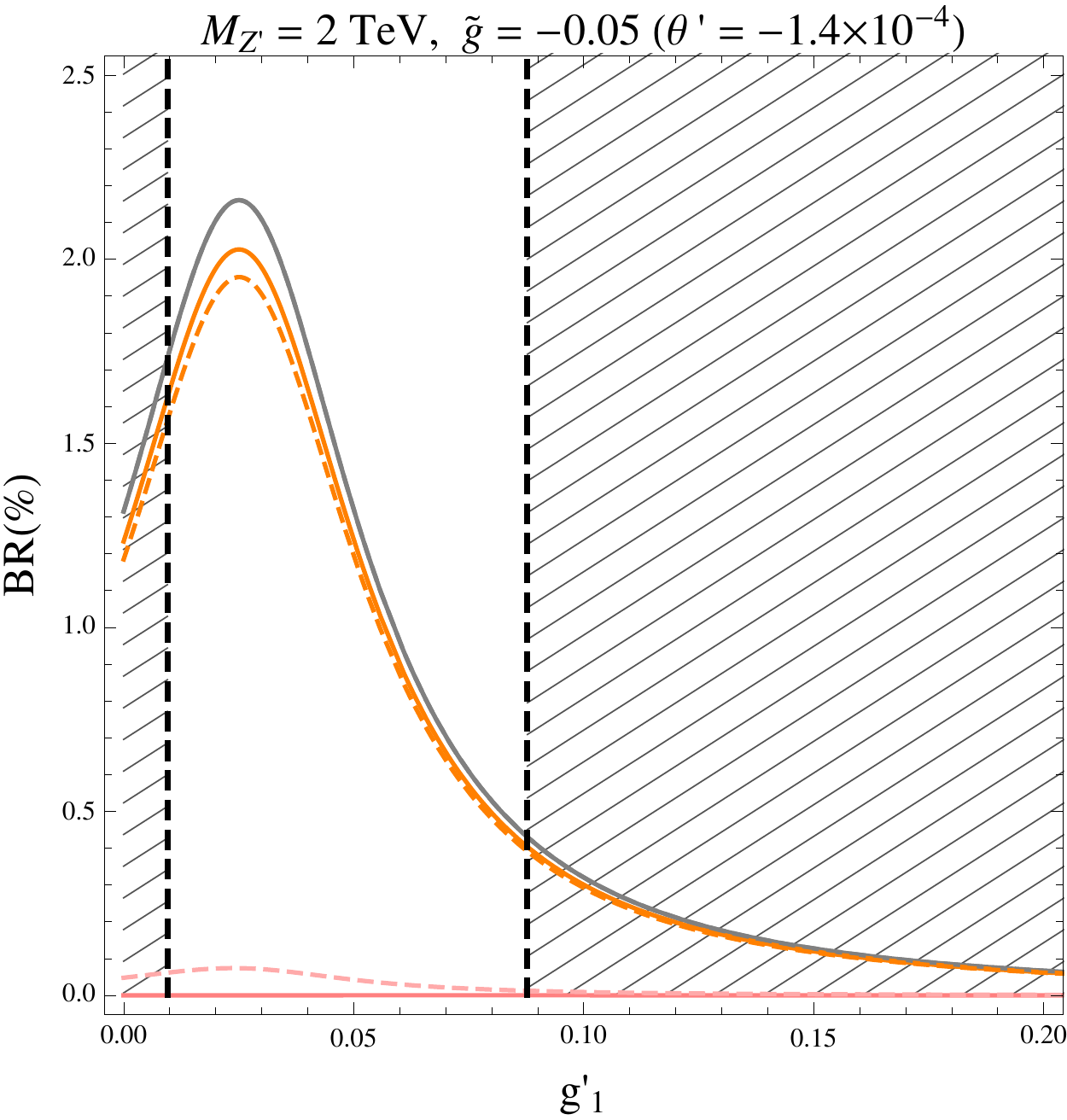}} 
\subfigure[]{\includegraphics[scale=0.39]{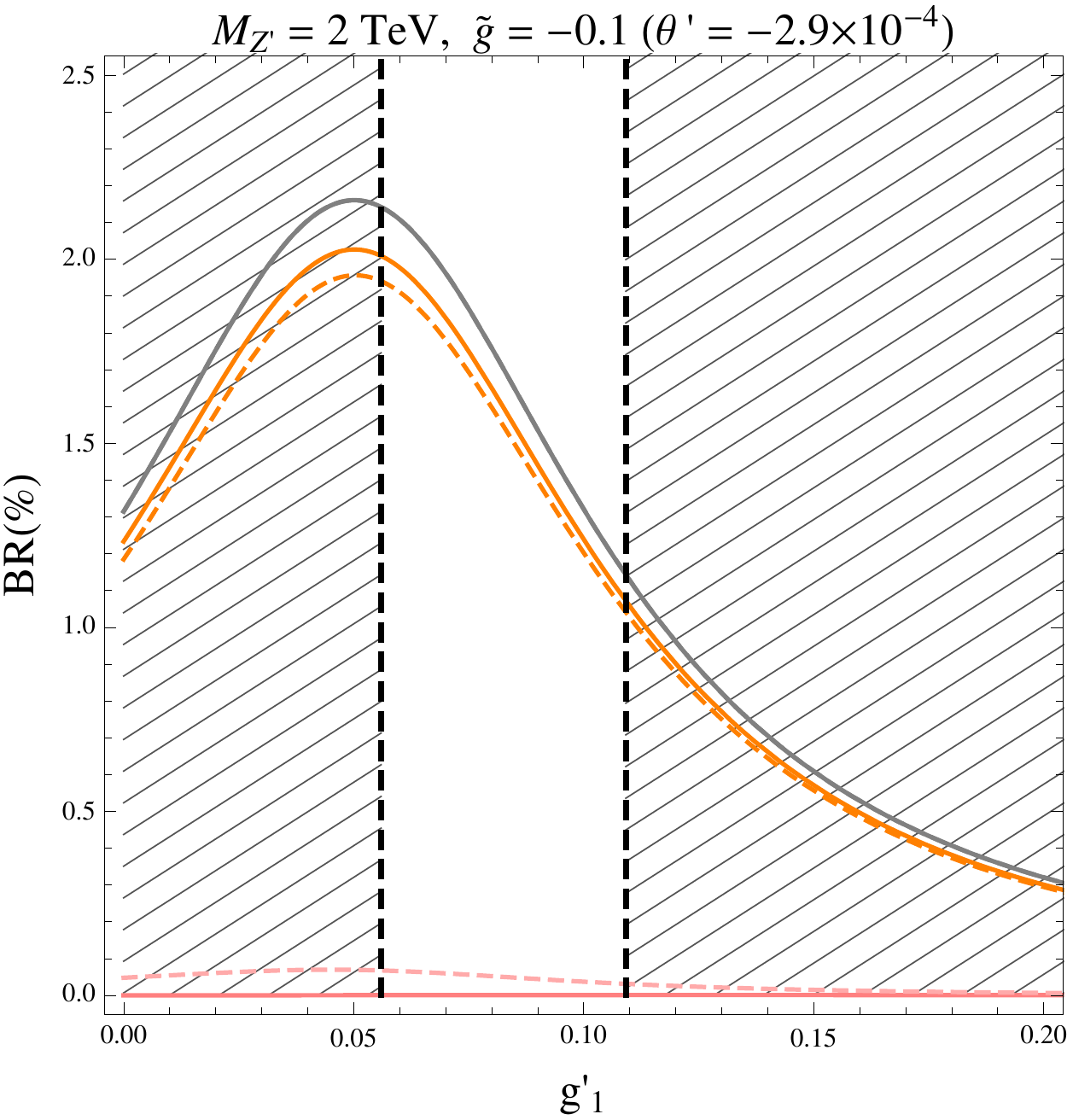}} \\
\subfigure[]{\includegraphics[scale=0.39]{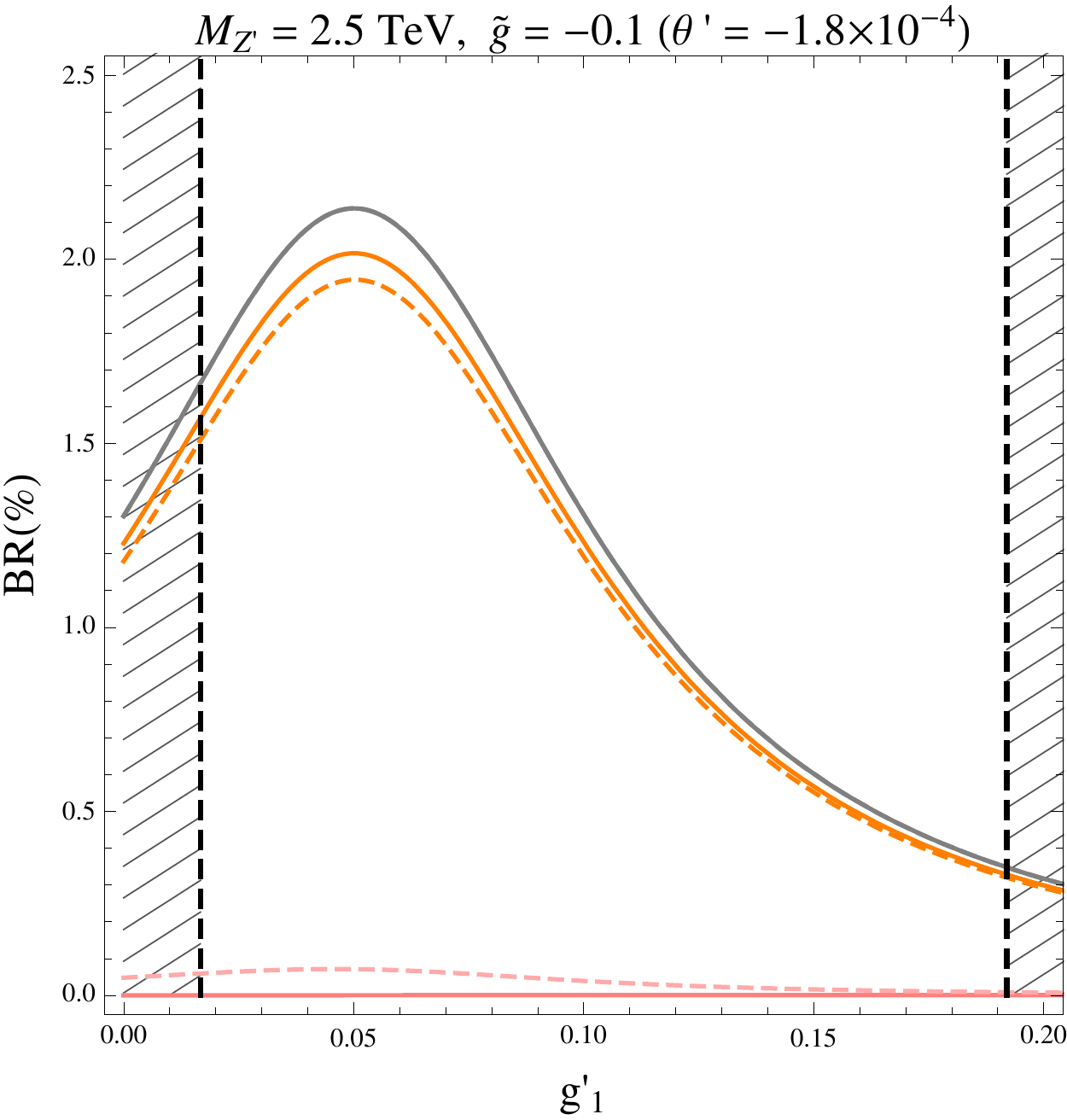}} 
\subfigure[]{\includegraphics[scale=0.39]{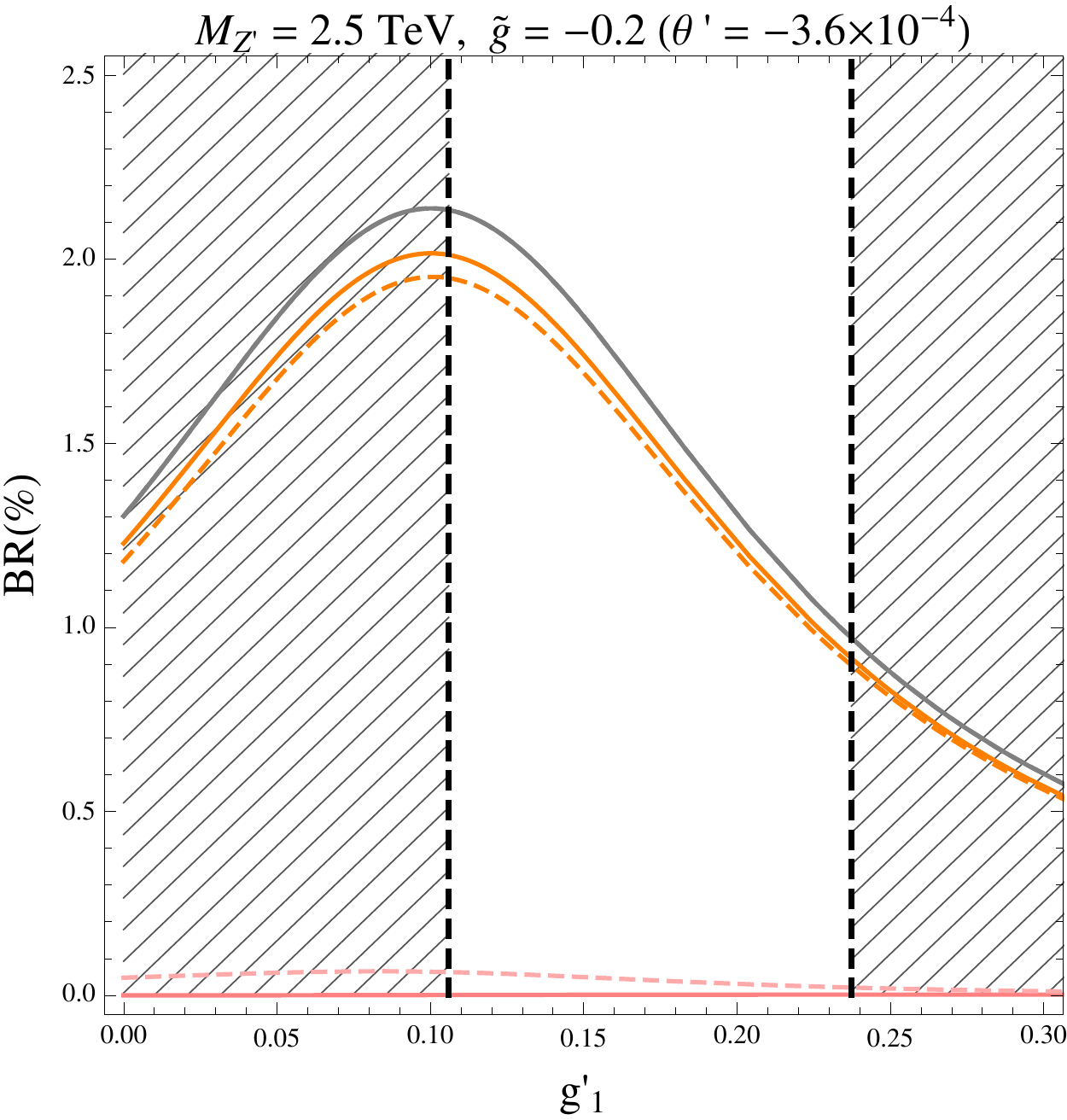}} \\
\subfigure[]{\includegraphics[scale=0.39]{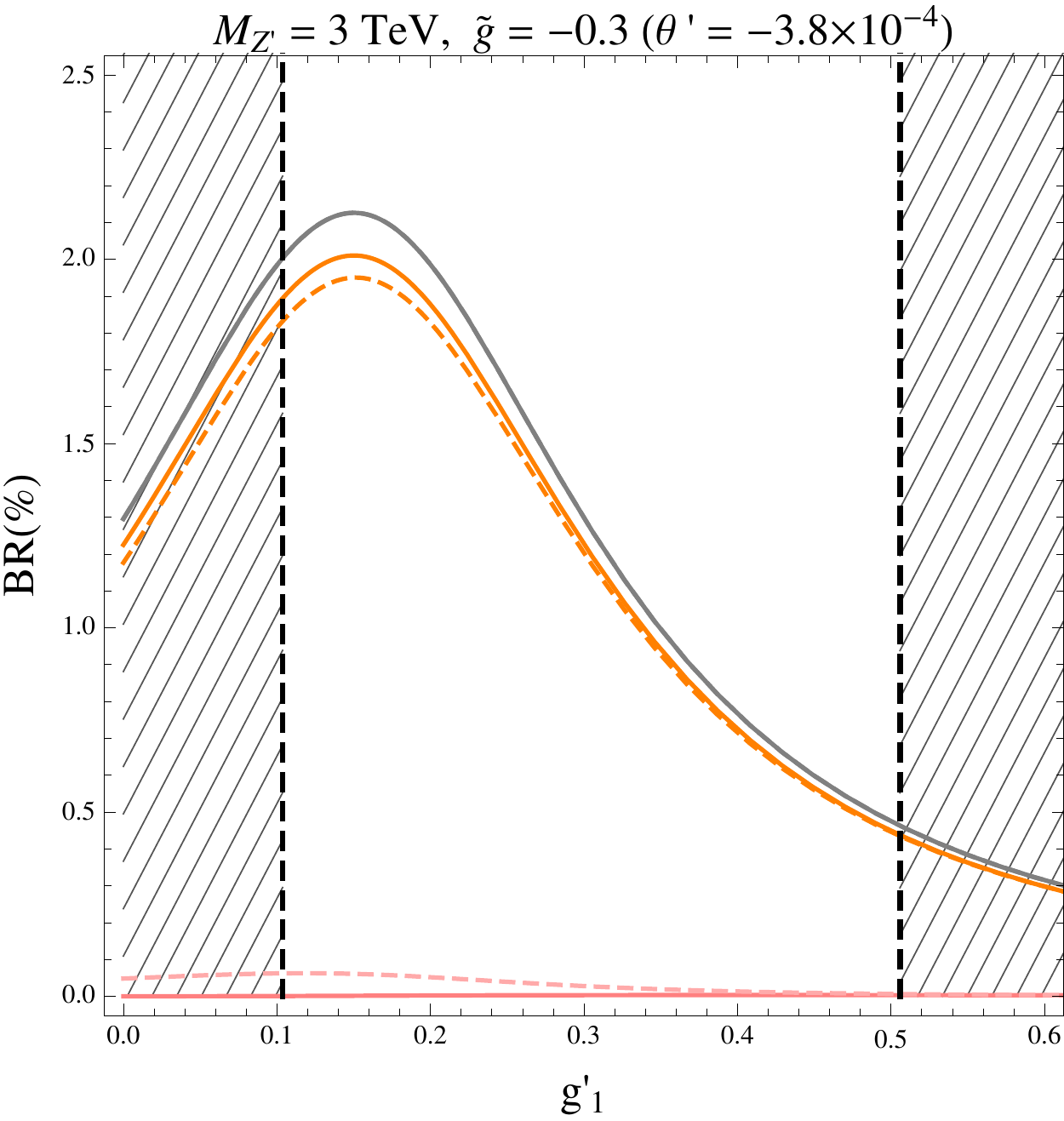}} 
\subfigure[]{\includegraphics[scale=0.39]{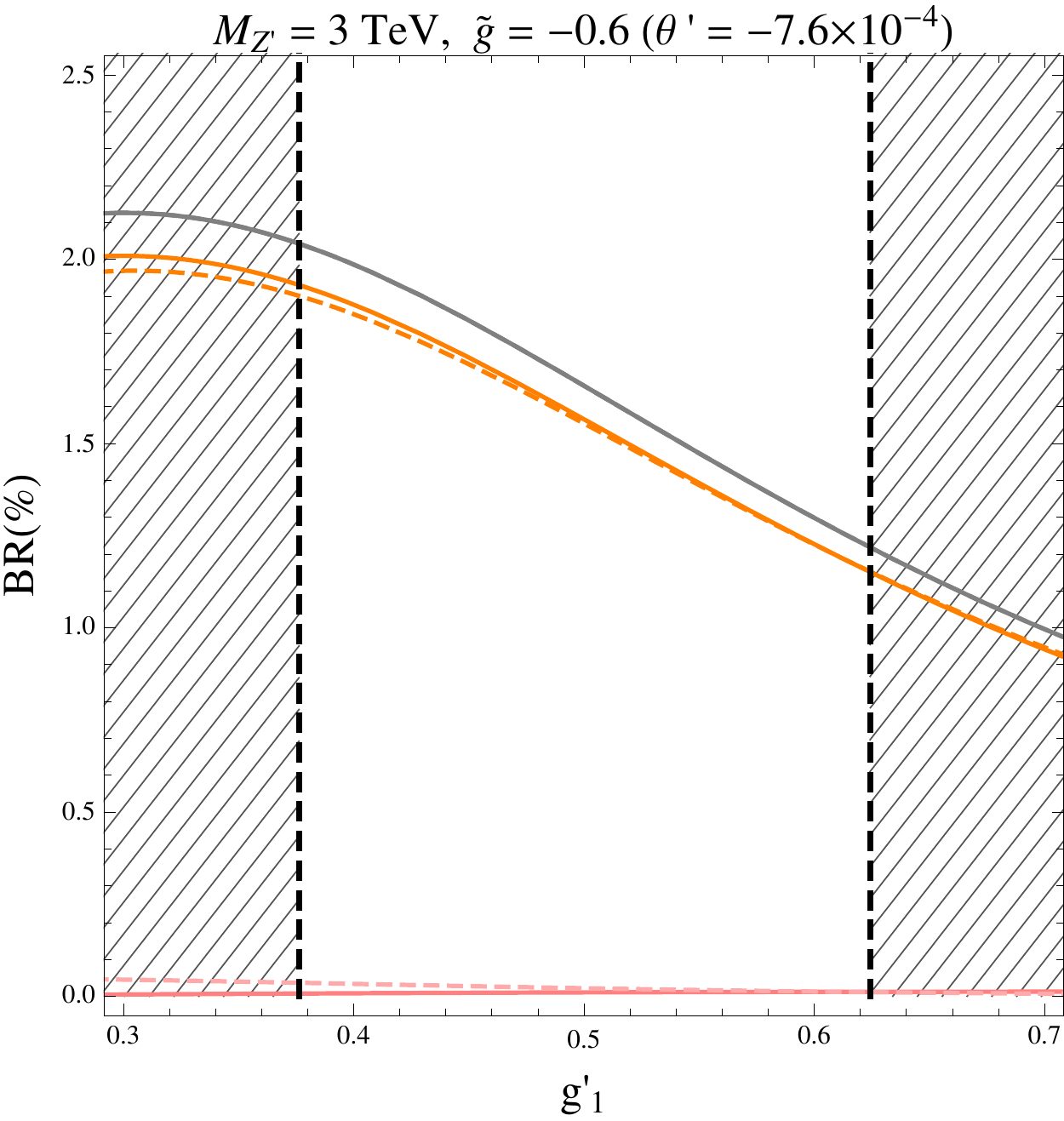}} 
\caption{$Z'$ BRs into $WW$ (gray), $ZH_1$ (orange) and $ZH_2$ (pink) as a function of $g_1'$ for several values of $\tilde g$ and for $M_{Z'} = 2, 2.5, 3$ TeV. 
  In the pure $B-L$ case these decay channels are absent. Solid (dashed) lines correspond to $\alpha = 0$ ($\alpha = 0.2$). 
  Dashed regions are excluded by LHC Run 1 at 95\% CL \label{Fig.BRs2}}
\end{figure}

\begin{figure}
\centering
\subfigure[]{\includegraphics[scale=0.83]{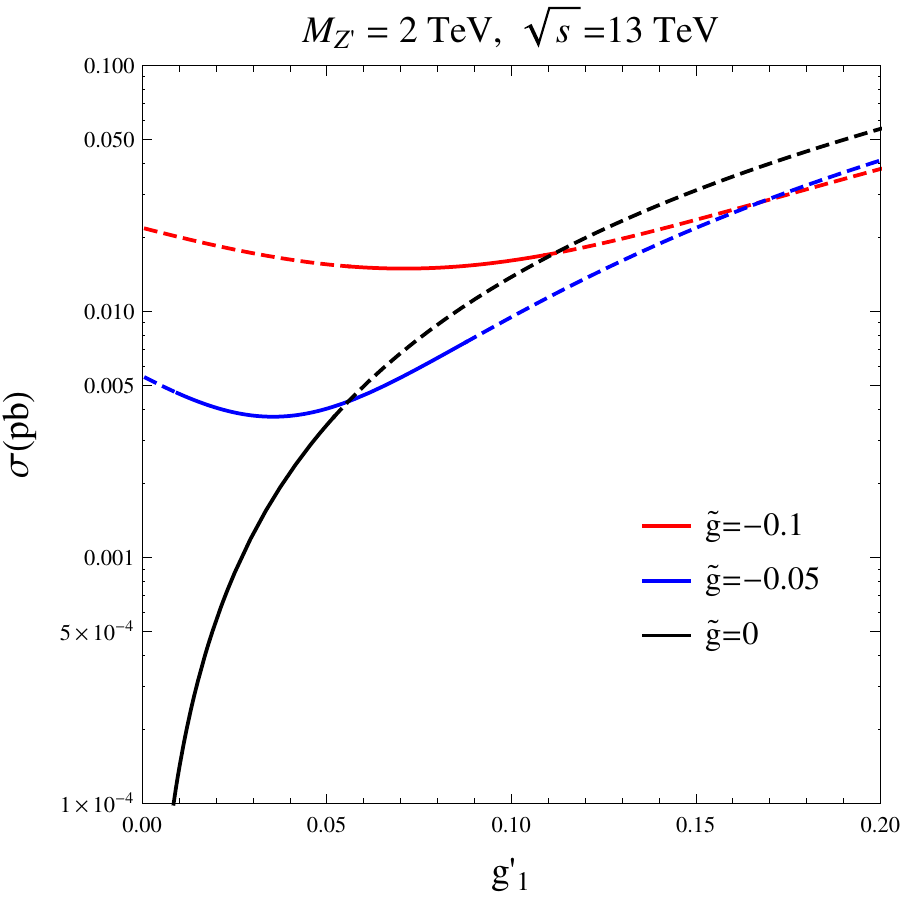}}
\subfigure[]{\includegraphics[scale=0.8]{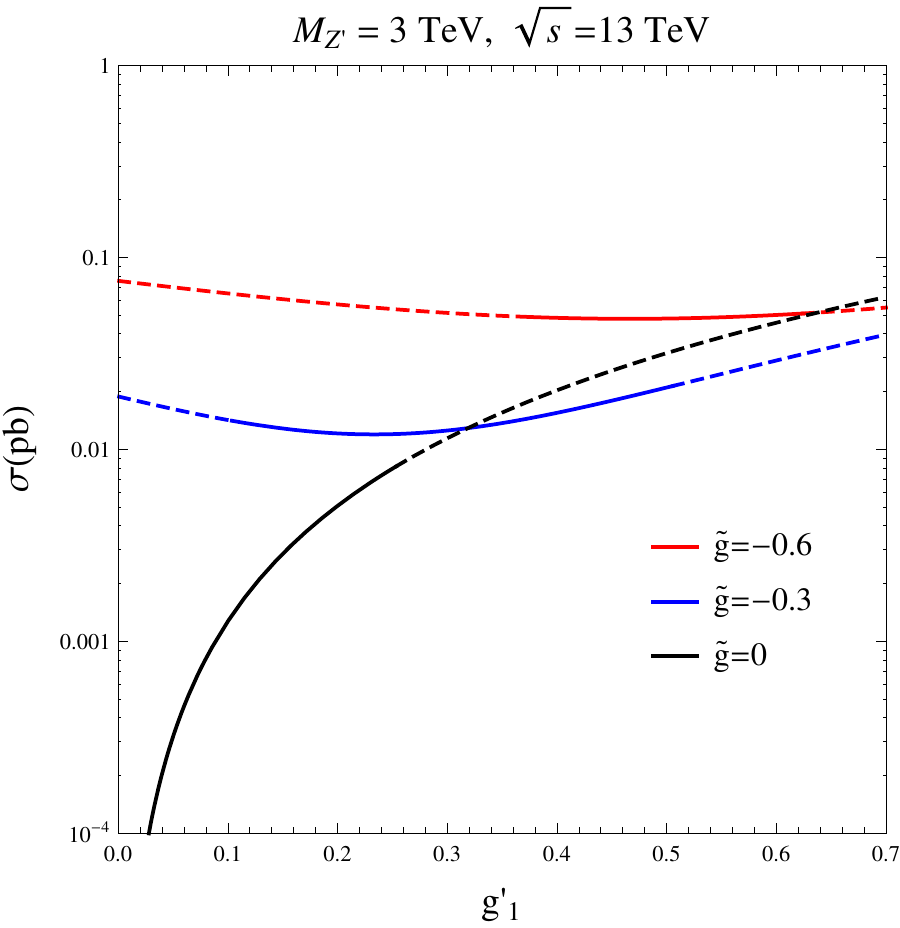}}
\caption{$Z'$ hadro-production cross sections at the LHC as a function of $g_1'$ for different values of $\tilde g$. The dotted parts of the lines refer to values of $g_1'$ 
  excluded by LHC Run 1.\label{Fig.ZpXSLHC13}}
\end{figure}

The possibility to explore different $Z'$ model configuations is enabled by gauge mixing, which opens new decay channels into SM bosons, which are absent in the pure $B-L$, namely, $Z'\to W^+W^-$, $Z\,H_1$ and $Z\,H_2$. The corresponding partial decay widths are given by
\bea
\label{widthsZptogaugeb}
\Gamma(Z' \rightarrow W^+W^-) &=& \frac{1}{48 \pi}  \frac{e^2 \, c_W^2 }{s_W^2} s'^2 \, M_{Z'} \sqrt{1- \frac{4 M_W^2}{M_{Z'}^2}} \left[ \frac{1}{4} \frac{M_{Z'}^4}{M_W^4} + 4 \frac{M_{Z'}^2}{M_W^2} -17 -12 \frac{M_W^2}{M_{Z'}^2} \right],  \nn \\
\Gamma(Z' \rightarrow ZH_1) &=& \frac{1}{96 \pi M_{Z'}^2}  \left[ c_\alpha \, v (c' g_Z - s' \bar g) (c' \bar g + s'  g_Z) + 4 s_\alpha \, x  \, z_\chi^2 \, g_1'^2 \, s' \, c'  \right]^2 p_Z \left( \frac{E_Z^2}{M_Z^2} + 2\right),  \nn \\
\Gamma(Z' \rightarrow ZH_2) &=& \frac{1}{96 \pi M_{Z'}^2} \left[ s_\alpha \, v (c' g_Z - s' \bar g) (c' \bar g + s'  g_Z) - 4 c_\alpha \, x  \, z_\chi^2 \, g_1'^2 \, s' \, c'  \right]^2 p_Z \left( \frac{E_Z^2}{M_Z^2} + 2\right),  \nn \\
\eea
where $p_Z$ and $E_Z$ are the momentum and the energy of the $Z$ boson in the $Z'$ rest frame. Moreover, in the previous equations, we have defined $s_\alpha \equiv \sin \alpha$, $c_\alpha \equiv \cos \alpha$, $g_Z = e / (s_W c_W)$, $\bar g = \tilde g + 2 g_1' \, z_\Phi$ where $z_\Phi$ and $z_\chi$ are, respectively, the $U(1)'$ charges of the SM $SU(2)$ doublet and singlet scalar which, in our case, are $z_\Phi = 0$ and $z_\chi = 2$. \\
The interplay between the mixing in the Abelian and scalar sectors is visible in the corresponding BRs as given in  fig.~\ref{Fig.BRs2}. The decays into charged gauge bosons and  $Z\,H_1$ represent the main patterns 
regardless of the value of the scalar mixing angle in the range $0 \leq \alpha \leq 0.2$ ($\alpha = 0.2$ is a very conservative choice, larger values are possibile depending on the $H_2$ mass, see fig.~\ref{HiggsBounds1}) with  kinematics accounting for the main differences. 
The non-zero scalar mixing also clears the way for a $Z \, H_2$ channel but with a highly dumped BR not exceeding the 0.1\% value. To understand these features, it is instructive to study the partial widths in eq.~(\ref{widthsZptogaugeb}) in the $M_{Z'} \gg M_Z, M_W, M_{H_{1,2}}$ regime taking into account the smallness of the gauge mixing angle $\theta'$ through Eq.~(\ref{thetapexpandend}). In this limit we obtain
\bea
\Gamma(Z' \rightarrow W^+W^-) = \frac{1}{c_\alpha^2} \Gamma(Z' \rightarrow ZH_1) = \frac{1}{s_\alpha^2} \Gamma(Z' \rightarrow ZH_2) =  \frac{1}{192 \pi } \frac{e^2 \, s'^2}{s_W^2 c_W^2} \frac{M_{Z'}^5}{M_{Z}^4}, 
\eea
which clearly describes the behaviour depicted in fig.~\ref{Fig.BRs2}. 
We concentrate now on the on-shell production of a $Z'$ gauge boson through DY mode to accomodate the discovery/exclusion opportunities of our model in LHC Run 2. The computation has been performed using \texttt{CalcHep} \cite{Belyaev:2012qa} and the corresponding $U(1)'$ model file implementation \cite{Basso:2010jm, Basso:2011na} on the
High Energy Physics Model Data-Base (HEPMDB) \cite{hepmdb}.
From this perspective, we present in fig.~\ref{Fig.ZpXSLHC13} the corresponding cross section at 13 TeV as a function of $g'_1$ and for different values of $\tilde{g}$ 
and $Z'$ mass. We consider the bounds coming from the previous significance analysis from DY at LHC Run 1 and highlight the excluded $g'_1$ with dotted lines. 
The $Z'$ of the pure $B-L$ model, which is strongly constrained in terms of $g'_1$, is characterised by a cross section up to $\sigma=5$ fb for $M_{Z'}=2$ 
TeV and up to $\sigma=10$ fb for $M_{Z'}=3$ TeV. Increasing $\tilde{g}$ may increase the $Z'$ coupling to quarks and also allow higher values of $g'_1$  and consequently more sizeable cross sections but 
without exceeding the $\sigma=100$ fb, a value approached at $M_{Z'}=3$ TeV and $\tilde{g}=-0.6$.\\

\subsubsection{Hallmark LHC signatures from a $U(1)'$ $Z'$}

\begin{figure}
\centering
\subfigure[]{\includegraphics[scale=0.8]{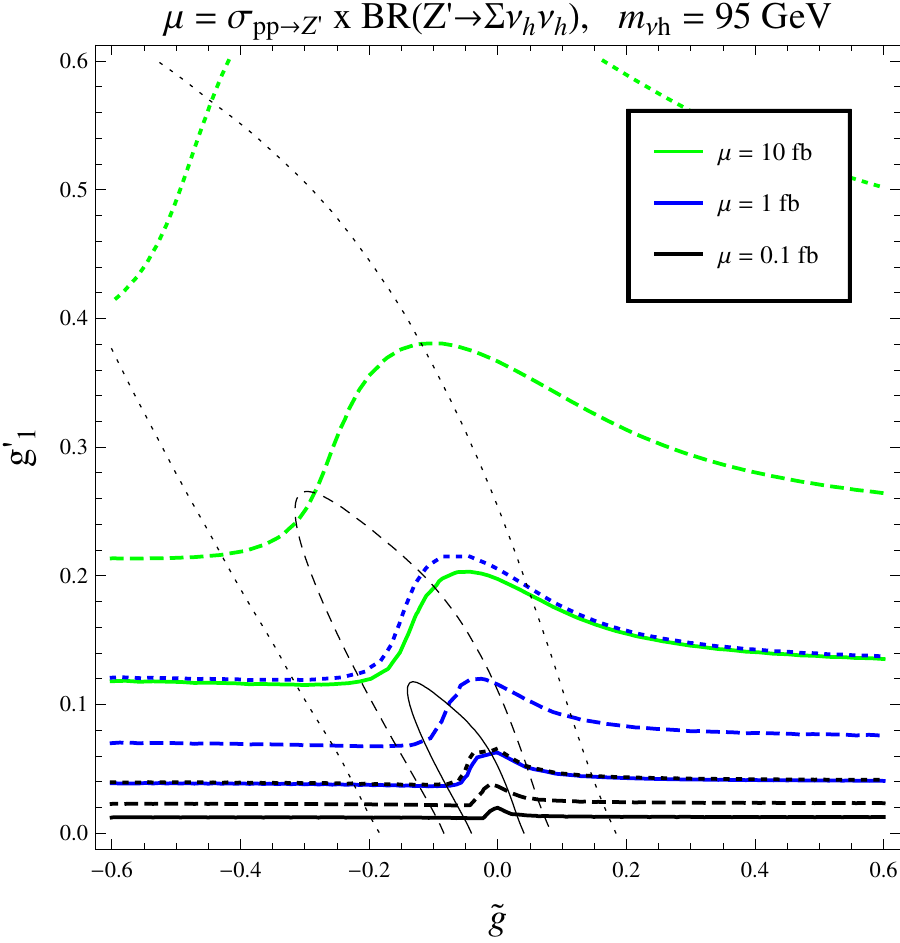}}
\subfigure[]{\includegraphics[scale=0.8]{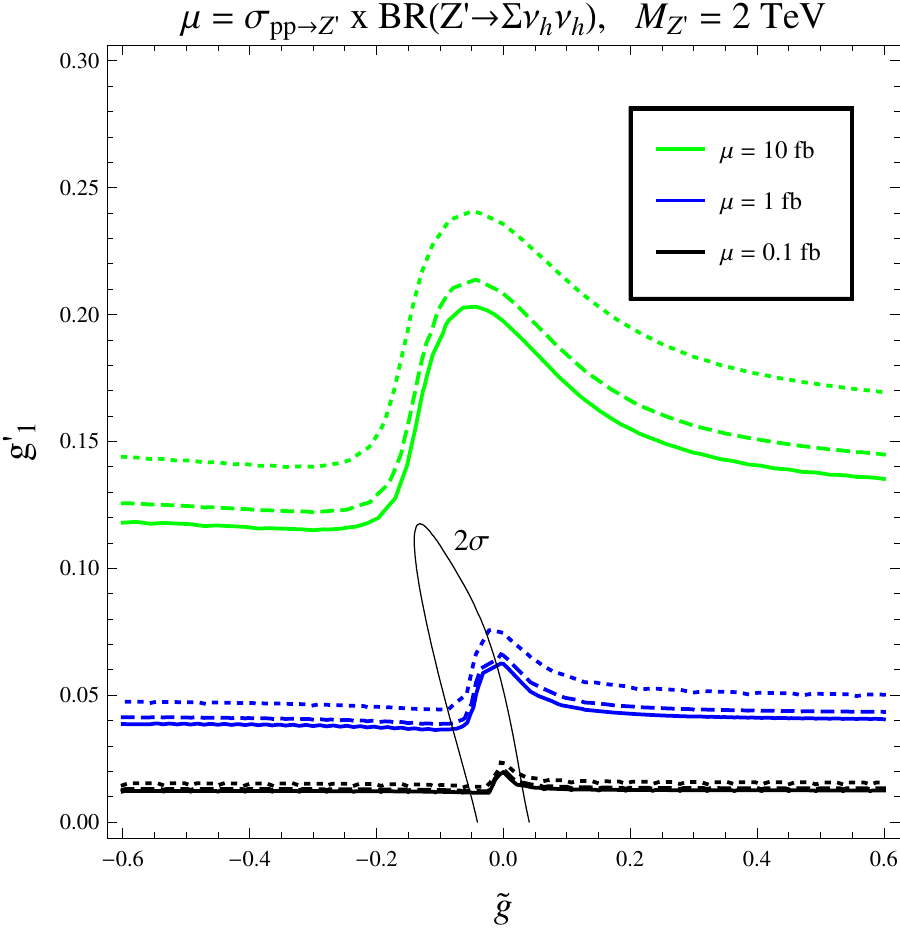}}
\caption{Contour plots of the cross section times BR for the process $pp \rightarrow Z' \rightarrow \nu_h \nu_h$ at the LHC with $\sqrt{s} = 13$ TeV in 
  the $(\tilde g, g_1')$ plane. (a) Solid, dashed and dotted lines refer  to $M_{Z'} = $ 2, 2.5, 3 TeV, respectively, for $m_{\nu_h} =$ 95 GeV. 
  (b) Solid, dashed and dotted lines refer  to $m_{\nu_h} = $ 95, 300, 500 GeV, respectively, for $M_{Z'} =$ 2 TeV. \label{Fig.ZpXSBR}}
\end{figure}

The production of heavy neutrinos from $Z'$ decay is a smoking-gun signal of the particular minimal class of models considered, where an extended fermion sector is
required to cure the anomalies of the new gauge boson. The successive decays of the heavy neutrino may result in distinctive multi-lepton signatures which have been under 
recent investigation (see, for instance, \cite{Emam:2007dy} for the 2-lepton, \cite{Basso:2008iv} for the 3-lepton and  
\cite{Huitu:2008gf,Khalil:2015naa} for the 4-lepton channel).
We explore here the role played, in this process, by the new Abelian couplings and different assignments of $Z'$ and $\nu_h$ masses computing the cross section 
for the production of heavy neutrinos from a decaying $Z'$. The results are plotted in fig.~\ref{Fig.ZpXSBR} with contour plots computed  for a Centre-of-Mass (CM) energy 
 of 13 TeV. 
\subsection{Higgs production and decay}
In this section we address the collider perspectives for a  scalar signal of $B-L$ origin at the LHC. In our setup the parameter space defining the new scalar
sector consists of the mass of the physical heavy scalar $m_{H_2}$ and the related scalar mixing angle $\alpha$.
The mixing angle plays, as expected, a central role scaling all the interactions with SM-like particles by $\cos(\alpha)$ ($\sin(\alpha)$) when involving a $H_1$ ($H_2$) 
and with the complementary angle when involving particles in the peculiar spectrum of the $U(1)'$ model (as $Z'$ and heavy neutrinos). 
Also, the case with $m_{H_2} > 2 \,m_{H_1}$ offers the chance to investigate new decay channels with multi-scalars, an important 
hallmark of the mechanism responsible for our extended spontaneous  EWSB. Here, we build on the results presented in \cite{Basso:2010yz} for a pure $B-L$ scenario taking into account the new exclusion data from Higgs searches and the impact of the gauge mixing coupling $\tilde g$.
\subsubsection{Standard production mechanisms}
\begin{figure}
\centering
\subfigure[]{\includegraphics[scale=0.8]{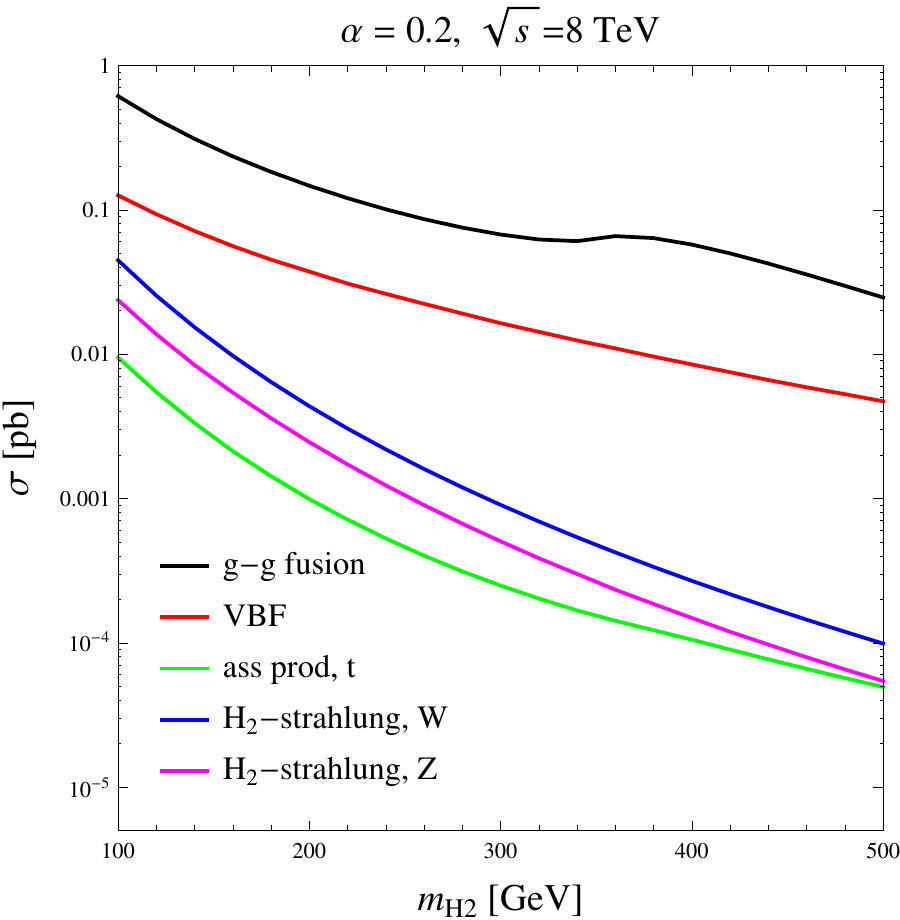}}
\subfigure[]{\includegraphics[scale=0.8]{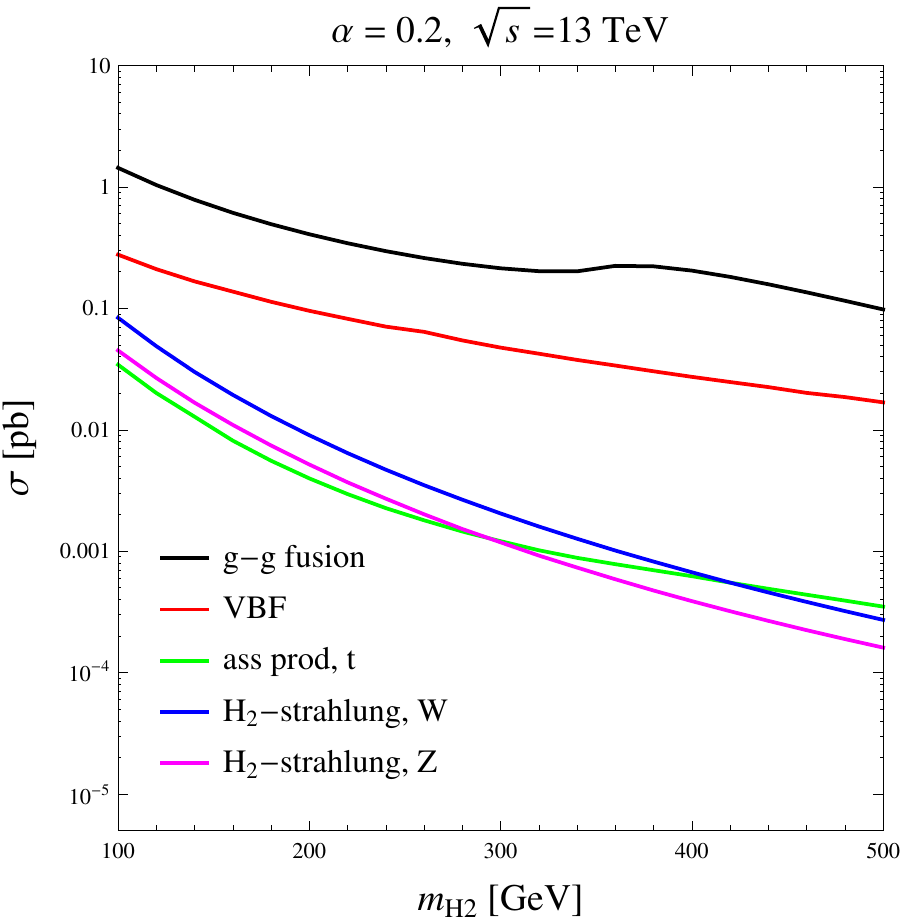}}
\caption{ Heavy-Higgs production cross sections at the LHC for $\sqrt{s} = 8$ TeV (a) and $\sqrt{s} = 13$ TeV (b) CM energy as a function of the $H_2$ mass for $\alpha = 0.2$. \label{Higgs2XS}}
\end{figure}

The most important set of mechanisms exploited to reveal the SM-like $125.09$ GeV Higgs boson at LHC involve gluon-gluon fusion, vector-boson fusion, 
$t \bar{t}$ associated production and Higgs-strahlung. 
The cross sections for these standard production channels of the light scalar $H_1$, which we assume to coincide with the 125.09 GeV Higgs boson, can be simply obtained from the SM results by a rescaling with a $\cos^2 \alpha$ factor. Here, instead, we present in fig.~\ref{Higgs2XS} the cross sections related to such processes for the case of the heavy scalar Higgs ($H_2$)
production as function of its mass and with the benchmark value of the mixing angle $\alpha=0.2$ for $\sqrt{s} = 8$  and $\sqrt{s} = 13$ TeV as CM energy at the LHC. 
The hierarchy of the cross sections is the same as for the SM Higgs case, the $H_2$ couplings to SM particles being rescaled by a factor of $\sin \alpha$. 
\subsubsection{Non-standard production mechanisms}
\begin{figure}
\centering
\subfigure[]{\includegraphics[scale=0.81]{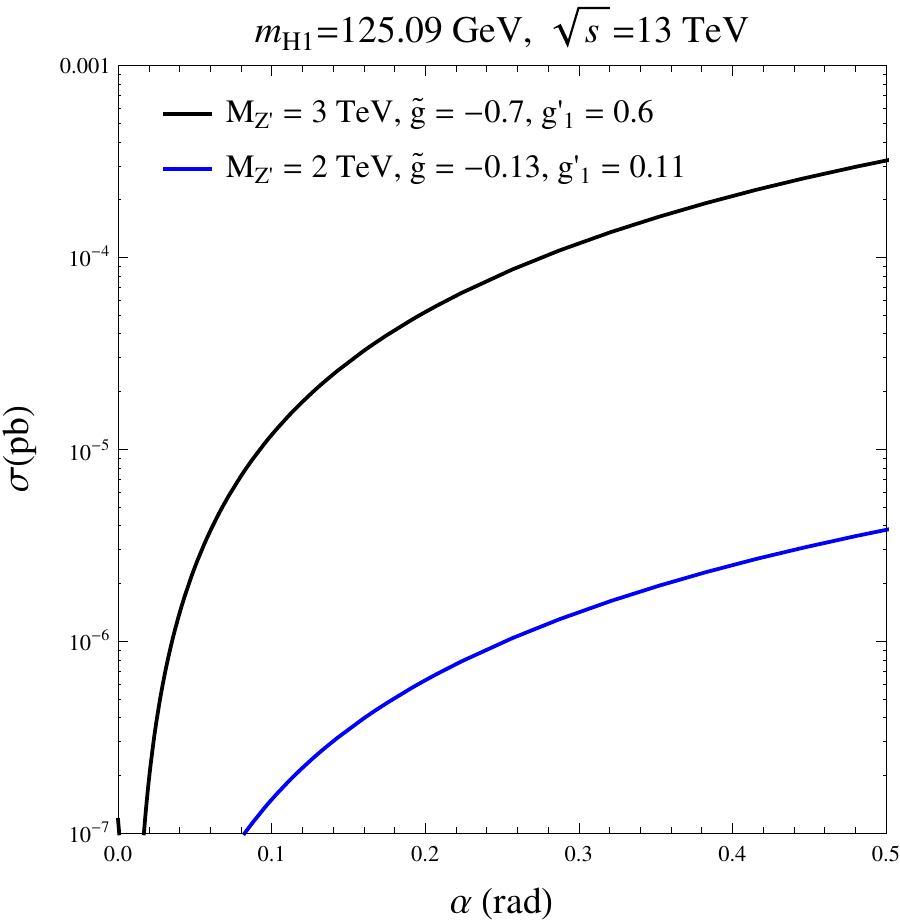}}
\subfigure[]{\includegraphics[scale=0.59]{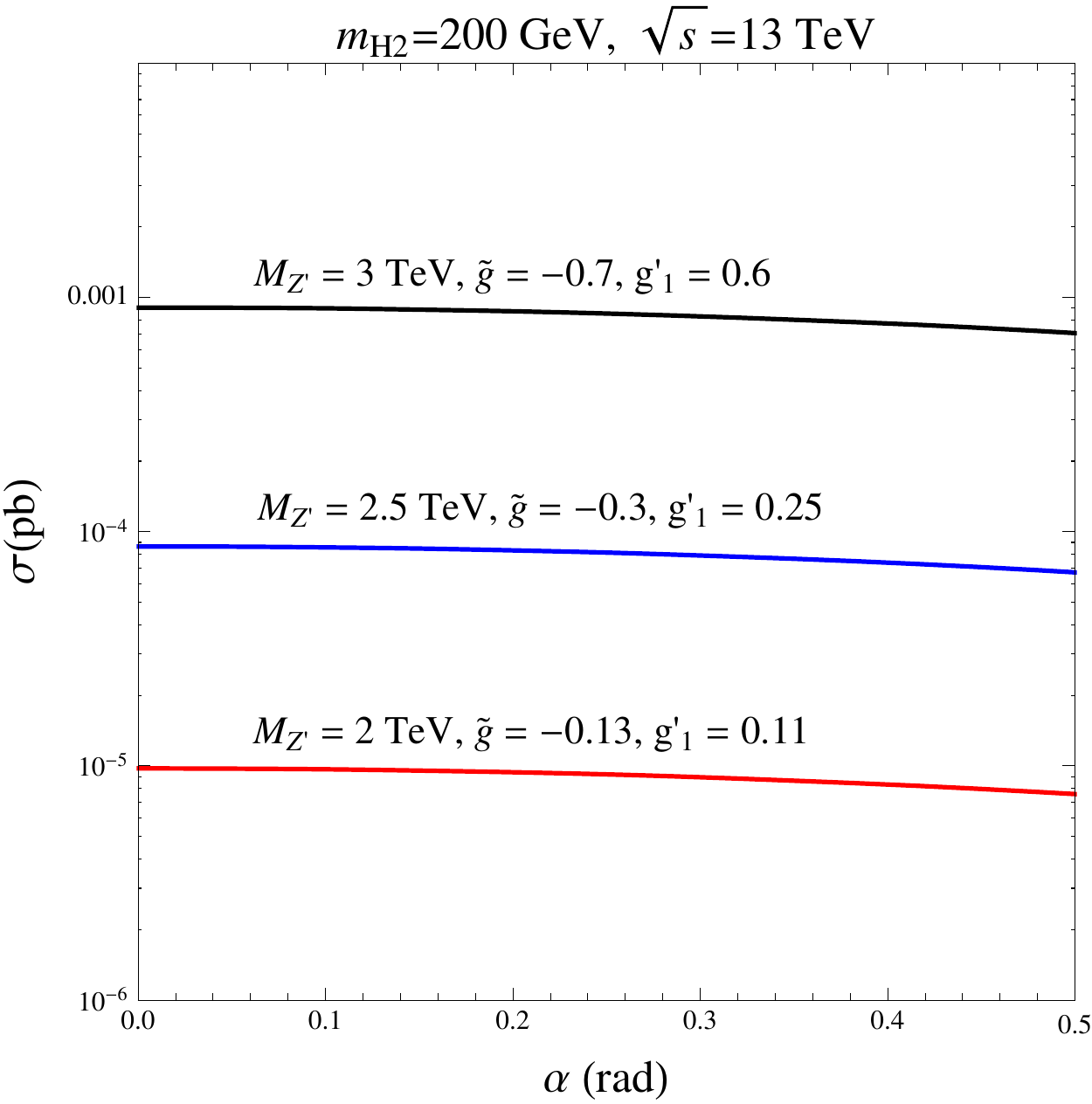}}
\caption{ Coss sections for associated production of the light  (a) and heavy (b) Higgs boson with the $Z'$ as a function of the scalar mixing angle $\alpha$ in the $0 \le \alpha \le 0.5$ range for different values of $M_{Z'}$ and  gauge couplings.  \label{Higgs2XSZpStrah}}
\end{figure}

The connection of the extended scalar sector with the remaining particles allows for new mechanisms for  heavy Higgs production. 
Among these, the associated production with the $Z'$ boson is of great importance, opening a window towards the $U(1)'$-specific spectrum.
In fig.~\ref{Higgs2XSZpStrah} we plot the variation of the cross section for the process $q \bar{q}\rightarrow Z'^*\rightarrow Z' H_{1,2}$ with respect to the scalar mixing angle. 
A fixed value of the heavy scalar mass has been taken and different benchmarks of $Z'$ mass and couplings have been considered. Notice that, due the $Z-Z'$ mixing, the same final state can be obtained with a $Z$ exchanged in the $s$-channel. (We have verified that this contribution and its interference with the $Z'$ diagram are non-negligible). 
The influence of the gauge sector in this production mechanism is translated in the enhancing effect from the Abelian gauge couplings and leads to a maximum value of $\sigma = 1$ fb.
Despite the small cross section, this is the only accessible production channel for $H_2$ when $\alpha = 0$. The ensuing
couplings have been chosen appropriately within the 95\% CL area of fig.~\ref{EWPTvsDY} and compensate for the dumping effect in the cross section due to the increasing $Z'$ mass.
\subsubsection{BRs and widths of the Higgs bosons}
\begin{figure}
\centering
\subfigure[]{\includegraphics[scale=0.59]{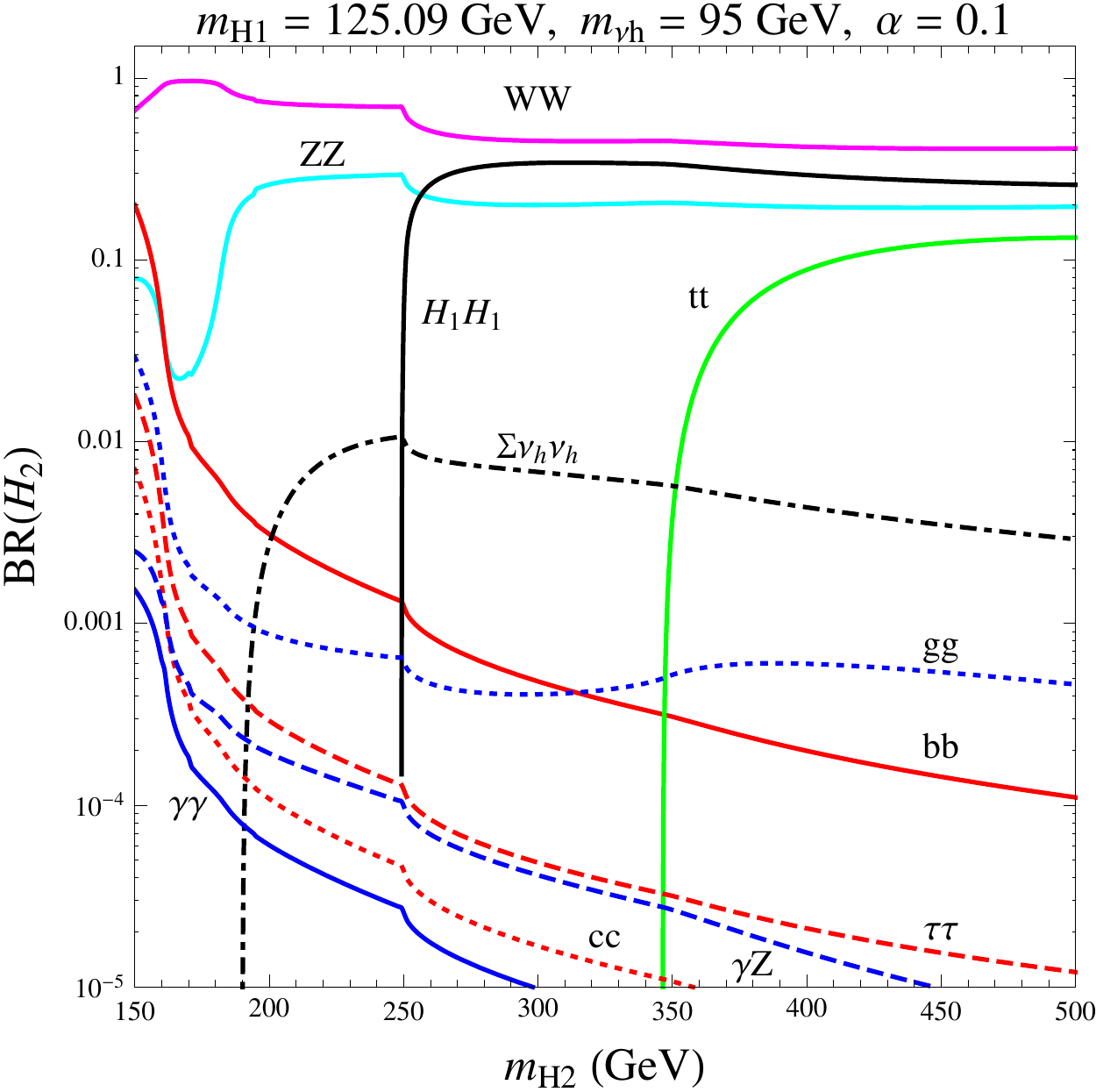}}
\subfigure[]{\includegraphics[scale=0.59]{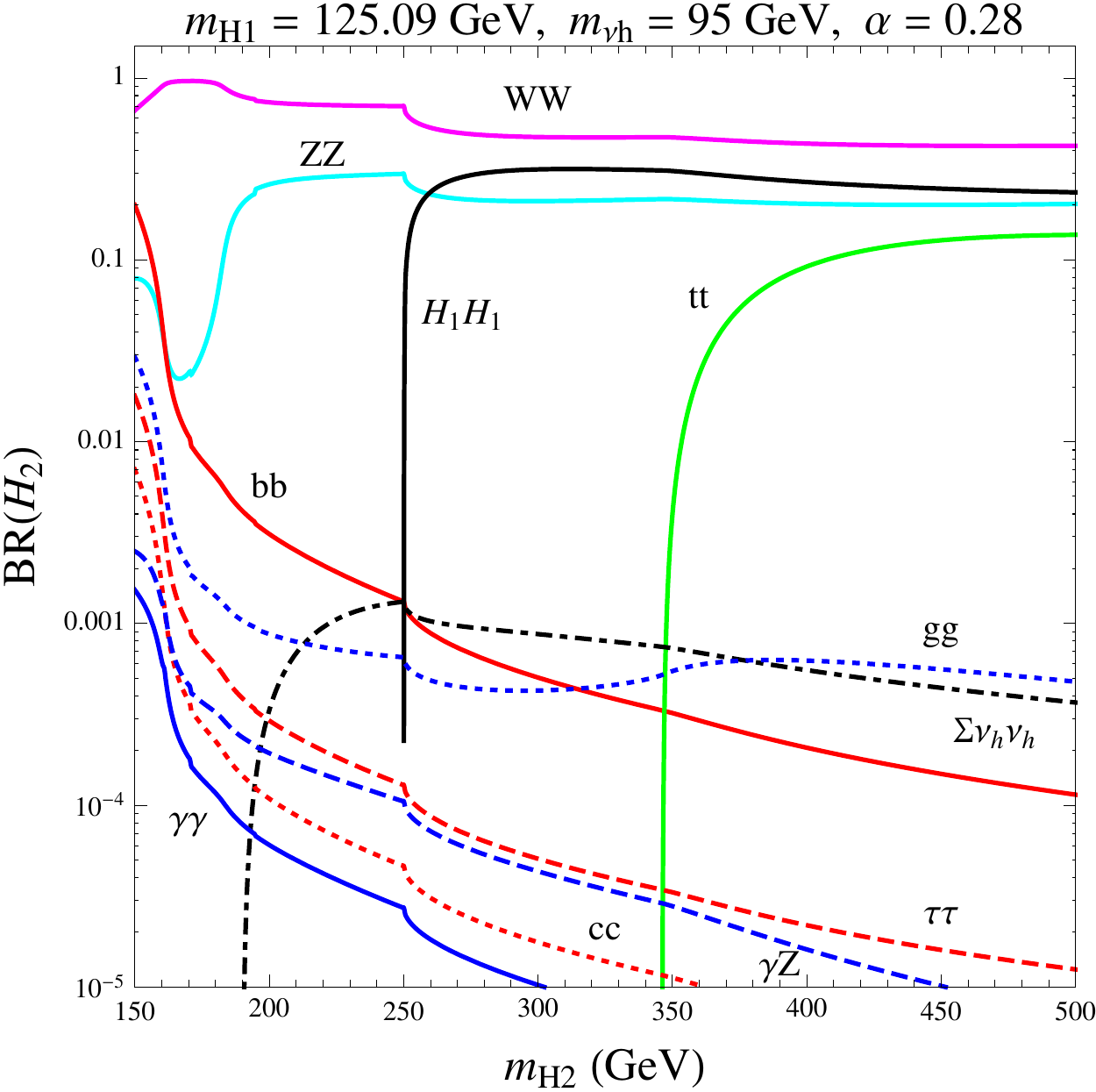}}
\caption{BRs of $H_2$ for (a) $\alpha = 0.1$ and (b) $\alpha = 0.28$. The other parameters are chosen as follow: $m_{H_1} = 125.09$ GeV, $m_{\nu_h} = 95$ GeV, $M_{Z'} = 2$ TeV, $g_1' = 0.11$ and $\tilde g = - 0.13$. \label{HiggsDecaysMass}}
\end{figure}

We now move to the investigation of the various decay modes of $H_2$ in two particle final states and the role played by the related unknown parameter space. 
We begin by  studying the variation of the branchings of $H_2$ for a change of its mass in the range 150--500 GeV. 
Two benchmark points have been considered with two assignments of the scalar mixing angle, consistent with the bounds extracted from Higgs searches, and a common value for the heavy neutrino and $Z'$ masses, as for the Abelian gauge couplings set at $g_1' = 0.11$ and $\tilde g = - 0.13$. The resulting BRs  are shown in fig.~\ref{HiggsDecaysMass}. 
With respect to the SM case, new decay channels are accessible, namely, the $H_2 \rightarrow H_1 H_1$ and $H_2 \rightarrow \nu_h \nu_h$, the former almost ubiquitous in many extensions of the scalar sector, the latter being a hallmark of  $U(1)'$ scenarios. For both  values of the mixing considered, $\alpha = 0.1$ and 0.28, the main channel is represented
by the decay into charged gauge boson, a predominance which is weakly challenged only by the decay in two $Z$s  and, when overcomes the threshold at $m_{H_2} = 250$ GeV, by the one in two light scalars.
Indeed, the hierarchy of the different decay modes in SM final states is the same as that of the SM Higgs, the partial decay widths being rescaled by a factor of $\sin^2 \alpha$.
The scalar mixing enters critically in the BRs into heavy neutrinos. When the corresponding kinematical region is allowed, it is evident that a heavy Higgs
$H_2$ mainly projecting onto the SM scalar singlet (for smaller values of $\alpha$) has, in our model, a stronger interaction with the heavy neutrinos, and, at the same time, a weaker coupling to SM particles. The corresponding BR endures a one order of magnitude suppression when $\alpha$ is raised to 0.28.

\begin{figure}
\centering
\subfigure[]{\includegraphics[scale=0.5]{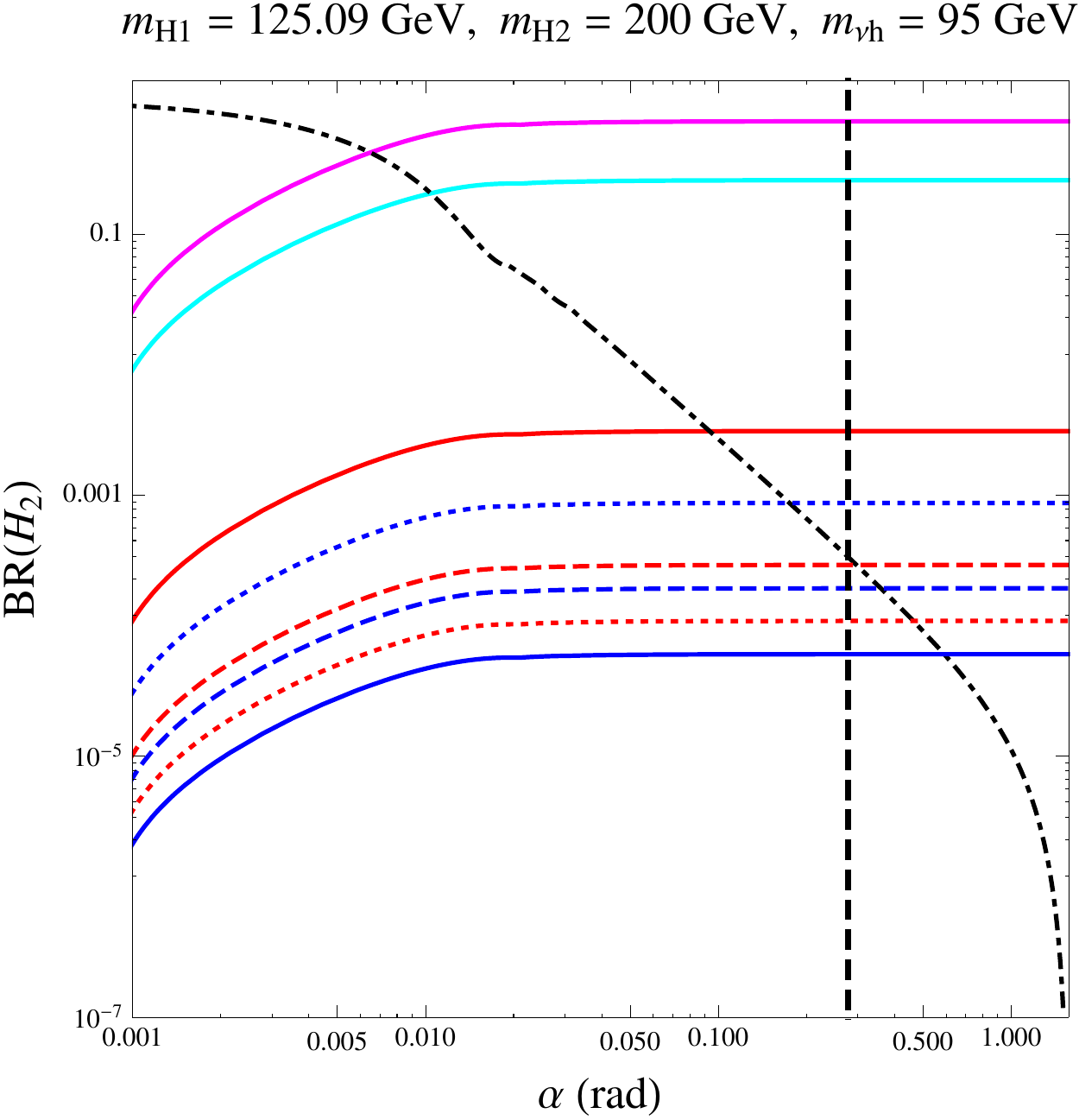}}
\subfigure[]{\includegraphics[scale=0.5]{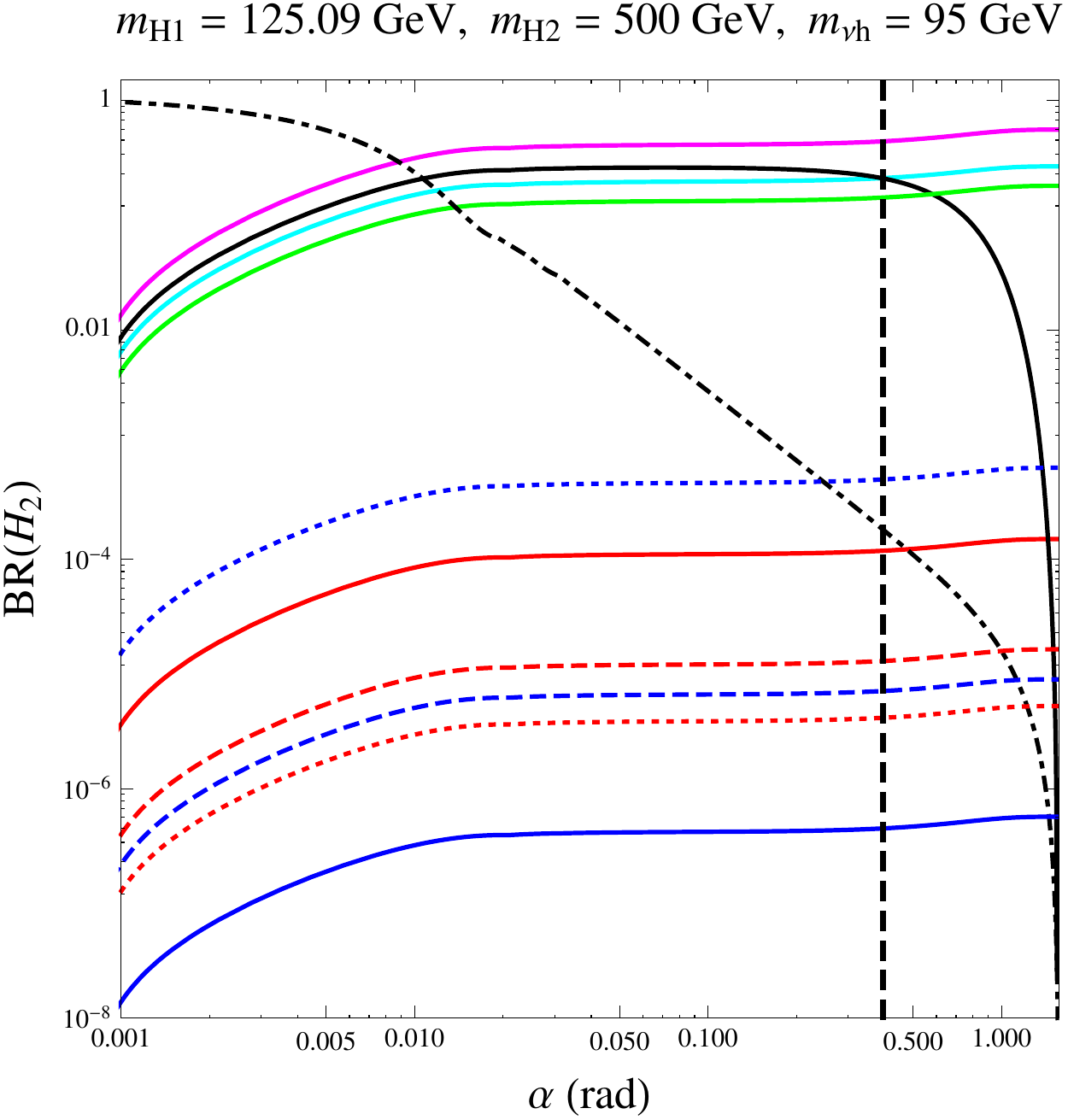}}
\subfigure[]{\includegraphics[scale=0.4]{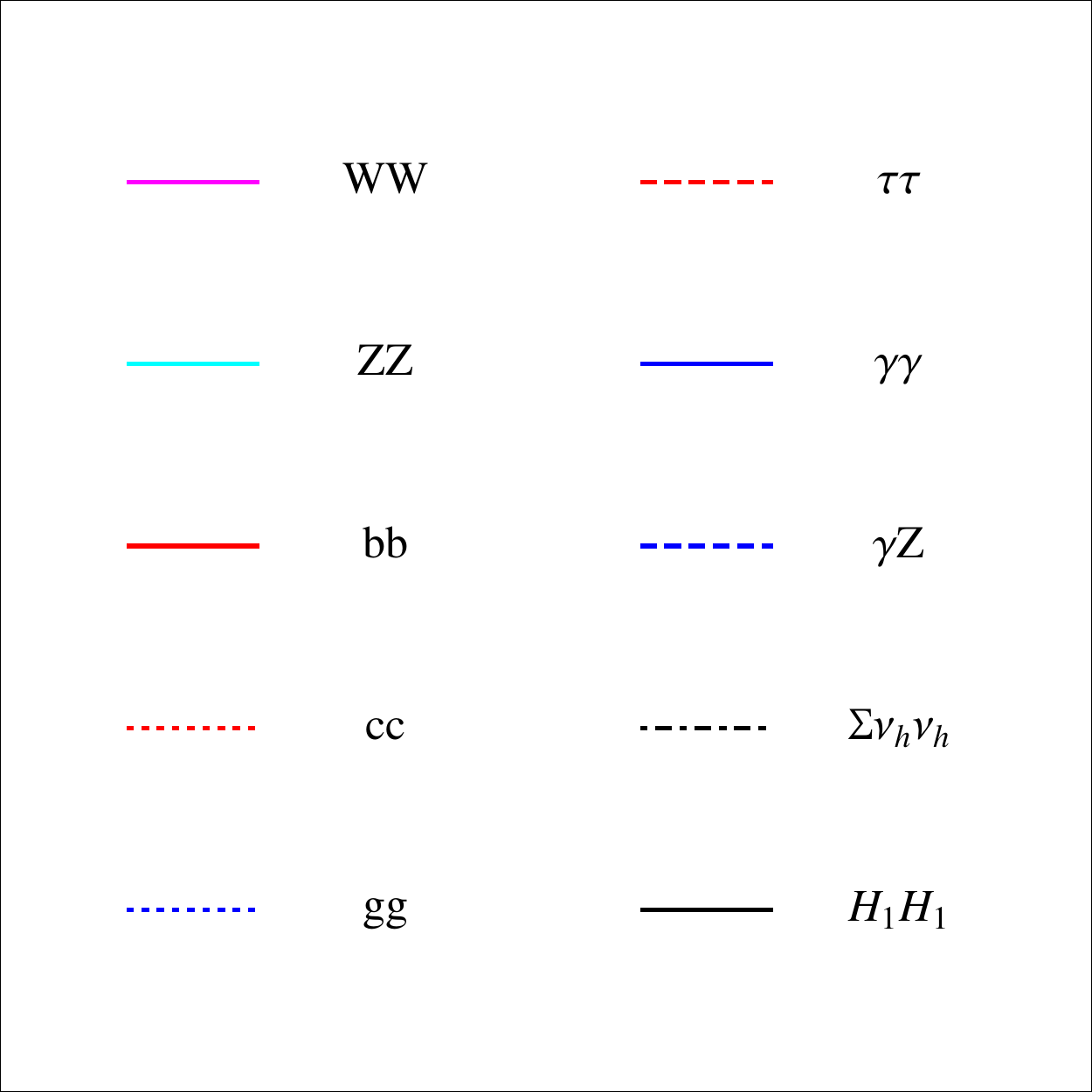}}
\caption{BRs of $H_2$ for (a) $m_{H_2} = 200$ GeV and (b) $m_{H_2} = 500$ GeV. The other parameters are chosen as follow: $m_{H_1} = 125.09$ GeV, $m_{\nu_h} = 95$ GeV, $M_{Z'} = 2$ TeV, $g_1' = 0.11$ and $\tilde g = -0.13$. The corresponding legend is depicted in (c). The regions on the right of the vertical dashed lines are excluded by \texttt{HiggsBounds}. \label{HiggsDecaysAlpha}}
\end{figure}

In fig.~\ref{HiggsDecaysAlpha} we show the $H_2$ BRs as a function of the scalar mixing angle for two values of its mass in order to 
explore different kinematical regions. Indeed, moving from the $m_{H_2} = 200$ GeV to the $m_{H_2} = 500$ GeV case, the decays in a top quark pair and in two $H_1$ become accessible. 
As mentioned before, the role of $\alpha$, for the interaction structure of our model, is clarified by the interplay between the decay in heavy neutrino and the other modes. 
In both  cases shown in fig.~\ref{HiggsDecaysAlpha} the increase in $\alpha$ causes the dropping of the heavy neutrino decay mode and a growth of the SM-like decay channels. Notice also that the $H_2 \rightarrow H_1 H_1$ mode does not have a trivial dependence on $\alpha$.

\begin{figure}
\centering
\subfigure[]{\includegraphics[scale=0.8]{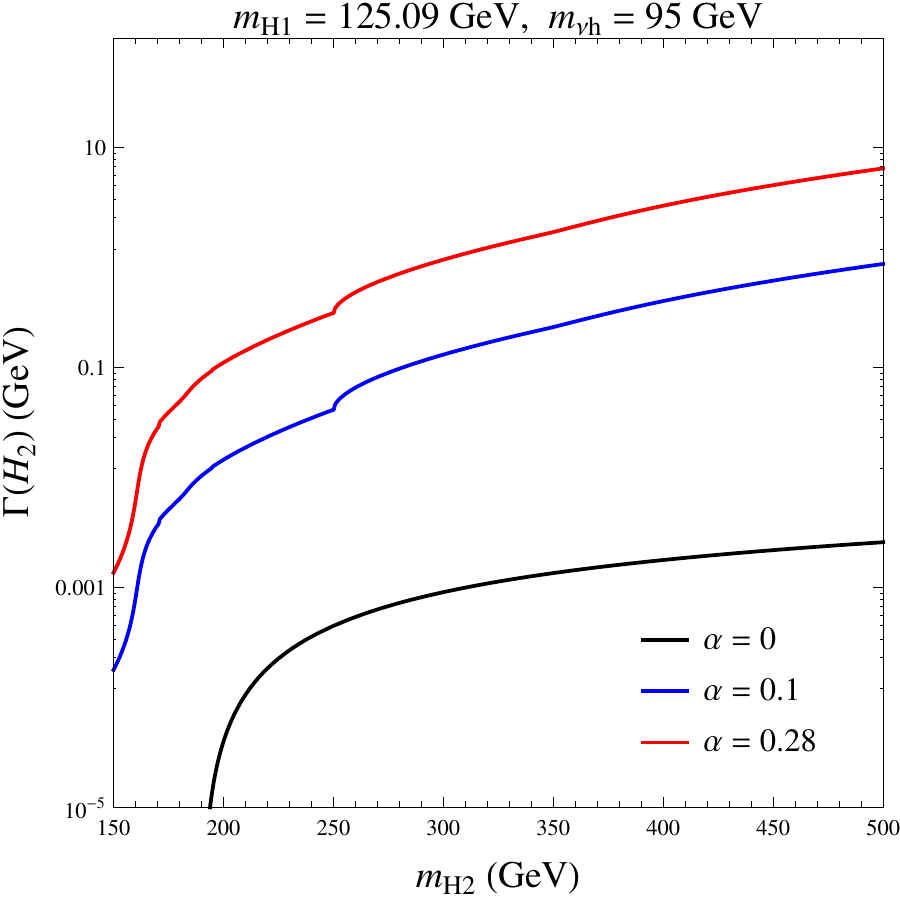}}
\subfigure[]{\includegraphics[scale=0.8]{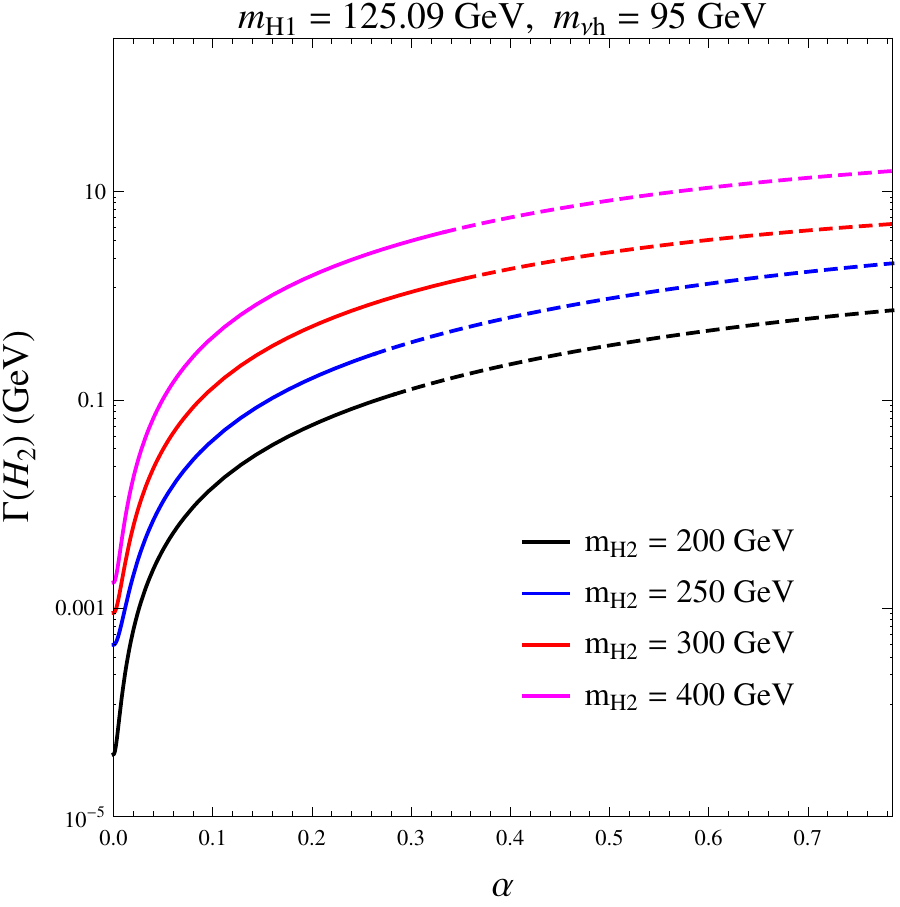}}
\caption{The $H_2$ total decay width as a function of $m_{H_2}$ for different values of $\alpha$ (a) and as a function of $\alpha$ for different values of $m_{H_2}$ (b). The other parameters are chosen as follow: $m_{H_1} = 125.09$ GeV, $m_{\nu_h} = 95$ GeV, $M_{Z'} = 2$ TeV, $g_1' = 0.11$ and $\tilde g = - 0.13$. In fig. (b) the dashed parts of the curves are excluded by the \texttt{HiggsBounds} analysis. \label{HiggsTotalWidth}}
\end{figure}

In fig.~\ref{HiggsTotalWidth} the dependence on $\alpha$ and $m_{H_2}$ of the heavy Higgs total width is illustrated. In fig.~\ref{HiggsTotalWidth}(a) the heavy scalar
masses were allowed to span in the range 150 GeV $\le m_{H_2} \le$ 500 GeV while three different assignments $\alpha=0, 0.1, 0.28$ have been considered. 
The case with zero mixing singles out in showing a recognisable threshold due to the heavy neutrino decay being the only allowed channel. 
The values of the width  rapidly grow when such threshold is exceeded reaching the MeV order. Further, with
the increase of the scalar mixing the width experiences another sizeable growth due to the now open SM decay channels. Also for such cases the channel $H_2 \rightarrow H_1 H_1$
is available resulting in a mild threshold in the width plot. We can appreciate how the non-zero mixing causes a large increment in the width allowing values of order GeV to be
reached for high $m_{H_2}$ values.
The critical role of the scalar mixing angle is more visible in fig.~\ref{HiggsTotalWidth}(b) where we considered the variation of the width respect
to $\alpha$ in the range $0 \le \alpha \le 0.8$. For the given choices of $m_{H_2}$, the constraints coming from Higgs searches at the LHC have been taken into account 
excluding a large sector (dashed lines) of the values of $\alpha$ in the plot.  

\begin{figure}
\centering
\subfigure[]{\includegraphics[scale=0.8]{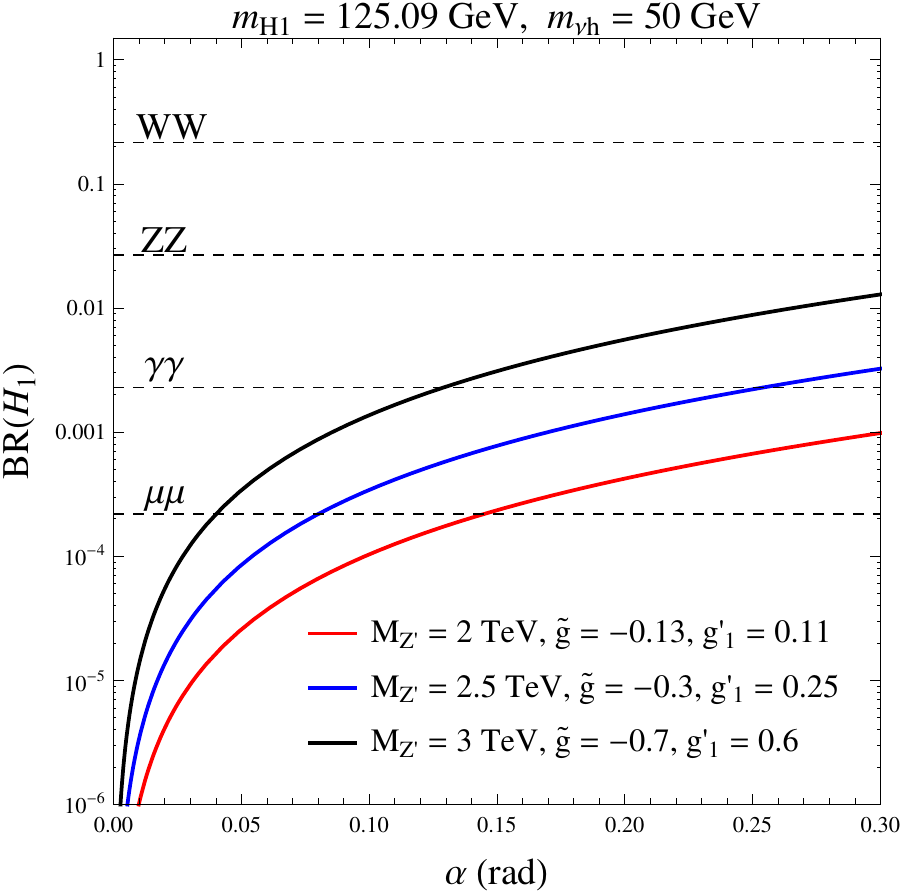}}
\subfigure[]{\includegraphics[scale=0.8]{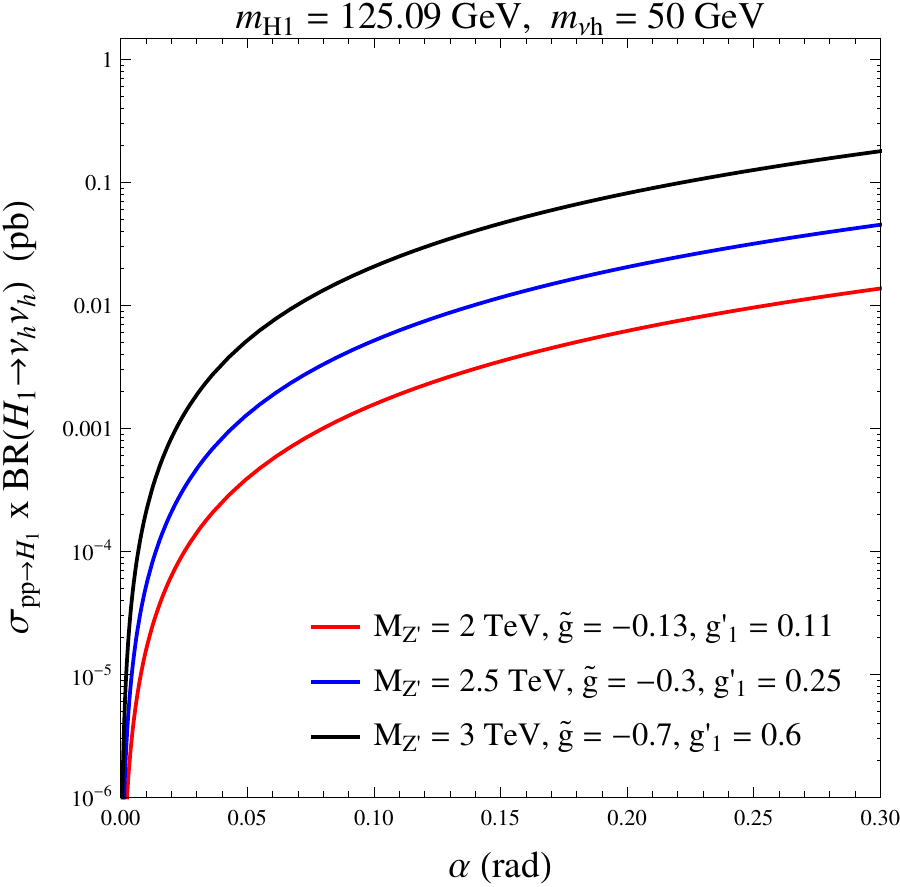}}
\caption{(a) Branching ratio of the $H_1 \rightarrow \nu_h \nu_h$ mode. For reference some of the SM Higgs branching ratios are shown with dashed lines. (b) Cross section times BR for the process $pp \rightarrow H_1 \rightarrow \nu_h \nu_h$ at the LHC with $\sqrt{s} = 13$ TeV. Only the gluon fusion channel has been considered. In both plots we have chosen $m_{\nu_h} = 50$ GeV and different assignments of $M_{Z'}$ and the gauge couplings. \label{BRH1heavyNu}}
\end{figure}

Let us now turn to the decay patterns of the SM-like Higgs state, $H_1$.
When $m_{H_1} > 2 m_{\nu_h}$ a new interesting channel become accessible to it, $H_1\to \nu_h \nu_h$ (into heavy neutrinos), otherwise it behaves as the SM Higgs boson, with the same BRs and a total width rescaled by a factor of $\cos^2 \alpha$. We show in fig.~\ref{BRH1heavyNu}(a) the light Higgs decay mode into a pair of heavy neutrinos for $m_{\nu_h} = 50$ GeV and for three different benchmark points. For comparison we also show the BRs of some decay channel of the SM Higgs boson. Quite interestingly the neutrino BR spans from 0.1\% to 1\% becoming comparable to,
or even exceeding, the $\gamma\gamma$ mode of the SM Higgs. The behaviour of the depicted curves can be understood
by  scrutinising the structure of the $H_1 \nu_h \nu_h$ vertex. This is  proportional to $\sin \alpha ( m_{\nu_h} / x) \sim \sin \alpha \, g'_1 ( m_{\nu_h} / M_{Z'})$ and therefore, for fixed $m_{\nu_h}$, can be increased by growing the ratio $g'_1/M_{Z'}$. Taking into account the LHC limits on the Abelian gauge couplings discussed in section ~\ref{sec:bounds}, which are obviously more constraining for lower $Z'$ masses, we find a bigger ratio for $M_{Z'} = 3$ TeV, in which case $g'_1$ is allowed to vary up to 0.6. For completeness, we depict in fig.~\ref{BRH1heavyNu}(b) the $\sigma \times$ BR
values for the process $pp \rightarrow H_1 \rightarrow \nu_h \nu_h$ at the LHC with $13$ TeV CM energy, which can reach 100 fb. Notice that the $H_1$ production cross section scales with a factor of $\cos^2 \alpha$ with respect to the SM case, which is reproduced by a vanishing scalar mixing angle. In such case $\sigma({gg \rightarrow H_1}) = 44.08$ pb \cite{LHCHXSWG} which has been used to normalise our cross section.
\subsubsection{Hallmark LHC signatures from $U(1)'$ Higgs states}
\begin{figure}
\centering
\subfigure[]{\includegraphics[scale=0.8]{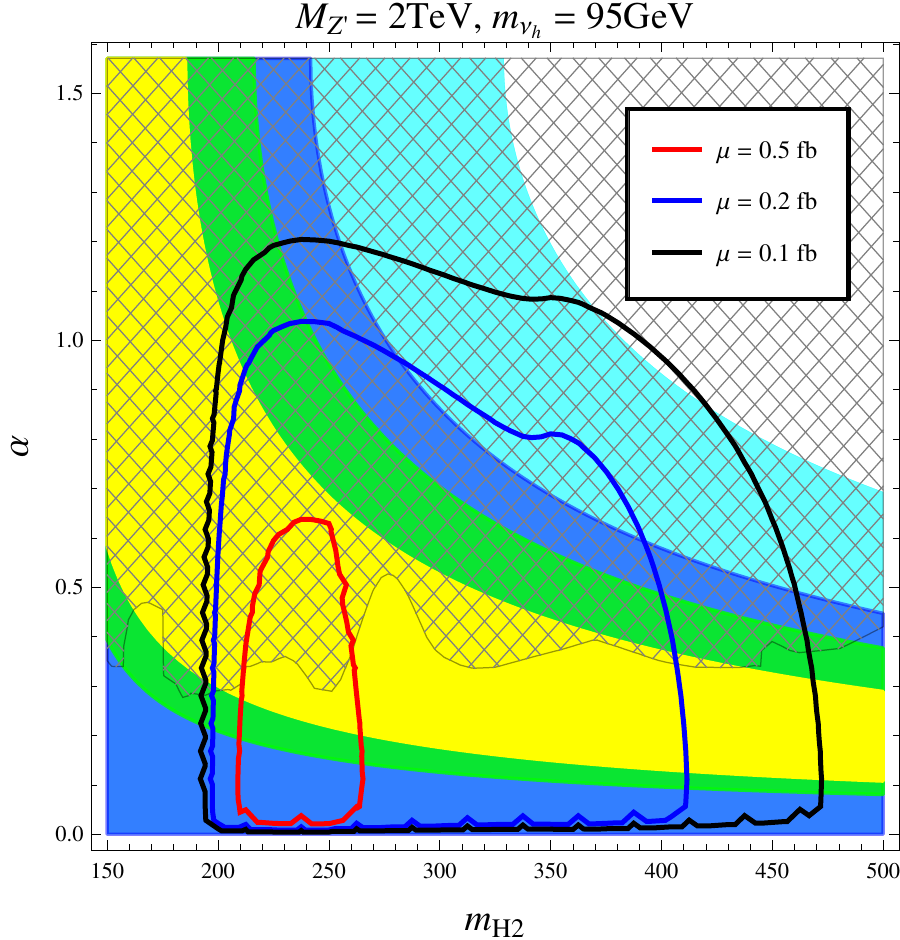}}
\subfigure[]{\includegraphics[scale=0.8]{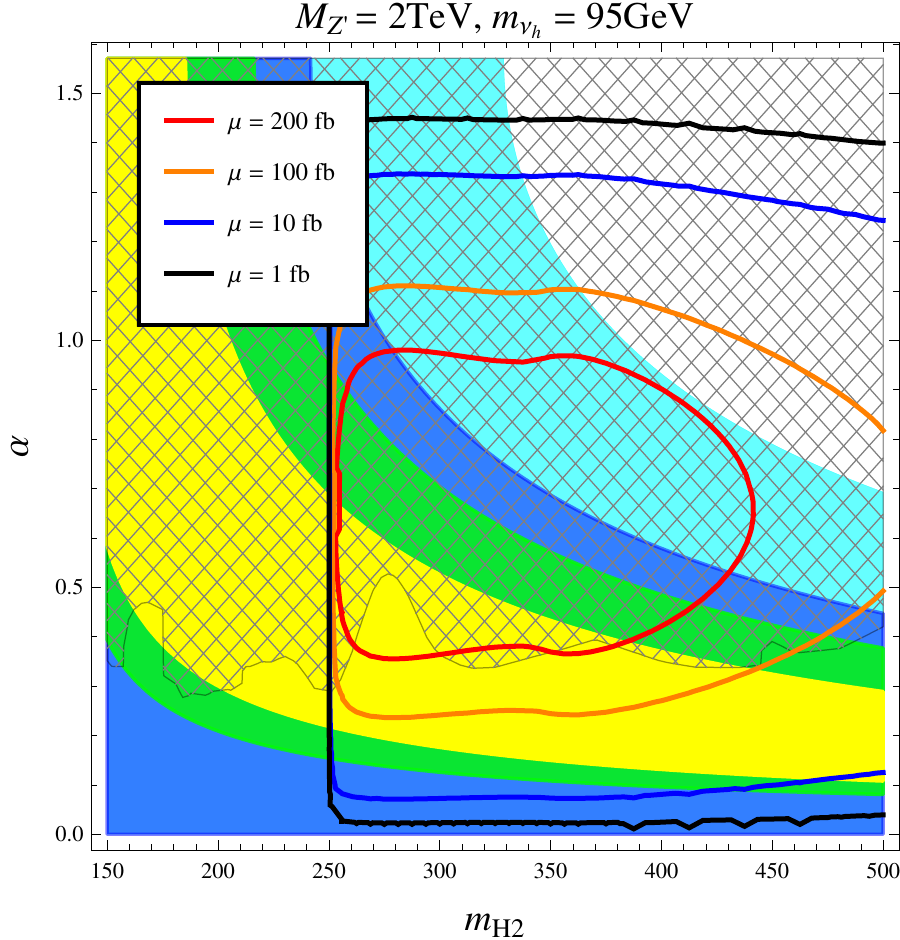}}
\caption{Contour plots of the cross section times BR for the processes $pp \rightarrow H_2 \rightarrow \nu_h \nu_h$ (a) and $pp \rightarrow H_2 \rightarrow H_1 H_1$ (b) at the LHC with $\sqrt{s} = 13$ TeV in the $(m_{H_2}, \alpha)$ plane. Only the gluon fusion channel has been considered. The parameters have been chosen as follows: $M_{Z'} = 2$ TeV, $\tilde g = -0.13$ and $g_1' = 0.11$. \label{HiggsSignatures}}
\end{figure}

\begin{figure}
\centering
\subfigure[]{\includegraphics[scale=0.8]{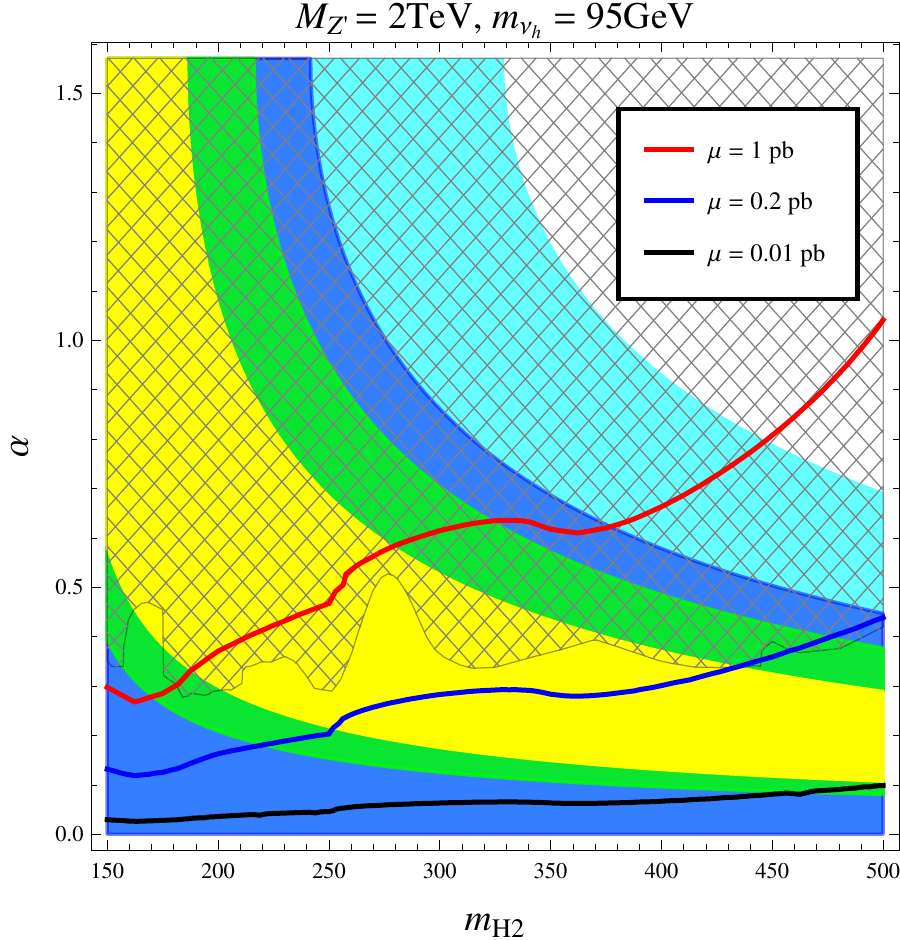}}
\subfigure[]{\includegraphics[scale=0.8]{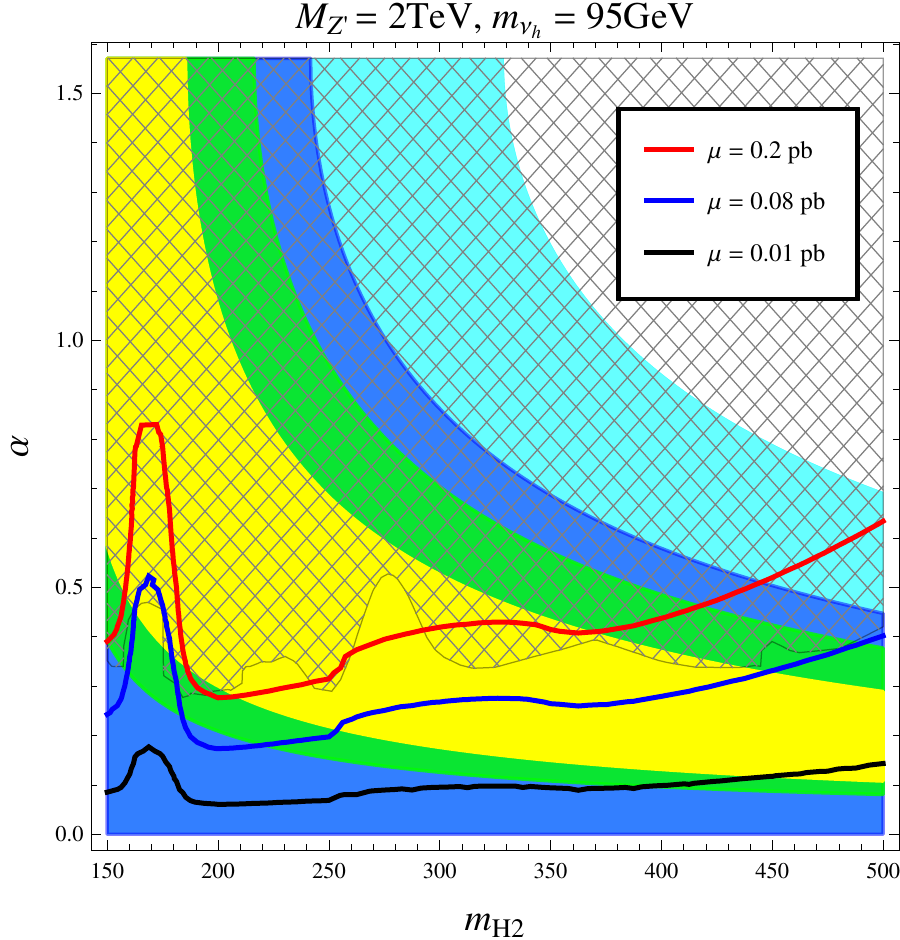}}
\subfigure[]{\includegraphics[scale=0.8]{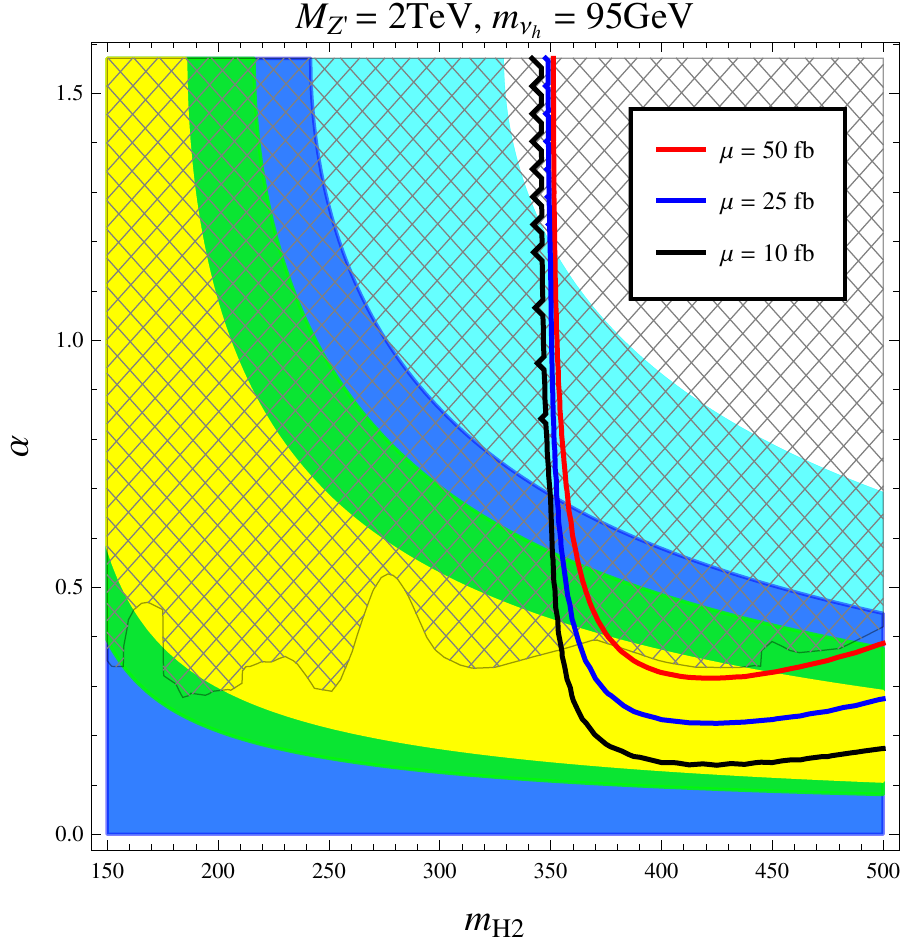}}
\caption{Contour plots of the cross section times BR for the processes $pp \rightarrow H_2 \rightarrow WW$ (a), $pp \rightarrow H_2 \rightarrow ZZ$ (b) and $pp \rightarrow H_2 \rightarrow t \bar{t}$ (c) at the LHC with $\sqrt{s} = 13$ TeV in the $(m_{H_2}, \alpha)$ plane. Only the gluon fusion channel has been considered. The parameters have been chosen as follows: $M_{Z'} = 2$ TeV, $\tilde g = -0.13$ and $g_1' = 0.11$.\label{HiggsWZt}}
\end{figure}

The production cross sections and decay BRs of $H_2$ can be combined with the recent limits, coming from LHC search on the extended Higgs sector, to probe realistic 
discovery opportunities. Our phenomenological scenario calls for a $\sqrt{\sigma} = 13$ TeV CM energy and an integrated luminosity of 100 fb$^{-1}$, as expected
to be collected at LHC. 
From what has been illustrated in the previous analysis, the heavy scalar decay can reveal its presence and that of the remaining beyond-SM spectrum 
through peculiar decay channels. Such distinctive signatures involve heavy neutrinos and light scalars. Considering production from gluon-gluon fusion we 
 project in the $(m_{H_2},\alpha)$ plane, fig.~\ref{HiggsSignatures}(a), the contour of equal value for the cross section times BR of the process 
$pp \rightarrow H_2 \rightarrow \nu_h \nu_h$. 
We kept the heavy neutrinos at a common degenerate mass of 95 GeV, summing the final state over generations, and considered the benchmark point in the 
extended gauge sector with $M_{Z'} =$  2 TeV, $\tilde g = -0.13$ and $g_1' = 0.11$. The values of $\sigma=0.1$,0.2 fb and 0.5 fb illustrate the magnitude involved and 
the number of neutrino events that can be expected. We crossed the results with the stability/perturbativity implications of a given choice of the parameter space.
We notice how the request to exceed 50 events selects a restricted area of the heavy scalar mass, 
roughly 200 GeV $\le m_{H_2} \le$ 250 GeV, with values of the scalar mixing not excluded (hatched area) by LHC data. The same area covers a region with a scale of 
stability/perturbativity breaking greater than the SM case.
A more generous response is obtained when the gluon-gluon cross section is multiplied for the branching of $H_2 \rightarrow H_1 H_1$. 
In fig.~\ref{HiggsSignatures}(b) are drawn, for the latter process and the same setting of masses and gauge parameters of the previous figure, 
the contours with $\sigma=$ 0.1, 1, 100 and 200 fb. Above the threshold $m_{H_2} = 250$ GeV,
the scalar mixing angle can critically raise the value of $\sigma$ leading potentially to $\sim$100 events. 
The LHC limits severely intervene to exclude large value of $\alpha$ with the resulting effect of an upper bound of $\sim$200 events in the space investigated. \\
The $H_2$ decay in light scalars or heavy neutrinos states represents a peculiar feature of our minimal class of $Z'$ models, nevertheless
a search aimed to a  heavy scalar discovery would favour different channels. 
From the previous analysis of the BRs (see figs.~\ref{HiggsDecaysMass}-\ref{HiggsDecaysAlpha}), $H_2$ decays in $WW$, $ZZ$ and $t\bar{t}$ are
the main candidates as search channels. Consequently, we proceed by testing the gluon-gluon induced cross section of such channels against the LHC exclusion limits in fig.~\ref{HiggsWZt}. 
The corresponding contours of equal value for the cross section of $pp \rightarrow H_2 \rightarrow WW$ and  $pp \rightarrow H_2 \rightarrow ZZ$ are illustrated
in figs.~\ref{HiggsWZt}a-b. The two cases share the absence of a threshold in the interval of $m_{H_2}$ considered  
and a cross section increasing with the scalar mixing. At the highest values of mixing allowed the $WW$ decay is more capable to get close to 1 pb while the $ZZ$ decay has a 
weaker growth as can be read off from the path of the line $\sigma = 0.2$ pb. The process $pp \rightarrow H_2 \rightarrow t \bar{t}$ completes our survey. The threshold is 
sufficiently high to concern only a small section of the $(m_{H_2}, \alpha)$ plane. The values of the cross section times BR depicted are for $\sigma =$ 10,25,50 fb. \\

\section{Conclusions and outlook}
\label{sec:conclusions}

In summary, we have shown how production and decay patterns peculiar to a class of $U(1)'$ models (of which we have taken the $B-L$ case as an example) involving the entirity of their additional particle spectrum, i.e.,  heavy neutrinos, a second Higgs state and a $Z'$, at times interplaying with each other in experimental signatures accessible at the second stage of the LHC, can be linked to the high scale behaviour of such scenarios. This has been made possible by combining the description of their low and high energy dynamics through an advanced 
RG analysis which specifically used as boundary conditions only those potentially
accessible by experiment at present and in the near future. 

Our study has  so far been confined to the inclusive level only, as no dedicated MC analysis has
been attempted yet to uncover the interesting $U(1)'$ signatures at the LHC and extract from these the values of the fundamental
parameters defining these scenarios. Indeed, building on the results obtained here, we intend to carry out this task in
a forthcoming publication which, once accomplished, will translate into the ability to use LHC data in order to open a window on high-scale physics, possibly providing circumstantial evidence of its ultimate structure. In fact, our initial study has already made clear that several $U(1)'$ signatures potentially accessible at the CERN machine during Run 2 cover a sizeable region of the parameter space which points towards a well-behaved scenario, stable and perturbative up to the GUT scale, thus supporting these models as viable extensions of the SM.

\acknowledgments

L.D.R. thanks L. Basso for useful discussions during the development of this manuscript.
E.A., J.F. and S.M. are supported in part through the NExT Institute. C.C. thanks The Leverhulme Trust for a
Visiting Professorship to the University of Southampton, where part of his work was carried out.
The work of L.D.R. has been supported by the "Angelo Della Riccia'' foundation.

\providecommand{\href}[2]{#2}\begingroup\raggedright\endgroup

\end{document}